\documentclass[12pt]{article}

\usepackage{latexsym}
\usepackage{amssymb,amsfonts,amsmath}
\usepackage{graphicx} 
\usepackage{indentfirst}
\usepackage{bbm}
\usepackage{amssymb}
\usepackage{verbatim}
\usepackage{amsmath, amsthm,amssymb}
\usepackage{mathrsfs}
\usepackage{hyperref}
\usepackage{amsfonts}
\usepackage{dsfont}
\usepackage{cite}
\usepackage{xcolor}
\usepackage[multiple]{footmisc}

\parskip=\medskipamount

\parskip=\medskipamount

\usepackage[mathscr]{euscript}

\newcommand{\de}{{\nabla}}

\numberwithin{equation}{section}

\newcommand {\cC}{{\cal C}}
\newcommand {\cD}{{\cal D}}

\newcommand {\cG}{{\cal G}}

\newcommand {\cK}{{\cal K}}

\newcommand {\cM}{{\cal M}}
\newcommand {\cN}{{\cal N}}
\newcommand {\cO}{{\cal O}}

\newcommand {\cR}{{\cal R}}

\newcommand {\cW}{{\cal W}}

\newcommand{\bD}{{\bf D}}

\newcommand{\bR}{{\bf R}}

\newcommand{\bT}{{\bf T}}

\newcommand{\scT}{{\mathscr{T}}}
\newcommand{\bbD}{\mathbb D}

\newcommand{\sRM}{\mathscr{R}(M)}
\newcommand{\sRD}{\mathscr{R}(\mathbb D)}

\newcommand{\sRJ}{\mathscr{R}(J)}
\newcommand{\sRS}{\mathscr{R}(S)}
\newcommand{\sRK}{\mathscr{R}(K)}

\def\a{\alpha}
\def\b{\beta}

\def\d{\delta}
\def\e{\epsilon}
\def\f{\phi}
\def\g{\gamma}
\def\G{\Gamma}

\def\m{\mu}

\def\o{\omega}

\def\q{\theta}
\def\r{\rho}
\def\s{\sigma}

\def\x{\xi}
\def\z{\zeta}

\def\F{\Phi}

\def\L{\Lambda}
\def\O{\Omega}

\def\U{\Upsilon}
\def\X{\Xi}

\def\ri{{\rm i}}
\def\re{{\rm e}}

\newcommand{\dbeta}{{\dot{\beta}}}
\newcommand{\dgamma}{{\dot{\gamma}}}
\newcommand{\ddelta}{{\dot{\delta}}}

\newcommand{\sSU}{\mathsf{SU}}

\newcommand{\sSO}{\mathsf{SO}}
\newcommand{\sU}{\mathsf{U}}

\newcommand{\sSpin}{\mathsf{Spin}}

\newcommand{\1}{{\underline{1}}}
\newcommand{\2}{{\underline{2}}}

\newcommand{\hal}{{\hat{\a}}}

\newcommand{\ve}{\varepsilon}

\newcommand{\pa}{\partial}
\newcommand{\hf}{\frac12}

\newcommand{\vf}{\varphi}

\newcommand{\be}{\begin{equation}}
\newcommand{\ee}{\end{equation}}
\newcommand{\bea}{\begin{eqnarray}}
\newcommand{\eea}{\end{eqnarray}}
\newcommand{\non}{\nonumber}
\newcommand{\ba}{\begin{array}}
\newcommand{\ea}{\end{array}}

\newcommand{\bm}[1]{\mbox{\boldmath$#1$}}

\def\double #1{#1{\hbox{\kern-2pt $#1$}}}

\newcommand{\bsubeq}{\begin{subequations}}
\newcommand{\esubeq}{\end{subequations}}

\newcommand{\rd}{\mathrm d}
\newcommand{\veps}{\varepsilon}

\newcommand{\eps}{\varepsilon}

\newcommand{\eol}{\notag \\}

\begin{document}
\begin{titlepage}
\begin{flushright}
UUITP-53/20\\
December, 2020\\
\end{flushright}

\begin{center}
{\Large \bf 
Symmetries
of $\bm{{\cal N} = (1,0)}$ 
supergravity backgrounds 
 in six dimensions}
\end{center}

\begin{center}

{\bf
Sergei M. Kuzenko${}^a$,
Ulf Lindstr\"om${}^{b,c}$,
Emmanouil S. N. Raptakis${}^{a}$ 
and 
Gabriele Tartaglino-Mazzucchelli${}^{d}$
} \\
\vspace{5mm}

\footnotesize{
${}^{a}${\it Department of Physics M013, The University of Western Australia,\\
35 Stirling Highway, Crawley W.A. 6009, Australia}}  
~\\
\vspace{2mm}
\footnotesize{
${}^{b}${\it
Department of Physics, Faculty of Arts and Sciences,\\
Middle East Technical University, 06800, Ankara, Turkey}}
~\\
\vspace{2mm}
\footnotesize{
${}^{c}${\it 
Department of Physics and Astronomy,\\
Division of Theoretical Physics, Uppsala University,\\
Box 516, SE-751 20 Uppsala, Sweden}}
~\\

\vspace{2mm}
\footnotesize{
${}^{d}${\it 
School of Mathematics and Physics, University of Queensland,
\\
 St Lucia, Brisbane, Queensland 4072, Australia}
}
\vspace{2mm}
~\\
\texttt{sergei.kuzenko@uwa.edu.au, 
ulf.lindstrom@physics.uu.se,
emmanouil.raptakis@research.uwa.edu.au,
g.tartaglino-mazzucchelli@uq.edu.au}\\
\vspace{2mm}

\end{center}

\begin{abstract}
\baselineskip=14pt
General
$\mathcal{N}=(1,0)$ supergravity-matter systems in six dimensions may be described using one of the two fully fledged superspace formulations for conformal supergravity: (i) $\sSU(2)$ superspace; and (ii) conformal superspace. With motivation to develop rigid supersymmetric field theories in curved space, this paper is devoted to the study of the geometric symmetries of supergravity backgrounds.
In particular, we introduce the notion of a conformal Killing spinor superfield $\e^\a$, which proves to generate extended superconformal transformations. Among its cousins are the conformal Killing vector $\xi^a$ and tensor $\z^{a(n)}$ superfields. The former parametrise conformal isometries of supergravity backgrounds, which in turn yield symmetries of every superconformal field theory. Meanwhile, the conformal Killing tensors of a given background are associated with higher symmetries 
of the hypermultiplet. By studying the higher symmetries of a non-conformal vector multiplet we introduce 
the concept of a Killing tensor superfield. We also analyse the problem of computing higher symmetries 
for the conformal d'Alembertian in curved space and demonstrate that, beyond the first-order case, these operators are defined only on conformally flat 
backgrounds.
\end{abstract}
\vspace{5mm}

\begin{center}
{\it Dedicated to Jim Gates on the occasion of his 70th birthday}
\end{center}

\vfill
\end{titlepage}

\newpage
\renewcommand{\thefootnote}{\arabic{footnote}}
\setcounter{footnote}{0}

\tableofcontents{}
\vspace{1cm}
\bigskip\hrule


\allowdisplaybreaks

\section{Introduction} \label{section1}

The  superconformal tensor calculus for $\cN=(1,0)$  supergravity 
in six dimensions was formulated by Bergshoeff, Sezgin and 
Van Proeyen in 1986 \cite{BSVanP}, as a natural generalisation of that for  
$d=4$,  $\cN=2$ supergravity
\cite{deWvHVP1,deRvHdeWVP,deWvHVP2,deWPVP,deWLPSVP,deWLVP}.
More recently it was further developed 
 \cite{CVanP,BCSVanP}, 
including the construction of the complete 
off-shell action for minimal Poincar\'e supergravity \cite{CVanP}
and a higher-derivative extension 
of chiral gauged supergravity \cite{BCSVanP}, see \cite{LVP} for a pedagogical review.

The  tensor calculus \cite{BSVanP} 
has found numerous applications, in particular the explicit construction of  
off-shell curvature squared supergravity actions 
  \cite{BSS1,BSS2,BR,BCSVanP}.
It  is a powerful approach to formulate supergravity-matter systems. However, similar to its $d=4$,  $\cN=2$ and $d=5$,  $\cN=1$ cousins, it has two limitations.
Firstly, it does not offer tools to describe off-shell charged hypermultiplets. 
Secondly, it is rather impractical from the point of view of constructing nonlinear supergravity actions such as invariants 
for conformal supergravity, see, e.g.,  \cite{VanP} for a related discussion. 
These limitations are avoided by resorting to superspace techniques. 
There exist two fully fledged superspace formulations for $\cN =(1,0) $
conformal supergravity and its general off-shell couplings to supersymmetric matter:
(i) $\sSU(2)$ superspace \cite{LT-M12}; and (ii) conformal superspace \cite{BKNT}. 
Both formulations have analogues in $d<6 $ dimensions. 

The $\sSU(2)$ superspace of \cite{LT-M12} is a particular $d=6$ realisation
of the general approach to formulate $\cN$-extended conformal supergravity 
in $3\leq d \leq 6$ dimensions
using the so-called $G_R {[d;\cN]}$ superspace, 
where $G_R {[d;\cN]}$ is the $R$-symmetry subgroup of the $\cN$-extended superconformal 
group in $d$ dimensions.\footnote{According to the Nahm classification \cite{Nahm}, 
superconformal algebras exist in spacetime dimensions 
$d \leq 6$. The   $d=5$ case is truly exceptional, 
for it allows  the existence of the unique superconformal algebra ${\mathfrak F}(4)$.} 
By definition, $G_R {[d;\cN]}$ superspace is a 
supermanifold $\cM^{d|\d_\cN}$, with $d$ bosonic and $\d_\cN$ fermionic dimensions.\footnote{Here
$\d_\cN = 2^{\left \lfloor{d/2}\right \rfloor }\cN$ for $d=3,4$ and 6, and $\d_\cN =8 $ for $d=5$.
We denote by  $z^M = (x^m, \q^{\hat \m}) $ the local coordinates
for  $\cM^{d|\d_\cN}$. Without loss of generality, we assume that 
the zero  section of $\cM^{d|\d_\cN}$  defined by $\q^{\hat \m} =0$ corresponds to 
the spacetime manifold $\cM^d$. }
Its  structure group is
$\sSpin(d-1,1)
\times G_R {[d;\cN]}$, 
where $\sSpin(d-1,1)$ is the double covering group of the connected Lorentz group 
$\sSO_0(d-1,1)$. This means that the  differential geometry of  $\cM^{d|\d_\cN}$ is realised 
in terms of covariant derivatives  of the form 
\bea
\cD_A = (\cD_a, \cD_{\hat \a}) = E_A -\O_A -\F_A ~.
\eea
Here $E_A = E_A{}^M \pa / \pa z^M $ denotes the inverse superspace vielbein,
$\O_A = \hf \O_A{}^{bc}  M_{bc} $ is the  Lorentz connection, 
and $\F = \F_A{}^I J_I $ the $R$-symmetry connection.
The index $\hat \a$ of the fermionic operator $\cD_{\hat \a}$ is, in general,  composite; 
it is comprised of a spinor index $\a$  and an $R$-symmetry index. 
The supergravity gauge group includes a subgroup  generated by local transformations
\bea
\d_\cK \cD_A &=& [\cK , \cD_A] ~, \qquad 
\qquad \cK:= \x^B  \cD_B + \hf K^{bc}  M_{bc} + K^I  J_I ~, 
\label{1.2a}
\eea
where the gauge parameters $\x^A$, $K^{bc}= -K^{cb}$ and $K^I$
obey standard reality conditions but are  otherwise  arbitrary. 
Given a tensor superfield $\vf$ (with suppressed Lorentz and $R$-symmetry 
indices), its transformation law under \eqref{1.2a} is
$\d_\cK \vf = \cK \vf$.  

In order to describe conformal supergravity, the superspace torsion $T_{AB}{}^C$
in 
\bea
[ \cD_A , \cD_B\} = -T_{AB}{}^C \cD_C - \hf R_{AB}{}^{cd} M_{cd} 
-R_{AB}{}^I J_I~, 
\eea
must  obey certain algebraic constraints, which may be thought of 
as generalisations of the torsion-free constraint in gravity. 
A fundamental requirement on the superspace geometry, in order to describe conformal supergravity,  
is that the constraints on the torsion be invariant under 
super-Weyl transformations of the form
\bea
\d_\s \cD_a &=& \s \cD_a + \cdots~, \qquad \d_\s \cD_\hal = \hf \s \cD_\hal + \cdots~, 
\label{1.4a}
\eea
where the scale parameter $\s$ is an arbitrary real superfield. The ellipsis 
in the expression for $\d_\s \cD_a $ includes, in general, 
a linear combination of the spinor covariant derivatives $\cD_{\hat \b}$ and 
the structure group generators $M_{cd}$ and $J_K$.  The ellipsis in 
$\d_\s \cD_\hal $ stands for a linear combination of the generators
of  the structure group. The resulting curved superspace will be denoted
$(\cM^{d|\d_\cN}, \cD)$. In many dynamical systems of interest, 
matter superfields may be chosen to be primary under the super-Weyl group, 
$\d_\s \vf= w_\vf \s \vf$,
where the parameter $w_\vf$  is the  super-Weyl weight of $\vf$.

The  approach sketched above was pioneered in $d=4$ by 
 Howe \cite{Howe1,Howe2} who put forward the concept of $\sU(\cN)$ superspace. 
In particular, he  introduced the $\sU(1)$ and $\sU(2)$ 
 superspace geometries  \cite{Howe2}, corresponding to 
  $\cN=1$ and $\cN=2$ conformal supergravity, respectively.  
 Howe's analysis was purely geometric in the sense that  he did not address the problem of constructing supergravity-matter actions. The full power of $\sU(1)$ superspace 
 was revealed in the book \cite{GGRS}, which provided a unified description of the off-shell 
 formulations for $\cN=1$ supergravity  and their couplings to matter. General off-shell $\cN=2$ supergravity-matter systems in $d=4$ were constructed in $\sU(2) $ superspace in \cite{KLRT-M2}, building on the concepts of rigid projective superspace \cite{KLR,LR1,LR2} and superconformal projective multiplets \cite{K06,K07}.
 The four-dimensional results of \cite{KLRT-M2} provided a natural extension of the earlier construction of the $\sSU(2) $ superspace formalism in five dimensions
  \cite{KT-M08}. The $d=3$ realisation of  $G_R {[d;\cN]}$ superspace
  is known as $\sSO(\cN)$ superspace. 
  Its geometry was developed in \cite{HIPT,KLT-M}. This formalism was used in \cite{KLT-M} to construct off-shell supergravity-matter couplings for $\cN \leq 4$.
  
As compared with the $d=6$, $\cN=(1,0)$ superconformal tensor calculus of \cite{BSVanP},  the important advantage of the $\sSU(2)$ superspace approach 
\cite{LT-M12} is that it  offered off-shell formulations for general supersymmetric nonlinear $\s$-models coupled to supergravity.\footnote{The component reduction of these locally supersymmetric $\s$-models can be carried out
using the techniques developed by Butter in the $d=4$, $\cN=2$ case \cite{Butter14,Butter15}.} 
This was achieved by making use of the concept of covariant projective supermultiplets.\footnote{The 
concept of covariant projective supermultiplets was introduced earlier in $d<6$ dimensions, 
first  in the framework of 
$d=5$, $\cN=1$  \cite{KT-M_5D2,KT-M_5D3,KT-M08}, followed by $d=4$, $\cN=2$ \cite{KLRT-M1,KLRT-M2}, 
then in $d=3$, $\cN=3$ and $\cN=4$ \cite{KLT-M}, and finally in $d=2$, $\cN=(4,4)$ supergravity \cite{GT-M_2D44}.}

The superspace formalism of \cite{BKNT} is a particular $d=6$ realisation of the universal approach to $\cN$-extended conformal supergravity in $d \leq 6$ dimensions,
 which is based on gauging the entire 
$\cN$-extended superconformal group,
of which 
$\sSpin(d-1,1)
\times G_R {[d;\cN]}$ is a subgroup. 
This approach, known as conformal 
superspace, was originally developed
for $\cN=1$  and $\cN=2$ supergravity theories in four dimensions
by Butter \cite{Butter4DN=1,Butter4DN=2}.
More recently, it has been extended to the cases of 
$d=3$, $\cN$-extended conformal 
supergravity  \cite{BKNT-M1},  $d=5$ conformal supergravity
\cite{BKNT-M5D}, and $d=6$, $\cN=(1,0)$ conformal supergravity \cite{BKNT}.
Conceptually, conformal superspace is a superspace analogue
of the famous formulation for conformal (super)gravity  as the gauge theory of the (super)conformal group pioneered 
by Kaku, Townsend and van Nieuwenhuizen \cite{KTvN1,KTvN2}, 
and further developed  by Kugo and Uehara \cite{KU}.

One of the important achievements of the conformal superspace approach \cite{BKNT}
 is that it provided the first ever construction of all the 
invariants for $\cN = (1, 0)$ conformal supergravity in six 
dimensions.\footnote{A simple by-product of the analysis in \cite{BKNT} was the 
first construction of the locally supersymmetric $F\Box F $ action coupled to conformal supergravity. 
In Minkowski space, the $\cN=(1,0)$ supersymmetric $F \Box F$ action was described for the first time in \cite{ISZ05} within the harmonic superspace approach.  }
Several months later, these invariants were reduced to components in \cite{BNT-M17}, which 
resulted in the first tensor calculus description of the conformal supergravity actions. 
Conformal superspace has also been used to describe the supersymmetric completion of several curvature-squared invariants
for $\cN = (1, 0)$ supergravity in six dimensions \cite{NOPT-M,BNOPT-M}.

Conformal superspace  is an ultimate formulation for 
 conformal supergravity in the sense that any different off-shell formulation is either equivalent to it  or is obtained from it by partially fixing the gauge freedom. 
 In particular, $G_R {[d;\cN]}$ superspace
 can be obtained from a partial gauge fixing of conformal superspace,
see \cite{Butter4DN=1,Butter4DN=2,BKNT-M1,BKNT-M5D} for the technical  details.
In the case of six dimensions, it was demonstrated in \cite{BNT-M17}
that the $\cN=(1,0)$ superconformal tensor calculus of \cite{BSVanP} is a gauged fixed 
version of the conformal superspace developed in \cite{BKNT}.

Recently, local supertwistor formulations for  $\cN=(1,0)$ 
and $\cN=(2,0)$ conformal supergravity in six dimensions have been constructed \cite{HL20}, 
and analogous formulations have been proposed in 
diverse dimensions 
\cite{Howe:2020xrg}.
Ref. \cite{HL20}  offered the first superspace 
description of the $\cN=(2,0)$ Weyl supermultiplet,
which was originally formulated using the superconformal tensor calculus \cite{BSVanP2}.
In accordance with the above discussion, the local supertwistor formulation should be equivalent 
to conformal superspace\footnote{Both constructions are based on  Cartan connections, first discussed in the superspace context in \cite{Lott:2001st}.}. The latter
is at present much more developed and is thus the one favoured in this paper.
We should also mention that the harmonic superspace formulation for $\cN=(1,0)$ conformal 
supergravity was briefly described in \cite{SokatchevAA}. 
Unfortunately, this approach has not been pursued for over thirty years.

The present work is devoted to new applications of the supergravity formulations  \cite{LT-M12,BKNT}. 
 Their fundamental property 
 is that they offer a universal setting to generate off-shell supersymmetric field theories in curved space. In particular, all $\cN=(1,0)$ supersymmetric theories that were originally constructed in terms of ordinary fields, 
 may be read off from a superfield theory upon elimination of the auxiliary fields.
 In order to develop supersymmetric field theory in a given supergravity background, one needs a formalism to determine the (conformal) isometries 
of the background superspace. 
 Such a formalism was developed long ago \cite{BK} within the framework of $d=4$, $\cN=1$ old minimal supergravity. 
The approach described in \cite{BK} is  universal, for  
in principle it may  be generalised to curved backgrounds associated with 
any supergravity theory formulated in superspace, see the discussion in \cite{K15Corfu}.
In particular, this approach has been properly generalised to study supersymmetric backgrounds
in $\cN=2$ supergravity  in three \cite{KLRST-M}
and four \cite{BIL} dimensions, and  $\cN=1$ supergravity in five dimensions \cite{KNT-M}. 
One of the goals of this paper is to work out the structure of (conformal) isometries of a given 
$\cN=(1,0)$ supergravity background in six dimensions. 

Within the $G_R {[d;\cN]}$ superspace formulation, there exists a universal 
description of all conformal isometries 
of a given curved background  $(\cM^{d |\d_\cN}, \cD)$. 
Following  the discussion in \cite{K15Corfu}, a real supervector field $\x= \x^B E_B$  
on $(\cM^{d |\d_\cN}, \cD)$ is called conformal Killing if 
\bea
 (\d_\cK + \d_\s) \cD_A = 0~,
 \label{1.5}
\eea
for some Lorentz $K^{bc}$, $R$-symmetry $K^I$ and super-Weyl 
$\s$ parameters. 
For any dimension $3\leq d\leq 6$ and any conformal supergravity,  
the following general properties are expected to hold:

\begin{itemize}

\item All parameters $K^{bc} $,  $K^I$ and $\s$ are uniquely determined 
in terms of $\x^B$, which allows us to write $K^{bc} = K^{bc}[\x]$, $K^I = K^I[\x]$ and 
$\s = \s[\x]$.

\item The spinor component $\x^{\hat \b}$ is uniquely determined in terms of $\x^b$.

\item The vector component $\x^b$ obeys a closed equation that contains all information 
about the  conformal Killing supervector field. 

\end{itemize}
The properties have been established for $d< 6$ in several publications
\cite{BK,KLRST-M,BIL,KNT-M}. The $d=6$, $\cN=(1,0)$ case will be studied in this paper.
By construction, the set of conformal Killing vectors on $(\cM^{d |\d_\cN}, \cD)$
is a Lie superalgebra with respect to the standard Lie bracket. This is 
the superconformal algebra of $(\cM^{d |\d_\cN}, \cD)$.
One may show that it is finite-dimensional. In the $d=6$, $\cN=(1,0)$ case, 
the proof will be given in section \ref{section3}.

Given a conformal Killing supervector field $\x^A$ on  $(\cM^{d |\d_\cN}, \cD)$, the first-order operator $\mathfrak{D}^{(1)}_{\xi} ={\cK[\x] } + \d_{\s[\x]}$ is a symmetry of any supersymmetric wave 
equation  $\cO \vf =0$, where  $\cO$ is the kinetic operator for some matter  supermultiplets $\vf$.
For every solution $\vf$ of the mass-shell equation, $\mathfrak{D}^{(1)}_{\xi}  \vf$ is also a solution.
It is of interest to study higher symmetries of supersymmetric wave equations, $n^{th}$-order operators $\mathfrak{D}^{(n)}_\z$ taking solutions to solutions, for instance in the context
of higher-spin superalgebras \cite{FV1,FV2,Vasiliev88,KV1,KV2}. 
Higher symmetries of relativistic wave equations have extensively been studied in the literature, 
see, e.g., \cite{BKM,Nikitin,NikitinP,BSSS,ShSh,ShV,Eastwood,Vasiliev2004} and references therein. 
In the supersymmetric case, however, the program of studying the higher symmetries of the so-called 
super-Laplacians and related geometric structures in diverse dimensions
has been initiated only a few years ago \cite{HL1,HL2,HL3}, 
mostly in Minkowski superspace $({\mathbb M}^{d |\d_\cN}, D)$. 
So far there has been only one publication \cite{KR} devoted to the higher symmetries of  
supersymmetric wave equations in curved supergravity backgrounds.
The present paper is aimed, in part, at a study of the higher symmetries of several on-shell supermultiplets  
in a background of $\cN=(1,0)$ conformal supergravity in six dimensions. Their non-supersymmetric analogues are also examined in diverse dimensions, and bring with them new insights for the supersymmetric story.

This paper is organised as follows. Section \ref{section2} reviews the $\sSU(2)$ and conformal superspace formulations for conformal supergravity. The conformal isometries of a fixed superspace are then studied in section \ref{section3}. In section \ref{section4}, we introduce the notion of a conformal Killing spinor superfield, which generates extended superconformal transformations. By a systematic study, it is shown that among its cousins are the conformal Killing vectors and tensors, which generate conformal isometries and higher symmetries, respectively. In section \ref{section5.1} we review the higher symmetries of the conformal d'Alembertian and present some new observations pertinent to the supersymmetric story. Following this, we study the higher symmetries of the hypermultiplet and vector multiplet in sections \ref{section6} and \ref{section7}, respectively. Section \ref{section8} is devoted to the study of $\mathcal{N} = (1,0)$ maximally supersymmetric backgrounds. Concluding comments are given in section \ref{section9}.

The main body of this paper is accompanied by several technical appendices. Appendix \ref{AppendixA} recounts our conventions. We review the conformal Killing supervector fields of $\mathcal{N}=(2,0)$ Minkowksi superspace in Appendix \ref{reductionAppendix}. In Appendix \ref{appendixB}, we detail a formalism for the study of supersymmetric backgrounds from a superspace perspective. Finally, in Appendix \ref{degauging}, we detail how to `degauge' from conformal to $\sSU(2)$ superspace.


\section{Conformal supergravity in superspace}
\label{section2}

As discussed in the introduction, there exist two fully fledged superspace 
formulations for $\cN =(1,0) $
conformal supergravity and its couplings to supersymmetric matter. In the literature they are 
referred to as (i) $\sSU(2)$ superspace \cite{LT-M12}; and (ii) conformal superspace \cite{BKNT}. 
Since both approaches will be used in the present paper, in this section we briefly review 
these formulations.


\subsection{${\sSU}(2)$ superspace}
\label{SU2-superspace}

We consider a 
supermanifold
$\cM^{6|8}$ parametrised by six bosonic ($x$) and eight fermionic ($\q$) 
coordinates $z^{M}=(x^{m},\q^{\mu}_\imath)$,
where $m=0,1,\cdots,5$, $\mu=1,\cdots,4$ and  $\imath=\1,\2$.
The name ``${\sSU(2)}$ superspace'' derives from the fact that
its structure group, 
$\sSpin(5,1)
\times \sSU(2)_R$,
includes the $R$-symmetry group $\sSU(2)_R$ in addition to the spin group.
Therefore, the superspace covariant derivatives, $\cD_{{} A}= (\cD_{{} a} , \cD_{{} \alpha}^{ i})$, 
have the form
\begin{eqnarray}
\cD_{{} A} = E_{{} A} -\frac12 \Omega_{{} A}{}^{{} b{} c} M_{{} b {} c} 
-\Phi_{{} A}{}^{jk} J_{jk}
~.
\label{cD0}
\end{eqnarray}
Here $E_{A}= E_{A}{}^{M} \partial_{M}$ is the frame field,
with $E_A{}^M$ being the inverse superspace vielbein,
$\Omega_{{} A} {}^{bc} $
the Lorentz connection,
and
$\Phi_{{} A}{}^{ij} $  the $\sSU(2)_R$ connection.
The Lorentz ($M_{ab}$) and the $R$-symmetry
$(J^{ij})$ generators are defined to act on Weyl spinors, vectors and isospinors 
as follows:
\bsubeq
\bea
\label{LorentzSpin}
M_\a{}^\b \psi^{\g}
= \d_{\a}^{\g} \psi^{\b} - \frac{1}{4} \d_{\a}^{\b} \psi^{\g}~&,&~~~
M_\a{}^\b \psi_{\g }=\frac{1}{4}\d_\a^\b\psi_{\g }-\d_\g^\b\psi_{\a}
~,
\\
M_{{} a {} b} V_c = 2 \eta_{{} c [ {} a} V_{b]} ~&,&~~~
J^{ij} \chi^{k} = \varepsilon^{k(i} \chi^{j)} ~,
\eea
\esubeq
where the Lorentz generator with spinor indices, $M_\a{}^\b$, is defined in accordance with the general rule \eqref{A.18}, $M_\a{}^\b=-\frac{1}{4}(\g^{ab})_\a{}^\b M_{ab}$.
For further details regarding our spinor conventions we refer the reader to appendix \ref{appendixA.1}.

The covariant derivatives are characterised by  graded commutation relations
\begin{eqnarray}
\label{TRF}
[ \mathcal D_{{} A}, \mathcal D_{{} B} \} &=& -T_{{} A{} B}{}^{{} C} \mathcal D_{{} C} -\frac12 R_{{} A {} B}{}^{{} c {} d} M_{{} c{} d} -R_{{} A {} B}{}^{kl} J_{kl}
~,
\end{eqnarray}
where $T_{AB}{}^C$ is the torsion, $R_{AB}{}^{cd}$ the Lorentz curvature, and
$R_{AB}{}^{kl}$ the $R$-symmetry curvature.
In order to describe conformal supergravity, 
the torsion must obey certain constraints 
\cite{LT-M12}
\bsubeq
\label{SU2constraints}
\begin{eqnarray}
\label{dim0}
{T_{{} \alpha}^{ i}{}_{\beta}^{  j}{}^{{} c}= 2\ri\ve^{ij} (\gamma^{{} c})_{{} \alpha {} \beta}}~, 
&& (\textrm{dimension-}0)\\
\label{dim1/2}
T_{{} \alpha}^{  i}{}_{\, {} \beta}^{  j}{}^{{} \gamma}_{  k} = 0 
~ ,~~  T_{{} \alpha}^{ i}{}_{ \, {} b}{}^{{} c}= 0~,
&& (\textrm{dimension-}\tfrac12)\\
\label{dim1} 
T_{{} a\, {} b}{}^{{} c}= 0 
~ ,~  T_{{} a}{}_{ \, {} \beta (j}{}^{{} \beta}_{ \, k)}= 0~.
&& (\textrm{dimension-}1)
\end{eqnarray}
\esubeq
Their general solution is given by the relations
\bsubeq
\bea
\label{Algebra-1}
\{ \mathcal D_{{} \alpha}^{ i}, \mathcal D_{{} \beta}^{ j} \}&=& 
	-2 \ri \varepsilon^{ij} (\gamma^{{} a})_{{} \alpha {} \beta}\mathcal D_{{} a}
	+2\ri C_{{} a}^{ ij} ({\gamma}^{{} a {} b {} c})_{{} \alpha {} \beta} M_{{} b {} c}
	-2\ri\varepsilon^{ij}W^{{} a{} b{} c} (\gamma_{{} a})_{{} \alpha {} \beta} M_{{} b{} c}
	\non\\
	&&
	-4\ri\varepsilon^{ij}N^{{} a{} b{} c} (\gamma_{{} a})_{{} \alpha {} \beta} M_{{} b{} c}
	+6\ri\varepsilon^{ij}C_{{} a}^{kl} (\gamma^{{} a})_{{} \alpha {} \beta} J_{kl} 
	- \frac{8\ri}3 N^{{} a {} b {} c} ({\gamma}_{{} a {} b {} c})_{{} \alpha {} \beta}  J^{ij}
	~,\\
\label{Algebra-2}
{[}\mathcal D_{{} a},\mathcal D_{{} \b}^{j} {]} &=& 
C^b{}^{j}{}_{k}(\g_{ab})_{\b}{}^{\d}\cD_{\d }^{k}
-\frac{1}{4} W_{acd}(\g^{cd})_\b{}^{\d}\cD_\d^j
-N_{acd}(\g^{cd})_\b{}^\d\cD_{\d}^{j}
\non\\
&&
-\hf {R}_a{}_\b^j{}^{cd}M_{cd}
-{R}_a{}_\b^j{}^{kl}J_{kl}
	~,
\eea
\esubeq
where the curvature tensors in the second line of \eqref{Algebra-2} have the following form:
\bsubeq
\bea
{R}_a{}_\b^j{}^{cd}
&=&
-\frac{1}{4}\Big[
(\g_{a}{}^{cd})_{\g\d}\d_\b^\r 
-2(\g_a)_{\b\g} (\g^{cd})_\d{}^\r\Big] \cW_\r{}^{\g\d}{}^j \non \\
&&
-\Big[
(\g_a{}^{cd})_{\b\g}
+4\d_{a}^{[c}(\g^{d]})_{\b\g}
\Big]
\Big(
 \frac{1}{12} \cW^{\g j}
-\cC^{\g j}
\Big)
\non\\
&&
+\Big[
(\g^{cd})_\b{}^\g\d_a^b
+2\d_\b^\g\d_{a}^{[c}\eta^{d]b}\Big]
\Big(
\cN_{b}{}_{\g}{}^{j}
-\cC_{b}{}_{\g}{}^{j} 
\Big)
~,\\
{R}_a{}_\b^j{}^{kl}
&=&
-(\g_a)_{\b\g}\cC^{\g}{}^{jkl}
-5(\g_a)_{\b\g}\Big(
\cC^{\g (k}
-\frac{1}{6}  \cW^{\g (k}\Big)
\ve^{l)j} \non \\
&&
-\Big(
4\cN_{a}{}_{\b}{}^{(k}
-3\cC_{a}{}_{ \b}{}^{(k} 
\Big)
\ve^{l)j}
~.
\eea
\esubeq
The algebra of the covariant derivatives 
is determined by three dimension-1 real tensors, 
$W_{ a b c} = W_{[a b c]} $, $N_{abc}=N_{[abc]}$ and $C_a^{ij}=C_a^{ji}$,
 and their covariant derivatives. 
 The 3-forms $N_{abc}$ and $W_{abc}$ are self-dual and anti-self-dual, respectively, 
 \bea
  \frac{1}{3!} \eps^{abcdef} N_{def} =  N^{abc} ~, \qquad
   \frac{1}{3!} \eps^{abcdef} W_{def} = - W^{abc} ~.
   \label{self}
 \eea
They are equivalently described  in terms of the symmetric chiral rank-2 spinors
$W^{\alpha \beta}:= \frac16 W_{abc}(\tilde \gamma^{abc})^{\alpha \beta}$ 
and $N_{\alpha \beta}:= \frac16 N_{abc}(\gamma^{abc})_{\alpha \beta}$.

The curvature tensors \eqref{Algebra-1} and \eqref{Algebra-2}
involve several dimension-3/2 descendants of $C_{a\, ij} $, $N_{\alpha \beta}$ and $W^{\alpha \beta}$,
defined by 
\bsubeq \label{2.6abc}
\begin{eqnarray}
\label{calC}
\mathcal D_{\gamma k } C_{a\, ij} &{=}&
 (\gamma_a)_{\gamma \delta}\mathcal C^{\delta}_{ijk}
	+\varepsilon_{k(i}\mathcal C_{a\, \gamma j)} + \varepsilon_{k(i}(\gamma_a)_{\gamma \delta}\mathcal C^{\delta}_{j)}
	\label{DC}~,\\ 
\mathcal D_{\gamma k} N_{\alpha \beta}&{=}&
\tfrac23\left( \mathcal D_{[\gamma}^{ k} N_{\alpha]\beta} 
+ \mathcal D_{[\gamma}^{k}N_{ \beta]\alpha}
\right)
:= (\gamma^a)_{\gamma(\alpha} \cN_{a\,\beta)}{}^{k}
~,~~~~~~
(\gamma^a)_{[\gamma\alpha} \cN_{a\,\beta]}{}^{k}=0
~,~~\\
\label{calW}
\mathcal D_{\gamma k} W^{\alpha \beta}&{=}& \mathcal W_{\gamma k}{}^{\alpha \beta} + \delta_\gamma^{(\alpha}\mathcal W^{\beta)}_k
~.
\end{eqnarray}
\esubeq
In accordance with the general discussion in section \ref{section1}, the curved superspace 
introduced above will be denoted $(\cM^{6|8}, \cD)$.

In $\sSU(2)$ superspace, the gauge group of conformal supergravity is generated by 
three types of local transformations:  (i) general coordinate transformations; (ii) structure group transformations; and (iii) super-Weyl transformations. 
An infinitesimal transformation of the combined type (i) and (ii) acts on the covariant derivatives as
\bea
\d_\cK\cD_A=[\cK,\cD_A]~,\qquad
\cK=K^C\cD_C+\hf K^{cd} M_{cd}+ K^{kl}J_{kl}
~.
\label{SUGRA-gauge-group1}
\eea
Given a tensor superfield $U$ (with its indices suppressed), its transformation law with respect to  \eqref{SUGRA-gauge-group1} is
\bea
\d_\cK U=\cK U ~.
\label{SUGRA-gauge-group2}
\eea
An infinitesimal super-Weyl transformation of the covariant derivative \cite{LT-M12} is
\bsubeq 
\label{WeylTransInf}
\bea
\d_\s \cD_\a^i&=&
\hf\s \cD_\a^i-2(\cD_\b^{ i}\s)M_{\a}{}^{\b}-4(\cD_{\a j}\s)J^{ij}~, \\
\d_\s \cD_{a}
&=& \s \cD_{a}
-\frac{\ri}{2}(\tilde{\g}_a)^{\a\b}(\cD_{\a}^{k}\s)\cD_{\b k}
-(\cD^b\s)M_{ab}
-\frac{\ri}{8} (\tilde{\g}_{a})^{\a\b}(\cD_{\a}^{k}\cD_{\b}^{l}\s)J_{kl} ~,
\eea
\esubeq
where the real parameter $\s$ is unconstrained. 
The crucial  feature of these transformations
is that they preserve the supergravity constraints \eqref{SU2constraints}.

A tensor superfield $U$ is said to be {\emph{primary}}  of Weyl weight (or dimension) $w$ 
if it transforms homogeneously 
under \eqref{WeylTransInf}
\bea
\label{sWPrimary}
\d_\s U  =w \s  U ~.
\eea
The torsion $W_{abc}$ proves to be a primary superfield of dimension $+1$.
It is the $\cN=(1,0)$  supersymmetric extension of the Weyl tensor \cite{LT-M12,Gates6D}.

In what follows, we will need a finite form of the super-Weyl transformations \eqref{WeylTransInf}.
Direct calculations lead to
\bsubeq 
\label{WeylTransWhole}
\bea
\cD'{}_\a^i&=&
\re^{\hf\s}\Big(\cD_\a^i-2(\cD_\b^{ i}\s)M_{\a}{}^{\b}-4(\cD_{\a j}\s)J^{ij}\Big)~,
\label{WeylTrans}
\\
\cD'{}_{a}
&=&
\re^{\s}\Big(\,
\cD_{a}
-\frac{\ri}{2}(\tilde{\g}_a)^{\a\b}(\cD_{\a}^{k}\s)\cD_{\b k}
-(\cD^b\s)M_{ab}
-\frac{\ri}{8} (\tilde{\g}_{a})^{\a\b}(\cD_{\a}^{k}\cD_{\b}^{l}\s)J_{kl} 
\non\\
&&~~~~
-\frac{\ri}{8}(\tilde{\g}_a{}^{cd})^{\a\b}(\cD_{\a}^k\s)(\cD_{\b k}\s) M_{cd}
-\frac{3\ri}{4} (\tilde{\g}_{a})^{\a\b}(\cD_{\a}^{k}\s)(\cD_{\b}^{l}\s)J_{kl} 
\Big)
~.
\label{WeylTransB}	
\eea
\esubeq
Such a transformation
 acts on 
 the dimension-1 torsion superfields as follows:
\bsubeq\label{WeylC}
\bea
W'_{abc}
&=&
\re^{\s}W_{abc}
~,
\label{WeylC.a}
\\
N'_{abc}
&=&
\re^{\s}\Big(
N_{abc}
-\frac{\ri}{32}(\tilde{\g}_{abc})^{\a\b}\big(
\cD_\a^k\cD_{\b k}\s
+4(\cD_{\a}^k\s)\cD_{\b k}\s
\big)
\Big)
~,
\\
C'_{a}{}^{ij}
&=&
\re^{\s}
\Big(
C_{a}{}^{ij} 
+\frac{\ri}{8}(\tilde{\g}_{a})^{\a\b}
\big(\,
\cD_{\a}^{(i}\cD_{\b}^{j)}\s
-2(\cD_{\a}^{(i}\s)\cD_{\b}^{j)}\s
\big)
\Big)
~.
\eea
\esubeq


\subsection{Conformal superspace}

In the conformal superspace approach of \cite{BKNT} (see also \cite{BNT-M17} for the component analysis)
the whole superconformal algebra is gauged in superspace 
by introducing covariant derivatives $ {\nabla}_A = (\nabla_a , \nabla_\a^i)$ of the following form
\bea
\nabla_A
= E_A - \hf  {\Omega}_A{}^{bc} M_{bc} - \Phi_A{}^{kl} J_{kl} - B_A \mathbb D
	-  {\mathfrak{F}}_A{}_B K^B \ .
	\label{conf-nabla}
\eea
The difference compared with the $\sSU(2)$ superspace covariant derivatives of \eqref{cD0} is the presence
of dilatation ($B_A$) and special conformal (${\mathfrak{F}}_A{}_B$) connections, 
where $\mathbb D$ is the dilatation generator and $K^A = (K^a, S^\a_i)$ are the special conformal
generators.
The complete list of graded commutation relations defining  the $\cN=(1,0)$ superconformal algebra are given in appendix \ref{AppendixA.2}.

To describe the standard $\mathcal{N} = (1,0)$ Weyl multiplet in conformal superspace,
one constrains the algebra of covariant derivatives
\begin{align}
[  {\nabla}_A ,  {\nabla}_B \}
&= - \mathscr{T}_{AB}{}^C  {\nabla}_C
- \frac{1}{2}  \mathscr{R}(M)_{AB}{}^{cd} M_{cd}
-  \mathscr{R}(J)_{AB}{}^{kl} J_{kl}
\non \\ & \quad
-  \mathscr{R}(\mathbb D)_{AB} \mathbb D
-  \mathscr{R}(K)_{AB}{}_C K^C ~,
\label{nablanabla}
\end{align}
to  be   completely determined   in terms of the super-Weyl tensor
 $W^{\a\b}$
\bea
K^AW^{\a\b} = 0 \ ,
 \quad \mathbb D W^{\a\b} = W^{\a\b} \ ,
\eea
which satisfies the constraints
\bsubeq\label{WBI}
\bea
\nabla_\a^{(i} \nabla_{\b}^{j)} W^{\g\d} &=& - \d^{(\g}_{[\a} \nabla_{\b]}^{(i} \nabla_{\r}^{j)} W^{\d) \r} \ , \\
\nabla_\a^k \nabla_{\g k} W^{\b\g} - \frac{1}{4} \d^\b_\a \nabla_\g^k \nabla_{\d k} W^{\g\d}
&=& 8 \ri \nabla_{\a \g} W^{\g \b} \ .
\eea
\esubeq
Additionally, we require that the algebra of covariant derivatives resembles a $d=6,~ \cN=(1, 0)$ super Yang-Mills theory
\bea
\{ \nabla_\a^i , \nabla_\b^j \} = - 2 \ri \eps^{ij} (\g^a)_{\a\b} \nabla_a \ , ~~~~~~
\left[ \nabla_a , \nabla_\a^i \right] = (\g_a)_{\a\b} \mathscr{W}^{\b i} \ .
\label{algb-000}
\eea
Here $\mathscr{W}^{\a i}$ is a primary dimension $3/2$ operator valued in the superconformal algebra. 
Moreover, one imposes that the structure group generators 
act on the covariant derivatives
$\de_A$ precisely as if they were the generators $P_A$. 

By solving the Bianchi identites, one obtains
\bea [\nabla_a , \nabla_b] = - \frac{\ri}{8} (\g_{ab})_\a{}^\b \{ \nabla_\b^k , \mathscr{W}^\a_k \}~,
\eea
and the additional constraints
\be \{ \nabla_\a^{(i} , \mathscr{W}^{\b j)} \} = \frac{1}{4} \d^\b_\a \{ \nabla_\g^{(i} , \mathscr{W}^{\g j)} \} \label{WalbeBI}
\ , \quad \{ \nabla_\g^k , \mathscr{W}^\g_k \} = 0 \ .
\ee
The operator $\mathscr{W}^{\a i}$ is then constrained to be
\bea
\mathscr{W}^{\a i} &=& 
W^{\a\b} \nabla_\b^i
+ \nabla_\g^i W^{\a\b} M_\b{}^\g
- \frac{1}{4} \nabla_\g^i W^{\b\g} M_\b{}^\a
+ \frac{1}{2} \nabla_{\b j} W^{\a\b} J^{ij}
+ \frac{1}{8} \nabla_\b^i W^{\a\b} \mathbb D \non\\
&&
- \frac{1}{16} \nabla_\b^j \nabla_\g^i W^{\a \g} S^\b_j
+ \frac{\ri}{2} \nabla_{\b\g} W^{\g\a} S^{\b i} \non\\
&&- \frac{1}{12} (\g^{ab})_\b{}^\g \nabla_b \big( \nabla_\g^i W^{\b \a} 
- \hf \d_\g^\a \nabla_\d^i W^{\b\d} \big) K_a~.
\label{def-W}
\eea
Results at mass-dimension higher than 3/2 can be found in \cite{BKNT}.

It is convenient to define the following
\bsubeq\label{XYfields}
\begin{align} 
X^{\a i} &:=-\frac{\ri}{10} \nabla_{\b}^i W^{\a\b}
~,\quad
X_\g^k{}^{\a\b} :=
-\frac{\ri}{4}\nabla_\g^k W^{\a\b} - \d^{(\a}_{\g} X^{\b) k}
~, \\
Y_\a{}^\b{}^{ij} &:= - \frac{5}{2} \Big( \nabla_\a^{(i} X^{\b j)} - \frac{1}{4} \d^\b_\a \nabla_\g^{(i} X^{\g j)} \Big)
= - \frac{5}{2} \nabla_\a^{(i} X^{\b j)} \ , \\
Y &:= \frac{1}{4} \nabla_\g^k X^\g_k \ , \\
Y_{\a\b}{}^{\g\d} &:=
\nabla_{(\a}^k X_{\b) k}{}^{\g\d}
- \frac{1}{6} \d_\b^{(\g} \nabla_\r^k X_{\a k}{}^{\d) \r}
- \frac{1}{6} \d_\a^{(\g} \nabla_\r^k X_{\b k}{}^{\d) \r} 
\ .
\end{align}
\esubeq
Due to the constraints \eqref{WBI}, these superfields
are the only independent descendants of $W^{\a \b}$.
As described in detail in \cite{BNT-M17}, the multiplet of superconformal field strengths of the standard Weyl multiplet
is described by the $\q=0$ projection of the previous superfields. A reduction to components is straightforward and discussed in
\cite{BNT-M17}.

In conformal superspace,
the gauge group  of conformal supergravity
is generated by
{\it covariant general coordinate transformations}, associated with a local superdiffeomorphism parameter $\xi^A$ and
{\it standard superconformal transformations}, associated with the following local superfield parameters:
the Lorentz $\L^{ab}=-\L^{ba}$, $\sSU(2)_R$ $\L^{ij}=\L^{ji}$, dilatation $\s$, and special conformal transformations $\L_A=(\L_a,\L_\a^i)$.
The covariant derivatives transform as
\bea
\d_{\mathfrak K} \nabla_A &=& [{\mathfrak K} , \nabla_A] 
~,~~~~~~
{\mathfrak K} = \xi^B  {\nabla}_B + \hf  {\L}^{bc} M_{bc} +  {\L}^{jk} J_{jk} +  \s \mathbb D +  {\L}_B K^B ~.
\label{TransCD}
\eea
While the transformation law for a tensor superfield $U$ is
\be
\d_{\mathfrak K} U =
{\mathfrak K} U
 ~.
\ee
The superfield $U$ is said to
be \emph{primary} and of dimension $w$ if
\bea
K^A U = 0~, \quad {\mathbb D} U = w U ~.
\eea 
It is important to point out that the dimension of $U$ coincides with its super-Weyl weight \eqref{sWPrimary}.

We conclude by mentioning that $\sSU(2)$ superspace is a gauge-fixed version of the conformal superspace geometry and thus their physical multiplets are equivalent. We refer the reader to appendix \ref{degauging} for a description of the degauging procedure.


\section{Conformal isometries}
\label{section3}
\setcounter{equation}{0}

In this paper a central role is played by the conformal isometries 
of  a given supergravity background  $(\cM^{6|8}, \cD)$
 and their extensions.\footnote{The conformal isometries of $\cN=(1,0) $ Minkowski 
superspace in six dimensions were studied in \cite{Park98,BKNT}.} 
We will say that a real supervector field $\xi = \xi^{B} E_{B}$ on  $(\cM^{6|8}, \cD)$
is conformal Killing if there exist Lorentz ($K^{bc} [\xi]$), $R$-symmetry ($K^{jk} [\xi]$) and super-Weyl ($\s [\xi]$) parameters such that
\bea
\d \cD_{A} =\big(\d_{\mathcal{K}[\xi]} + \d_{\s[\xi]} \big)\cD_{A} 
= \Big[ \xi^{B} \cD_{B} + \frac{1}{2} K^{bc} [\xi] M_{bc} + K^{jk}[\xi] J_{jk} , \cD_A \Big] 
+ \d_{\s[\xi]}  \cD_{A} = 0~,
\label{SCIsom}
\eea
where the super-Weyl transformation $\d_\s \cD_A$ is defined in \eqref{WeylTransInf}.
Such transformations render the superspace geometry invariant, in particular
\bea
\d C_a^{ij}= \d W_{abc}= \d N_{abc}=0
~,
\label{inv-torsions}
\eea
 and thus are said to be superconformal.
\subsection{Conformal Killing vector superfields}
The solution to \eqref{SCIsom} is:
\begin{subequations}
\label{ConfIsoParameters}
\bea
\xi^{\a}_{i} &=& \frac{\ri}{12} \cD_{\b}{}_{i} \xi^{\b \a}=-\frac{\ri}{12}(\tilde{\g}^a)^{\a\b} \cD_{\b}{}_{i} \xi_a~, \\
K_{\a}{}^{\b} [\xi] &=& \frac{1}{2} \big( \cD_{\a}^{i} \xi^{\b}_{i} - \frac{1}{4} \d_{\a}^{\b} \cD_{\g}^{i} \xi^{\g}_{i} \big) + \hf \xi^{a} ( W_{abc} + 2 N_{abc} ) (\g^{bc})_{\a}{}^{\b} = - \frac{1}{4} (\g^{ab})_\a{}^\b \cD_{a}\x_{b} ~, \\
K_{ij} [\xi] &=&
 \frac{1}{4} \cD_{\a (i} \xi^{\a}_{j)}
 =-\frac{\ri}{48}(\tilde{\g}^a)^{\a\b} \cD_{\a (i} \cD_{\b}{}_{j)} \xi_a 
 ~, \\
\s [\xi] &=& \frac{1}{4} \cD_{\a}^{i} \xi_{i}^{\a} = \frac{1}{6} \cD_{a} \xi^{a}~,
\eea
\end{subequations}
where $\xi^{a}$ obeys
\bea
\label{ConfKillingVec}
\cD_{\a}^{i} \xi^{a} = - \frac{1}{5} (\g^{ab})_{\a}{}^{\b} \cD_{\b}^{i} \xi_{b} ~.
\eea
We have shown that every infinitesimal 
conformal  transformation,
$\d_{\mathcal{K}[\xi]} + \d_{\s[\xi]} $,  of $(\cM^{6|8}, \cD)$
 is parametrised by the vector superfield $\xi^a$,
and eq.~\eqref{ConfKillingVec} is the fundamental constraint defining this transfromation. 
All other conditions are implications of \eqref{ConfIsoParameters}
and \eqref{ConfKillingVec}.
For instance, the latter
implies the usual conformal Killing equation for the superfield $\xi_a$:
\bea
\label{3.5}
\cD_{(a} \xi_{b)} = \frac{1}{6} \eta_{a b} \cD_{c} \xi^{c}~.
\eea

While the analysis above was carried out in the $\sSU(2)$ superspace setting,  equivalent results can be derived using  the conformal superspace approach. Here we will say that $\xi = \xi^B E_B$ is conformal Killing if there exist Lorentz ($\L ^{bc}[\xi]$), $R$-symmetry ($\L^{jk} [\xi]$), dilatation ($\s[\xi]$) and special conformal ($\L_B[\xi]$) parameters such that
\bea
\label{CSSConfIso}
\d_{\mathfrak{K}[\xi]} \nabla_A = [\xi^B  {\nabla}_B + \hf  {\L}^{bc}[\xi] M_{bc} 
+  {\L}^{jk} [\xi] J_{jk} +  \s[\xi] \mathbb D +  {\L}_B [\xi] K^B , \nabla_A] = 0
~.
\eea
Since this transformation preserves the superspace geometry, it must also leave the super-Weyl tensor invariant,
\bea
\d_{\mathfrak{K}[\xi]} W^{\a \b} = \mathfrak{K}[\xi] W^{\a \b} = 0 ~.
\eea

The solution to \eqref{CSSConfIso} is:
\begin{subequations} \label{3.8parameters}
\bea
\xi^{\a}_{i} &=& \frac{\ri}{12} \nabla_{\b}{}_{i} \xi^{\b \a}=-\frac{\ri}{12}(\tilde{\g}^a)^{\a\b} \nabla_{\b}{}_{i} \xi_a~, \\
\L_{\a}{}^{\b} [\xi] &=& \frac{1}{2} \big( \nabla_{\a}^{i} \xi^{\b}_{i} - \frac{1}{4} \d_{\a}^{\b} \nabla_{\g}^{i} \xi^{\g}_{i} \big) + \xi_{\a \g} W^{\b \g}  = - \frac{1}{4} (\g^{ab})_{\a}{}^{\b} \nabla_{a} \xi_b  + \xi_{\a \g} W^{\b \g} \\
\L_{ij} [\xi] &=&
\frac{1}{4} \nabla_{\a (i} \xi^{\a}_{j)}
=-\frac{\ri}{48}(\tilde{\g}^a)^{\a\b} \nabla_{\a (i} \nabla_{\b}{}_{j)} \xi_a 
~, \\
\s [\xi] &=& \frac{1}{4} \nabla_{\a}^{i} \xi_{i}^{\a} = \frac{1}{6} \nabla_{a} \xi^{a}~, \\
\L_{\a}^i [\xi] &=& \frac{1}{2} \nabla_{\a}^{i} \s[\xi] - \frac{1}{16} \xi_{\a \b} \nabla_{\g}^{i} W^{\b \g} = \frac{1}{12} \nabla_{\a}^i \nabla^a \xi_a - \frac{1}{16} \xi_{\a \b} \nabla_{\g}^i W^{\b \g} ~, \\
\L_{a} [\xi] &=& 2 (\tilde{\g}_a)^{\a \b} \nabla_{\a}{}^{i} \L_{\b i} = \frac{4}{3} \nabla_{a} \nabla^b \xi_b - \frac{1}{8} (\tilde{\g}_a)^{\a \b} \nabla_{\a}^i (\xi_{\b \g} \nabla_{\d i} W^{\g \d}) ~,
\eea
\end{subequations}
where $\xi^{a}$ obeys the conformal Killing vector equation
\bea
\label{CKVCSS}
\nabla_{\a}^{i} \xi^{a} = - \frac{1}{5} (\g^{ab})_{\a}{}^{\b} \nabla_{\b}^{i} \xi_{b} ~.
\eea
This equation is conformally invariant provided $\xi^{a}$ is primary and of dimension
 $-1$, 
\bea
K^B \x^a = 0 ~,
 \qquad \mathbb D \x^a = -\x^a~.
 \label{3.10}
\eea
These relations
determine the superconformal properties of the parameters 
in \eqref{3.8parameters}.
 An important corollary of \eqref{CKVCSS} is
\bea
\nabla_{(a} \xi_{b)} = \frac{1}{6} \eta_{a b} \nabla_{c} \xi^{c} ~.
\eea

In what follows, we will often make use of the first-order operator
\bea
\mathfrak{D}^{(1)}_{\xi} = \xi^b  {\nabla}_b + \x^\a_i  \nabla_\a^i + \hf  {\L}^{bc}[\xi] M_{bc} 
+  {\L}^{jk} [\xi] J_{jk} +  \s[\xi] \mathbb D +  {\L}_B [\xi] K^B ~,
\label{3.12}
\eea
where $\x^a$ is characterised by the superconformal properties \eqref{3.10}, and the remaining parameters are given by \eqref{3.8parameters}.
The operator $\mathfrak{D}^{(1)}_{\xi}$ is superconformal and of dimension 0 in the sense that it takes 
every primary superfield $U$ of dimension $w$ to a primary superfield of the same dimension, 
\bea
K^A U = 0~, \quad {\mathbb D} U = w U \quad \implies \quad 
K^A \mathfrak{D}^{(1)}_{\xi} U = 0~, \quad 
{\mathbb D} \mathfrak{D}^{(1)}_{\xi} U = w \mathfrak{D}^{(1)}_{\xi} U ~.
\eea
If $\x^a$ is a solution to \eqref{CKVCSS}, then $\mathfrak{D}^{(1)}_{\xi} $
generates a conformal isometry, 
\bea
\nabla_{\a}^{i} \xi^{a} = - \frac{1}{5} (\g^{ab})_{\a}{}^{\b} \nabla_{\b}^{i} \xi_{b} 
\quad \implies \quad \big[ \mathfrak{D}^{(1)}_{\xi} , \nabla_A
\big] =0~.
\eea

Since $\sSU(2)$ superspace is a gauge fixed version of conformal superspace, it must necessarily be true that \eqref{CSSConfIso} reproduces \eqref{SCIsom} upon degauging. In particular, it is trivial to see that \eqref{CKVCSS} degauges to \eqref{ConfKillingVec}.

It is clear from \eqref{SCIsom} that the commutator of superconformal transformations must result in another such transformation
\begin{subequations}
	\bea
	\left[ \d_{\mathcal{K}[\xi_{2}]} + \d_{\s_{2}} , \d_{\mathcal{K}[\xi_{1}]} + \d_{\s_{1}} \right] \cD_{A} &=& \left( \d_{\mathcal{K}[\xi_3]} + \d_{\s_3} \right) \cD_{A} = 0 ~,\\
	\mathcal{K}[\xi_3] &:=& \Big[ \mathcal{K}[\xi_{2}] , \mathcal{K}[\xi_{1}] \Big] ~,
	\eea
\end{subequations}
From this we may extract the form of $\xi_3$ and $\s_3$
\begin{subequations}
	\bea
	\x^{a}_3 &=& \xi^{b}_1 \cD_{b} \xi^{a}_2 - \xi^{b}_2 \cD_{b} \xi^{a}_1 - \frac{\ri}{48} (\tilde{\g}^a)^{\a\b} \cD_{\a}^{i} \xi_1^b \cD_{\b i} \xi_{2b} + \frac{\ri}{48} (\tilde{\g}^{a}{}_{bc})^{\a \b} \cD_{\a}^i \xi_1 ^b \cD_{\b i} \xi_2 ^c
	~, \label{superLieBracket} \\
	\s_3 &=& \xi_{2}^{A} \cD_{A} \s_{1} - \xi_{1}^{A} \cD_{A} \s_{2} ~.
	\eea
\end{subequations}
This analysis implies that the conformal Killing supervector fields generate a finite dimensional super Lie algebra.
The independent parameters are set of superfields 
$\U = ( \x^{B}, K^{ab} , K^{jk} , \s, \cD_{C} \s)$
and one can prove that applying any number of covariant derivatives to $\U$  gives
a linear combination of $\U$. 
 
The statement above is most easily proven when working in Minkowski superspace, ${\mathbb M}^{6|8}$. Here the superspace covariant derivatives $D_A = (\partial_a , D_{\a}^i)$ take the form
\bea
\partial_a = \frac{\partial}{\partial x^a} ~, \quad D_{\a}^i = \frac{\partial}{\partial \theta^\a_i} - \ri (\g^a)_{\a \b} \theta^{\b i} \partial_a ~,
\eea
and satisfy the algebra
\bea
\{ D_{{} \alpha}^{ i}, D_{{} \beta}^{ j} \} =
-2 \ri \ve^{ij} \partial_{\a \b} ~, \quad \big[ \partial_a , D_{\a}^i \big] = 0 ~, \quad \big[ \partial_a , \partial_b \big] = 0 ~.
\eea
In this context, one may readily derive the following constraints
\begin{subequations}
\bea
D_\g^k K_\a{}^\b [\xi] &=& 
2 \d_\g^\b D_\a^k \s [\xi] - \hf \d^\b_\a D_\g^k \s [\xi]
\ , \\
D_\a^i K^{jk} [\xi] &=& 4 \eps^{i(j} D_\a^{k)} \s [\xi] \ , \\
D_\a^i D_\b^j \s [\xi] &=& -  \ri \eps^{ij} \partial_{\a\b} \s [\xi] \ , \\
D_\a^i \partial_b \s [\xi] &=& 0 \ .
\eea
\end{subequations}
Thus, our claim holds for this geometry.
 
The proof above readily generalises to curved superspace. For example, by analysing the invariance $\d C_a{}^{ij}=0$ and $\d N_{abc}=0$ one can derive the following
 relations for the second spinor derivative of $\s$
\bea
\label{3.17}
- \frac{\ri}{8} (\tilde\g_a)^{\g\d} \cD_{\g}^{(i}\cD_{\d}^{j)} \s
&=&
\xi^c \cD_c C_a{}^{ij}
+ \xi^\g_k \cD_\g^k C_a{}^{ij}
+ K_a{}^b C_b{}^{ij}
+ 2 K^{(i}{}_k C_a{}^{j) k}
+ \s C_a{}^{ij}
~,
\\
\frac{\ri}{32} (\tilde\g_{abc})^{\g\d} \cD_\g^k \cD_{\d k} \s 
&=&
 \xi^d \cD_d N_{abc}
+ \xi^\g_k \cD_\g^k N_{abc}
+ 3 K_{[a}{}^d N_{bc] d}
+ \s N_{abc}
~.
\eea
Another implication of \eqref{ConfIsoParameters} and \eqref{ConfKillingVec} is 
\bea
\cD_a \xi^\g_k 
&=&
\frac{\ri}{2} (\tilde\g_a)^{\b\g} \cD_{\b k} \s
+ (\g_{ab})_\b{}^\g \xi^{\b j} C^b{}_{jk}
+ \hf (\g^{bc})_\b{}^\g \xi^\b_k ( W_{abc} + 2 N_{abc} )
- \xi^b T_{ab}{}^\g_k
~,
\label{conf-Killing-0}
\eea
which implies the following expression for the spinor derivative of the super-Weyl parameter
\bea
\cD_{\a k} \s
= - \frac{\ri}{3} \cD_{\a\b} \xi^\b_k
+ \frac{4 \ri}{3} \xi^{\d j} C_{\a\d jk}
- \frac{\ri}{6} (\g^{abc})_{\d\a} \xi^\d_k ( W_{abc} + 2 N_{abc} )
- \frac{\ri}{3} \xi^b (\g^a)_{\a\b} T_{ab}{}^\b_k
~.
 \label{3.38}
\eea
Note that equation \eqref{conf-Killing-0} plays a fundamental role 
in the study
of supersymmetric spacetimes.

We also note that by imposing the invariance of the super-Weyl tensor $\d W_{abc}=0$ one obtains
\bea
\xi^d \cD_d W_{abc}
+ \xi^\g_k \cD_\g^k W_{abc}
+ 3 K_{[a}{}^d W_{bc] d}
+ \s W_{abc}
&=&0
~,
\eea
which hints at the fact that superspace backgrounds admitting non-trivial conformal isometries are
in general constrained.


\subsection{Conformally related superspaces}\label{section3.2}

By definition a superspace  $(\cM^{6|8}, \widetilde \cD)$
is 
said to be 
conformally related to
 $(\cM^{6|8}, \cD)$
if the corresponding covariant derivatives $\widetilde{\cD}_{A} $
and 
$\cD_{A} $ 
are related by a finite 
super-Weyl transformation, 
\begin{subequations}\label{CRSG}
\bea
\widetilde{\cD}_\a^i&=&
\re^{\hf \r}\Big(\cD_\a^i
-2(\mathcal D_{\beta}^{ i} \r)M_\alpha{}^\beta 
-4(\mathcal D_{\alpha j}\r) J^{ij} \Big) \ ,
\label{D_hal-ConfFlat}
\\
\widetilde{\cD}_{a}&=&
\re^{\r}\Big(
  \cD_{a}
-\frac{\ri}{2}(\tilde{\g}_a)^{\a\b}(\cD_{\a}^{k}\r)\cD_{\b k}
-( \cD^b\r)M_{ab}
-\frac{\ri}{8} (\tilde{\g}_{a})^{\a\b}(\cD_{\a}^{k}\cD_{\b}^{l}\r)J_{kl} 
 \non\\
&&~~~~
-\frac{\ri}{8}(\tilde{\g}_a{}^{cd})^{\a\b}(\cD_{\a}^k\r)(\cD_{\b k}\r) M_{cd}
-\frac{3\ri}{4} (\tilde{\g}_{a})^{\a\b}(\cD_{\a}^{k}\r)(\cD_{\b}^{l}\r)J_{kl} 
\Big)
~,~~~~~~
\label{D_a-ConfFlat-2}
\eea
\end{subequations}
for some super-Weyl parameter $\r$. 
The torsion superfields are then mapped from a curved superspace to the other according to \eqref{WeylC} with $\s$  
replaced by $\r$.
The two superspaces  $(\cM^{6|8}, \widetilde \cD)$ and  $(\cM^{6|8}, \cD)$
prove to have the same conformal Killing vector superfields. In fact an efficient way to analyse conformal isometries is 
by mapping their conformal Killing supervector fields from one superspace to its conformally related one -- see for instance the case of
conformally flat superspaces.
Given such a supervector field $\x = \x^{A} E_{A} = \widetilde{\x}^{A} 
\widetilde{E}_{A}$, it may be shown that 
\begin{subequations}
\bea
\cK [ \widetilde{\x} ] &:=&  \widetilde{\x}^{B} 
\widetilde{\cD}_{B} + \hf K^{bc}[\widetilde{\x}] M_{bc} 
+ K^{kl} [\widetilde{\x} ]J_{kl}= \cK[\x ]~, \\
\s[ \widetilde{\x} ] &=& \s[\x] - \x \r \ . 
\eea
\end{subequations}


\subsection{Isometries}
\label{Isometries}

In order to describe Poincar\'e supergravity in $3\leq d \leq 6$ dimensions, 
the Weyl multiplet of conformal supergravity has 
to be coupled to some compensating multiplets $\X$. Two compensators are required for theories with eight supercharges such as $\cN=(1,0)$ supergravity in six dimensions. The conceptual setup is actually universal, which is why it is suitable to start with a general discussion of  the situation in $d$ dimensions where conformal supergravity is described using 
$G_R {[d;\cN]}$ superspace 
$(\cM^{d|\d_\cN}, \cD)$, see section \ref{section1}.

In general, the compensators are Lorentz scalars,  and at least one of them 
must have a  non-zero super-Weyl 
weight $w_\X \neq 0$,
\bea
\d_\s\X=w_\X\s\X~.
\label{GenCom}
\eea
They may also transform in some  
representations of the $R$-symmetry group.
The compensators are required to be nowhere vanishing in the sense that 
the $R$-symmetry singlets $|\X|^2 $ should be strictly positive. 
Different off-shell supergravity theories correspond to
different choices of $\X$. The superspace corresponding to Poincar\'e supergravity 
is identified with a triple $(\cM^{d|\d_\cN}, \cD,\X)$. 
The notion of conformally related superspaces, which was introduced in section  
\ref{section3.2}, is naturally generalised to the case under consideration. 
Specifically, two curved superspaces $(\cM^{d|\d_\cN}, \widetilde \cD, \widetilde \X)$ and 
$(\cM^{d|\d_\cN}, \cD,\X)$ are conformally related if their covariant derivatives
related to each other according to 
\eqref{CRSG}, and the compensators $\widetilde \X$ and $\X$ 
are connected by the same finite super-Weyl transformation, 
\bea
\widetilde \X = \re^{w_\X\r}\X~.
\eea
 
Once $\X$ has been fixed, 
the off-shell supergravity multiplet is completely described in terms of the following data:
(i) a superspace geometry for conformal supergravity; and (ii) the conformal compensators. 
Given a supergravity background, its isometries should preserve both of these inputs. 
This leads us to the concept of Killing supervector fields. 

Let $\x= \x^B E_B$ be a conformal Killing supervector field on $(\cM^{d |\d_\cN}, \cD)$,
\begin{subequations}
\bea
(\d_{\cK [\x]} + \d_{\s [\x]}) \cD_A = 0~,  \label{3.7a}
\eea
for uniquely determined parameters
 $K^{bc}[\x]$,  $K^I[\x]$ and $\s[\x]$. 
 It is called a Killing supervector field on $(\cM^{d |\d_\cN}, \cD ,\X)$ if the compensators
are invariant, 
\bea
( {\cK [\x]} + w_\X \s[\x] ) \X =0~.
\label{3.7b}
\eea
\end{subequations}
 The set of Killing vectors on $(\cM^{d |\d_\cN}, \cD,\X)$
is a Lie superalgebra. 
The Killing equations \eqref{3.7a} and \eqref{3.7b} are super-Weyl invariant
in the sense that 
they hold for all conformally related superspace geometries.  

Using the compensators $\X$ we can always construct a 
superfield ${\bm \X} = f(\X)$ 
that is a singlet under the structure group and has the properties: 
(i) it is an algebraic function of $\X$; 
(ii)  it is nowhere vanishing; and 
(iii) it has a  non-zero super-Weyl weight $w_{\bm \X}$, 
$\d_\s {\bm \X} = w_{\bm \X}  \s {\bm \X}$.
It follows from \eqref{3.7b} that  
\bea
(\x^{B} \cD_{B} + w_{\bm \X} \s[\x] ) {\bm \X} =0~.
\label{3.9}
\eea
The super-Weyl invariance may be used to  impose the gauge condition 
\bea
{\bm \X}=1 \ ,
\label{Xgauge1}
\eea
Then eq. \eqref{3.9} reduces to
\bea
\s [\x]&=&0
~,
\label{constr-sigma}
\eea
and the Killing equations \eqref{3.7a} and \eqref{3.7b}  take the following form: 
\begin{subequations}
\bea
\big[{\cK [\x]} ,  \cD_A \big] &=& 0~,  \\
 {\cK [\x]}  \X &=&0~.
\eea
\end{subequations}

Now we specialise the Killing equations \eqref{3.7a} and \eqref{3.7b} to the case  
of $\cN=(1,0)$ supergravity in six dimensions. The equations read
\begin{subequations} 
\bea
\Big[\x^{B} \cD_{B} + \hf K^{bc}[\x] M_{bc} + K^{kl}J_{kl}, \cD_A \Big]  + \d_{\s [\x]} \cD_A &=&0~ ,
\\ 
\Big(\x^{B} \cD_{B}  + K^{kl}[\x]J_{kl}   + w_\X \s [\x] \Big) \X& =&0~.
\label{3.26b}
\eea
\end{subequations}
The most convenient set of compensators for $\cN=(1,0)$ Poincar\'e supergravity \cite{BSVanP} 
consists of  a tensor multiplet $\F $ and a linear multiplet $G^{ij}= G^{ji}$. 
The former is a primary real  scalar of super-Weyl weight $w_\F=2$, which 
obeys the constraint \cite{HSierraT,BSSokatchev,LT-M12}
\bea
\big(\cD_\a^{(i}\cD_{\b}^{j)}+4\ri C_{\a \b}^{ij}\big)\Phi=0
\label{3.35}
\eea
and is nowhere vanishing in the sense that $\F^{-1}$ exists. 
 The latter is  a real $\sSU(2)$ triplet (that is, 
 $\overline{G^{ij}}= G_{ij} =\ve_{ik} \ve_{jl} G^{kl}$), which is a primary 
 superfield of super-Weyl weight $w_{G}=4$ and obeys the constraint  \cite{HSierraT,LT-M12}
  \bea
 \cD_\a^{(i}G^{jk)}=0
\label{3.36}
 ~.
 \eea
 The linear compensator is required to be nowhere vanishing in the sense 
 that $G^{-1}$ exists for $G:=\sqrt{\frac{1}{2} G^{ij}G_{ij}}$.
 There are two natural choices for $\bm \X$: either $\F$ or $G$.

The above formalism will be employed in Section \ref{section8} and Appendix \ref{appendixB}
to study supersymmetric spacetimes in the superspace setting.
Now we will turn to describing the extension of (conformal) Killing vector superfields 
to the case of (conformal) Killing tensor superfields and 
higher symmetries of $\cN=(1,0)$ supermultiplets.


\section{Conformal Killing spinor superfields and their higher rank cousins}
\label{section4}

In this section we introduce various cousins of the conformal Killing vector superfields
$\x^a$, eq. \eqref{ConfKillingVec}. Some of them can be used to describe extended superconformal transformations (conformal Killing spinor superfields) and higher symmetries of $\cN=(1,0)$ supermultiplets (conformal Killing tensor superfields).

In $\sSU(2)$ superspace, a conformal Killing spinor superfield $\e^{\a}$
is defined to satisfy the constraint
\bea
\label{ConfKillingSpinor}
\cD_{\a}^i \e^{\b} = \frac{1}{4} \d_{\a}^{\b} \cD_{\g}^i \e^{\g} ~.
\eea
This equation is super-Weyl invariant provided the super-Weyl transformation of $\e^\a$ is 
\bea
\d_\s \e^\a = - \hf \s \e^\a~.
\eea
In conformal superspace, $\e^\a$ is required to 
(i) be primary and of dimension $-1/2$;
and (ii) obey  the constraint obtained from \eqref{ConfKillingSpinor}
by replacing $\cD$'s with $\nabla$'s. 

Equation \eqref{ConfKillingSpinor}
imposes significant restrictions on the component content of $\e^\a$. 
In particular, the following corollary of \eqref{ConfKillingSpinor}
\bea
\label{CKSCorr}
\cD_{\a}^i \cD_{\b}^j \e^{\b} = - \frac{8 \ri}{3} \ve^{ij} \cD_{\a \b} \e^{\b} + 16 \ri C_{\a \b}^{ij} \e^{\b} + 16 \ri \ve^{ij} N_{\a \b} \e^{\b}~,
\eea
implies that $\e^{\a}|_{\q=0} $ and $\cD_{\a}^i \e^\a |_{\q=0} $ are the only independent component fields. 
Making further use of \eqref{CKSCorr} leads to
\begin{subequations}
\bea
\cD_{\a \b} \e^{\g} &=& - \frac{2}{3} \d^{\g}{}_{[\a} \cD_{\b]\d} \e^{\d} - \ve_{\a \b \s \d} W^{\s \g} \e^{\d}~, \\
\cD_{\a \b} \cD_{\g}^i \e^\g &=& \ve_{\a \b \g \d} \bigg[ \frac{5\ri}{12} \mathcal{W}^{\g i} - 16 \mathcal{C}^{\g i} \bigg] \e^\d + 8 \big[\mathcal{N}_{\a \b , \g}{}^i - \mathcal{C}_{\a \b , \g}^i \big] \e^{\g}~,
\label{444b}
\eea
\end{subequations}
where the torsion superfields on the right-hand side of \eqref{444b} are defined in \eqref{2.6abc}.

Associated with $\e^\a$ is its conjugate $\bar \e^\a$ defined by \eqref{A.12}. 
The latter is also a  conformal Killing spinor superfield. 
We can combine $\e^\a$ and $\bar \e^\a$ into 
a symplectic Majorana spinor $\e^{\a \hat i}$ 
that carries a new  $\sSU(2) \neq \sSU(2)_{R}$ index. 
Such objects naturally arise from an $\mathcal{N} = (2,0) \longrightarrow \mathcal{N} = (1,0)$ superspace reduction, see appendix \ref{reductionAppendix} for more details. Given an $\mathcal{N} = (2,0)$ superconformal theory realised in $\mathcal{N} = (1,0)$ superspace, 
$\e^{\a \hat i}$ describes 
extended superconformal transformations.  

Let $\e_1^{\a}$ and $\e_2^{\a}$ be two conformal Killing spinor superfields. 
Associated with them is a vector superfield
\bea
\label{CKTVec}
\xi^a = \e_1^\a (\g^{a})_{\a\b} \e_2^{\b} ~,
\eea
which is primary and of dimension $-1$. As follows from \eqref{ConfKillingSpinor},
$\x^a$
 satisfies the conformal Killing vector equation \eqref{ConfKillingVec}. As was shown in the previous section, these generate the conformal isometries of superspace.

Given $n$ conformal Killing vector superfields $\xi_1^{a_1}, \dots  , \xi_n^{a_n}$, we find that their {\it symmetric and traceless} product
\bea
\z^{a(n)} = \xi_1^{\{a_1} \dots  \xi_n^{a_{n} \}}:=  \xi_1^{(a_1} \dots  \xi_n^{a_{n} )}
- \text{traces}
\label{traces}
\eea
has the following super-Weyl transformation law
\bea
\d_{\s} \z^{a(n)} = - n \s \z^{a(n)}~,
\label{4.5}
\eea
and satisfies the constraint
\bea
\label{ConfKillingTensor}
\cD_{\a}^{i} \z^{a(n)} = \frac{n}{n+4} (\g^{b (a_1})_{\a}{}^{\b} \cD_{\b}^i \z^{a_2 \dots a_n)}{}_{b} ~.
\eea
We will say that any solution $\z^{a_1\dots a_n} = \z^{\{ a_1 \dots a_n\} } $
to \eqref{ConfKillingTensor} is a conformal Killing tensor superfield.\footnote{Our definition is equivalent to the one proposed in \cite{HL1}, although the $\sSU(2)$ superspace formulation was not used. 
} 
It is clear from \eqref{traces} 
and  \eqref{ConfKillingTensor} that the symmetric and traceless product of two such tensors is also conformal Killing. As will be shown shortly, such tensors generate higher symmetries of the kinetic operators of superconformal field theories with at most two derivatives, in accordance with \cite{HL2}. An immediate consequence of \eqref{ConfKillingTensor} is the usual conformal Killing tensor equation
\bea
\label{ConfKillingEquation}
\cD_{\{a_1} \z_{a_2 \dots a_{n+1} \}} = 0~.
\eea

The above definition of the conformal Killing tensor superfield can be recast in conformal superspace.
A symmetric traceless tensor superfield $\z^{a_1\dots a_n} = \z^{\{ a_1 \dots a_n\} } $
is called conformal Killing if it has the superconformal properties
\bea
K^B \z^{a(n)}  = 0 ~,
 \qquad \mathbb D \z^{a(n)} = -n\z^{a(n)}
\label{SCKTSCProperties}
\eea
and solves the equation 
\bea
\label{ConfKillingTensor2}
\nabla_{\a}^{i} \z^{a(n)} = \frac{n}{n+4} (\g^{b (a_1})_{\a}{}^{\b} \nabla_{\b}^i \z^{a_2 \dots a_n)}{}_{b} ~.
\eea

For a given curved superspace, 
the set of conformal Killing tensor superfields 
may be endowed with an additional algebraic structure. Let $\z^{a(m)}_1$ and $\z^{a(n)}_2$ be two such tensors, then
\bea
[\z_1 , \z_2]^{a(m+n-1)} &=& m \z_1^{\{a_1 \dots a_m-1|b|} \cD_{b} \z_2^{a_m \dots a_{m+n-1} \} } - n \z_2^{\{a_1 \dots a_n-1|b|} \cD_{b} \z_1^{a_n \dots a_{m+n-1} \} } \non \\
&-& \frac{\ri m n}{8(m+n+2)} (\tilde{\g}^{\{a_1})^{\a \b} \cD_\a ^i \z_1^{a_2 \dots a_m |b|} \cD_{\b i} \z_2^{a_{m+1} \dots a_{m+n-1} \}}{}_{b} \non \\
&+& \frac{\ri m n}{8(m+n+2)} (\tilde{\g}^{\{a_1}{}_{bc})^{\a \b} \cD_\a ^i \z_1^{a_2 \dots a_m |b|} \cD_{\b i} \z_2^{a_{m+1} \dots a_{m+n-1} \} c}~,
\label{SNb}
\eea
is a conformal Killing tensor superfield. This generalises the Lie bracket for conformal Killing vector superfields \eqref{superLieBracket} and coincides with the one presented in \cite{HL1}, where it was called the supersymmetric even 
Schouten-Nijenhuis bracket.\footnote{The supersymmetric extension of the even Schouten-Nijenhuis bracket was proposed for the first time 
in the framework of $\cN=1$ AdS supersymmetry in four dimensions
\cite{GKS}, although no mention of the even Schouten-Nijenhuis bracket was made.}

Having investigated the structure of conformal Killing tensors, we now return to the master equation \eqref{ConfKillingSpinor}. This constraint admits non-trivial generalisations\footnote{The case $n=2,~ m =0$ describes a conformal Killing-Yano tensor superfield, introduced in a flat superspace context in \cite{HL3}.}
\bea
\label{CKYT}
\cD_{\a}^{(i_1} \e^{\b(n) i_2 \dots i_{m+1})} = \frac{n}{n+3} \d_{\a}^{(\b_{1}} \cD_{\g}^{(i_1} \e^{\b_2 \dots \b_n) \g i_2 \dots i_{m+1})} ~.
\eea
It is conformally invariant provided
\bea
\d_{\s} \e^{\b(n) i(m)} = \bigg[ 2m - \frac{n}{2} \bigg] \s \e^{\b(n) i(m)}~.
\eea
Given two solutions $\e^{\b(n_1) i(m_1)}$ and $\e^{\b(n_2) i(m_2)}$ to \eqref{CKYT}, one can show that 
\bea
\e^{\b(n_1 + n_2) i(m_1 + m_2)} = \e_1^{(\b_1 \dots \b_{n_1} (i_1 \dots i_{m_1}} \e_2^{\b_{n_1 +1} \dots \b_{n_1 + n_2}) i_{m_1 +1} \dots i_{m_1 + m_2})}~,
\eea
also satisfies this constraint.

Let $\e_1^\a$, $\e_2^\a$ and $\e_3^\a$ be conformal Killing spinors \eqref{ConfKillingSpinor}. It is clear from the analysis above that their totally symmetric product is a solution to \eqref{CKYT}, while their antisymmetric product is dual to a right-handed spinor
\bea
\chi_{\a} = \ve_{\a \b \g \d} \e_1^\b \e_2^\g \e_3^\d ~,
\eea
which satisfies
\bea
\cD_{(\a}^i \chi_{\b)} = 0~.
\eea
This constraint may be immediately generalised
\bea
\label{LHCKS}
\cD_{(\a_1}^{(i_1} \chi_{\a_2 \dots \a_{n+1})}^{i_2 \dots i_{m+1})} = 0~,
\eea
and is conformally invariant provided
\bea
\d_\s \chi_{\a(n)}^{i(m)} = \bigg[ 2m - \frac{3 n}{2} \bigg] \s \chi_{\a(n)}^{i(m)}~.
\eea
If $\chi_{\a(n_1)}^{i(m_1)}$ and $\chi_{\a(n_2)}^{i(m_2)}$ are solutions to this constraint then
\bea
\chi_{\a(n_1+n_2)}^{i(m_1+m_2)} = \chi_{(\a_{1} \dots \a_{n_1}}^{(i_1 \dots i_{m_1}} \chi_{\a_{n_1+1} \dots \a_{n_1 + n_2})}^{i_{m_1 + 1} \dots i_{m_1} + i_{m_2})} ~,
\eea
also solves \eqref{LHCKS}.

We may also construct a hook field from our three spinors
\bea
\ell^{\a \b , \g} = \frac{1}{2} ( \e_1^{[\a} \e_2^{\b]} \e_3^\g - \e_1^{\g} \e_2^{[\a} \e_3^{\b]}) = - \ell^{ \b \a, \g}  ~,
\eea
which satisfies
the Young condition
\bea
\ell^{\a \b , \g} + \ell^{\b \g , \a} + \ell^{\g \a , \b} &=& 0~.
\eea
Further, it satisfies the conformally invariant constraint
\bea
\cD_\a^i \ell^{\b \g , \d} &=& \frac{1}{3}\big( \d_\a^\b \cD_{\e}^i \ell^{\e (\g , \d)} - \d_\a^\g \cD_{\e}^i \ell^{\e (\b , \d)} \big) + \frac{1}{5} \big( \d_\a^\b \cD_{\e}^i \ell^{\e [\g , \d]} - \d_\a^\g \cD_{\e}^i \ell^{\e [\b , \d]} \big)\non \\
&& - \frac{2}{5} \d_\a^\d \cD_\e^i \ell^{\e [\b, \g]} ~.
\eea


\section{Higher symmetries of the conformal d'Alembertian}\label{section5.1}

There has been extensive study of the higher symmetries of the conformal d'Alembertian in dimensions $d>2$, including the important publications \cite{ShSh,Eastwood}. Here we will review known results and present some new observations. The outcomes of the non-supersymmetric analysis in this section will guide our study in the next two sections. 

Let  $\phi$ be a solution to the conformal wave equation in $d$ dimensions
\bea
\Box \phi = \nabla^a \nabla_a \phi = 0~,
\label{5.1}
\eea
where $\nabla_{a}$ is the conformally covariant derivative (compare with \eqref{conf-nabla})
\bea
\nabla_a
= e_a - \hf  {\omega}_a{}^{bc} M_{bc}  - b_a \mathbb D
	-  {\mathfrak{f}}_a{}^b K_b \ ,
\eea
with the commutation relations \cite{BKNT,BKNT-M1}
\bea
\big[ \nabla_{a} , \nabla_{b} \big] = -\frac{1}{2} C_{ab}{}^{cd} M_{cd} + \frac{1}{2(d-3)} \nabla_{c} C_{ab}{}^{cd} K_d ~.
\eea
Equation \eqref{5.1} is known to be conformally invariant if $\f$ has the transformation properties 
\bea
K_a \f  = 0 ~,
 \qquad \mathbb D \f = \hf (d-2) \f ~.
\eea
A differential operator $\mathfrak{D}$ is called a symmetry of the 
conformal d'Alembertian, $\Box$,
if it obeys 
the following conditions: 
\begin{subequations} 
\bea
\Box \mathfrak{D} \f &=& 0~, \label{5.4a}\\
K_a \mathfrak{D} \f &=&0~, \qquad \mathbb D \mathfrak{D} \f =  \hf (d-2) \mathfrak{D} \f ~.
\label{5.4b}
\eea
\end{subequations}
Condition \eqref{5.4b} means that $\mathfrak D$ is a conformal dimension-0 operator. 
The symmetry operators of $\Box$ naturally form an associative algebra. 

In the algebra of all symmetry operators of $\Box$, it is natural to 
introduce 
the equivalence relation
\bea
\label{dAlembertianEQR}
\mathfrak{D}_1 \sim \mathfrak{D}_2 
\quad \Longleftrightarrow \quad
\big( \mathfrak{D}_1 - \mathfrak{D}_2 \big)\phi = 0~.
\eea
Utilising \eqref{dAlembertianEQR}, it is possible to show that every  symmetry operator $\mathfrak D$  
of order $n$ can be reduced to the canonical form
\bea
\label{HSdAlembertianCanonicalForm}
\mathfrak{D}^{(n)}_{\z} = \sum_{k=0}^{n}\z^{a(k)} \nabla_{a_1} \dots \nabla_{a_k} ~, \quad n \geq 0~,
\eea
where the parameters $\z^{a(k)}$ are symmetric and traceless. 
Making use of the condition \eqref{5.4a}, one observes that  
$\z^{a(n)}$ satisfies the conformal Killing tensor equation
\bea
\label{ConfKillingEquation2}
\nabla_{\{a_1} \z_{a_2 \dots a_{n+1} \}} = 0~.
\eea
Due to  \eqref{5.4b}, $\z^{a(n)}$ is primary, $K_b \z^{a(n)} =0$,  and of dimension $-n$.

Let us first study the $n = 0$ and $1$ cases in more detail. It is easily seen that zeroth-order symmetry operator $\mathfrak{D}_{\z}^{(0)}$ is  a constant,
\bea
\Box \mathfrak{D}_{\z}^{(0)} \phi = 0 \quad  \iff \quad \nabla_{a} \z = 0~. 
\eea
Given a conformal Killing vector field $\x^a$, the following first-order operator
\bea
\label{HSLaplacianFirstOrder}
\mathfrak{D}^{(1)}_{\xi} = \xi^a \nabla_a + \frac{d-2}{2d} \nabla_{a} \xi^a~,
\eea
is a symmetry  of the conformal d'Alembertian,
\bea
\Box \mathfrak{D}^{(1)}_{\xi} \phi = 0 ~.
\eea
The second term on the right-hand side of \eqref{HSLaplacianFirstOrder}
is uniquely determined by each of the conditions \eqref{5.4a} and  \eqref{5.4b}. 

Actually, the operator  \eqref{HSLaplacianFirstOrder} is simply a special case of the conformal isometry
\begin{subequations}\label{ConfIso}
\bea
\mathfrak{D}^{(1)}_{\xi} &=& \xi^a \nabla_a + \hf \nabla^a \xi^b M_{ab} + 
\frac{1}{d} \nabla^a \xi_a \mathbb{D} + \frac{1}{2d} \nabla^a \nabla^b \xi_b K_a ~, \label{5.11a}\\
\big[ \mathfrak{D}^{(1)}_{\xi} , \nabla_{a} \big] &= &0 ~,
\label{5.11b}
\eea
\end{subequations}
which reduces to \eqref{HSLaplacianFirstOrder} when acting on any primary scalar field
of dimension $\hf (d-2)$. In addition to \eqref{5.11b}, the other fundamental property of 
$\mathfrak{D}^{(1)}_{\xi} $ is the following: 
\bea
K_a T =0 ~, \quad {\mathbb D} T =w T \quad \implies \quad 
K_a \mathfrak{D}^{(1)}_{\xi} T =0 ~, \quad {\mathbb D} \mathfrak{D}^{(1)}_{\xi} T 
= w \mathfrak{D}^{(1)}_{\xi} T~,
\label{5.12} 
\eea
for every primary tensor field $T$ (with suppressed indices) of dimension $w$. 
The relations \eqref{5.11b} and \eqref{5.12} tell us that 
$\mathfrak{D}^{(1)}_{\xi}$ satisfies the conditions \eqref{5.4a} and 
\eqref{5.4b}, and therefore $\mathfrak{D}^{(1)}_{\xi}$ is a symmetry  of the conformal d'Alembertian. An immediate corollary of the above consideration is that, for any conformal Killing
vector fields $\x^a_1, \x^a_2, \dots \x^a_n$, the operator
\bea
 \mathfrak{D}^{(n)}:=  \mathfrak{D}^{(1)}_{\xi_1}  \mathfrak{D}^{(1)}_{\xi_2} 
 \dots  \mathfrak{D}^{(1)}_{\xi_n}
\label{5.13}
\eea
is a symmetry  of the conformal d'Alembertian. Therefore, the algebra of symmetries of $\Box$ includes the universal enveloping algebra of the conformal algebra of the background spacetime.

Our consideration above allows for important generalisations. Consider a dynamical system described by primary fields $\vf^i$ coupled to conformal gravity. 
We place this theory on a fixed gravitational background 
and consider a conformal Killing vector field, $\x^a$, on spacetime.   
Since the operator \eqref{5.11a} 
preserves the background geometry, the matter action $S[ \vf ]$ is invariant under the conformal 
transformation $\d_\x \vf^i  = \mathfrak{D}^{(1)}_{\xi} \vf^i$.
Consequently, $\mathfrak{D}^{(1)}_{\xi} $ is a symmetry of the corresponding equation 
of motion,  $S_{, i} [ \vf ]=0$. 

Let us now return to the general symmetry operator \eqref{HSdAlembertianCanonicalForm}
for $n>1$.
Similar to the first-order operator  \eqref{HSLaplacianFirstOrder}, we would like 
$\mathfrak{D}^{(n)}_{\z} $ to be determined by its top component, which is 
the conformal Killing tensor $\z^{a(n)}$. 
Imposing 
the condition \eqref{5.4b} 
leads to 
\begin{subequations}\label{symbol}
\bea
\z^{a(k)} = A_{k} \nabla_{b_1} \dots \nabla_{b_{n-k}} \z^{a(k) b(n-k)}~, \quad 0 \leq k \leq n ~,
\eea
where the constants $A_{k}$ are given by the solution to the recurrence relation
\bea
\frac{A_{k-1}}{A_k} = \frac{k(4-2k-d)}{2(k(k+d-3) - n(n+d-1) - d + 2)} ~, \quad A_n = 1~.
\eea
\end{subequations} 
In order for the constructed operator to be a symmetry of $\Box$, it turns out that
 the background geometry must be conformally flat. To prove this claim it suffices to 
 analyse the $n=2$ case. 

We assert that the second-order operator
\bea
\mathfrak{D}^{(2)}_{\z} = \z^{ab} \nabla_a \nabla_b + \frac{d}{d+2} \nabla_b \z^{ab} \nabla_a + \frac{d(d-2)}{4 (d+1)(d+2)} \nabla_{a} \nabla_b \z^{ab}~,
\eea
only results in a symmetry in backgrounds with vanishing Weyl tensor
\bea
\label{n=2Result}
\Box \mathfrak{D}^{(2)}_{\z} \phi = 0\quad  \Longleftrightarrow \quad C_{abcd} = 0~.
\eea
The direct computation necessary to verify \eqref{n=2Result} is tedious, thus here we will present a simpler proof. Consider two conformal Killing vectors $\xi_{1}^{a}$ and $\xi_{2}^{a}$
and the corresponding 
first-order symmetry operators $\mathfrak{D}^{(1)}_{\xi_1}$ and $ \mathfrak{D}^{(1)}_{\xi_2}$
defined 
 via \eqref{HSLaplacianFirstOrder}. Then their product $\mathfrak{D}^{(2)}: = \mathfrak{D}^{(1)}_{\xi_1} \mathfrak{D}^{(1)}_{\xi_2}$ is a second-order symmetry operator. Modulo the equivalence relation \eqref{dAlembertianEQR}, it may be expressed as a sum of operators of the form \eqref{HSdAlembertianCanonicalForm}:
\bea
\mathfrak{D}^{(2)} \sim \mathfrak{D}_\z^{(2)} + \frac{1}{2} \mathfrak{D}^{(1)}_{[\xi_1 , \xi_2]} 
- \frac{d-2}{4(d+1)} \mathfrak{D}^{(0)}_{\langle \xi_1 , \xi_2 \rangle} ~,
\eea
where
\begin{subequations}
	\bea
	\z^{ab} &=& \xi_1^{ \{ a } \xi_2^{ b \}} ~, \\
	\big[ \xi_{1} , \xi_{2} \big]^{a} &=& \xi_1^b \nabla_b \xi_2^a - \xi_2^b \nabla_b \xi_1^a ~, \\
	\langle \xi_1 , \xi_2 \rangle &=& \nabla_b \xi_1 ^a \nabla_a \xi_2 ^b - \frac{d-2}{d} \nabla_a \xi_1 ^a \nabla_b \xi_2 ^b - \frac{2}{d} \Big( \xi_1^a \nabla_a \nabla_b \xi_2 ^b + \xi_2^a \nabla_a \nabla_b \xi_2^b \Big) ~.
	\eea
\end{subequations}
As $\mathfrak{D}^{(2)}= \mathfrak{D}^{(1)}_{\xi_1} \mathfrak{D}^{(1)}_{\xi_2}$ is a symmetry operator by construction, we obtain
\bea
\label{Symmrank2+0}
\Box \mathfrak{D}^{(2)} \phi \sim \Box \mathfrak{D}^{(2)}_\z \phi 
- \frac{d-2}{4(d+1)}  \Box \mathfrak{D}^{(0)}_{\langle \xi_1 , \xi_2 \rangle} \phi = 0~,
\eea
which has the immediate consequence
\bea
\Box \mathfrak{D}^{(2)}_{\z} \phi = 0\quad  \iff \quad \nabla_{a} \langle \xi_{1} , \xi_2 \rangle = 0~.
\eea
Now a direct computation leads to
\bea
\nabla_{a} \langle \xi_1 , \xi_2 \rangle = - C_{abcd} \nabla^c \z^{bd} - \frac{d(d-2)}{d-3} \nabla^c C_{abcd} \z^{bd} ~.
\eea
Hence, $\mathfrak{D}^{(2)}_{\z}$ only results in a symmetry when the background Weyl tensor vanishes. We thus expect that for $n \geq 2$ the operator $\mathfrak{D}^{(n)}_{\z}$ is a symmetry only in conformally flat backgrounds.

A natural extension of the analysis above is to determine if there exists a symmetry operator of the form
\bea
\mathfrak{D}^{(2)} = \mathfrak{D}^{(2)}_{\z} + \mathcal{Z} ~, \quad \nabla_{a} \mathcal{Z} = - C_{abcd} \nabla^c \z^{bd} - \frac{d(d-2)}{d-3} \nabla^c C_{abcd} \z^{bd} ~,
\eea
when the conformal Killing tensor is irreducible ($\z^{ab} \neq \xi_1^{\{a} \xi_2^{b\}}$). This fails since $\mathcal{Z}$ must be primary, of dimension $0$ and first order in $\z^{ab}$. It is easily verified that no such term exists.

Our analysis in this section has lead to several non-trivial results regarding the structure of higher symmetry operators. 
The most important observation is that, beyond the first-order case, these operators are defined only on conformally flat 
backgrounds.\footnote{ This is in keeping with the fact that the  background geometry generally restricts Killing tensors, see, 
e.g., \cite{Weir,Thompson}.} 
It may be shown that our arguments immediately generalise to the higher symmetries of superconformal operators. In 
particular, such symmetries exist only for superspace backgrounds with vanishing super-Weyl tensor, $W^{\a \b} = 0$.


\section{Higher symmetries of the hypermultiplet}
\label{section6}

The study of symmetries of relativistic wave equations has a long-standing history in mathematical physics. More recently their supersymmetric generalisations have also been explored. Specifically, in flat superspace it was shown in \cite{HL2} that the higher symmetries of so-called `super-Laplacians' (superspace differential operators containing the spacetime Laplacian as their highest-dimensional component) are in one-to-one correspondence with conformal Killing tensor superfields. Further, the higher symmetries of the Wess-Zumino operator in curved $d=4$, $\cN=1$ superspace were analysed in \cite{KR}. It is now time to extend this analysis to the hypermultiplet.

The off-shell formulation for a hypermultiplet coupled to conformal supergravity is given
in  \cite{LT-M12}. 
On the mass shell, the hypermultiplet is described by an isospinor superfield $q^{i}$ satisfying 
the equation
\bea
\label{HM}
\nabla_{\a}^{(i} q^{j)} = 0~.
\eea
The constraint is conformally invariant provided $q^i$ is a primary superfield of dimension 2
\bea
\label{HMSW}
K_{A} q^i = 0~, \quad
\mathbb{D} q^i = 2 q^i~.
\eea
Additionally, \eqref{HM} yields the useful corollary
\bea
\label{HMCorollary}
\nabla_{\a}^{i} \nabla_{\b}^{j} q_{j} &=& - 4 \ri \nabla_{\a \b} q^i ~.
\eea

Here we will study the higher symmetries of this model. We will say that a differential operator $\mathfrak{D}$ is a symmetry operator (of the hypermultiplet) if 
\bea
\label{HigherSymmConstraint}
\nabla_{\a}^{(i} \mathfrak{D}q^{j)} = 0 ~.
\eea
It is useful to introduce an equivalence relation on the space of symmetries so that redundant structures can be discarded. Specifically, we say that two symmetry operators $\mathfrak{D}_{1}$ and $\mathfrak{D}_2$ are equivalent if 
\bea
\label{HMEQR}
\mathfrak{D}_1 \sim \mathfrak{D}_2 \quad 
 \Longleftrightarrow \quad
  \big( \mathfrak{D}_1 - \mathfrak{D}_2 \big) q ^{i} = 0 ~.
\eea
Owing to \eqref{HMSW}, we will also require
\bea
\label{HMSymmSW}
K_{A} \mathfrak{D} q^i = 0~, \quad
\mathbb{D}  \mathfrak{D} q^i = 2 q^i~,
\eea
which means
that $\mathfrak{D}$ is a superconformal dimension-0 operator.

Given a positive integer $n$, the most general $n^\text{th}$-order symmetry operator $\mathfrak{D}^{(n)}$ is
\bea
\label{HSGeneralForm}
\mathfrak{D}^{(n)} = \sum_{k=0}^{n} \z^{A_{1} \dots A_{k}} \nabla_{A_k} \dots \nabla_{A_1} + \sum_{k=0}^{n-1} \z^{A_{1} \dots A_{k},ij} \nabla_{A_k} \dots \nabla_{A_1} J_{ij}~,
\eea
where the coefficients may be chosen to be graded-symmetric in their superspace indices
\begin{subequations}
\bea
\z^{A_{1} \dots A_{m} A_{m+1} \dots A_{k}} &=& (-1)^{\ve_{A_{m}} \ve_{A_{m+1}}} \z^{A_{1} \dots A_{m+1} A_{m} \dots A_{k}} ~, \\ \quad \z^{A_{1} \dots A_{m} A_{m+1} \dots A_{k}, ij} &=& (-1)^{\ve_{A_{m}} \ve_{A_{m+1}}} \z^{A_{1} \dots A_{m+1} A_{m} \dots A_{k} , ij} ~.
\eea
\end{subequations}
The equivalence relation \eqref{HMEQR} allows us to bring $\mathfrak{D}^{(n)}$ to the canonical form
\bea
\label{HSCanonicalForm}
\mathfrak{D}^{(n)} &=& \sum_{k=0}^{n} \z^{a(k)} \nabla_{a_1} \dots \nabla_{a_k} + \sum_{k=0}^{n-1} \z^{a(k) \b}{}_i \nabla_{a_1} \dots \nabla_{a_k} \nabla_{\b}^{i} \non \\
&+&  \sum_{k=0}^{n-1} \z^{a(k)ij} \nabla_{a_1} \dots \nabla_{a_k} J_{ij} ~.
\eea
Here all parameters are symmetric and traceless in their vector indices, $\z^{a(k)\b}{}_i$ is gamma-traceless, $(\g_b)_{\a \b}\z^{a(k-1)b\b}{}_i = 0$ , and $\z^{a(k)ij}$ is symmetric in its isospinor indices.

Equation \eqref{HigherSymmConstraint} yields numerous constraints on the parameters of $\mathfrak{D}^{(n)}$, including
\begin{subequations}
\bea
\label{HSHMImplications}
\nabla_{\a}^{i} \z^{a(n)} &=& \frac{n}{n+4} (\g^{b (a_1)})_{\a}{}^{\b} \nabla_{\b}^i \z^{a_2 \dots a_n)}{}_{b} ~, \\
\z^{a(n-1) \b i} &=& \frac{\ri}{4(n+2)} \nabla_{\a}^i \z^{a(n-1)b} (\tilde{\g}_b)^{\a \b}~, \\
\z^{a(n-1) ij} &=& - \frac{\ri n (n+1)}{8(n+2)(n+3)} \nabla^{(i} \tilde{\gamma}_{b} \nabla^{j)} \z^{a(n-1) b} ~.
\eea
\end{subequations}
Hence, we obtain expressions for $\z^{a(n-1) \b i}$ and $\z^{a(n-1) ij}$ in terms of $\z^{a(n)}$, which is necessarily a conformal Killing tensor \eqref{SCKTSCProperties}, \eqref{ConfKillingTensor2}. Further, if $\mathfrak{D}^{(n)}$ is completely determined in terms of $\z^{a(n)}$, we will denote it by $\mathfrak{D}^{(n)}_\z$.

If the supergravity background admits a conformal Killing vector superfield $\x^a$, it may be shown that the corresponding conformal isometry \eqref{3.12} yields the unique first-order symmetry operator
\bea
\nabla_{\a}^{(i} \mathfrak{D}^{(1)}_{\xi} q^{j)} = \nabla_{\a}^{(i} \Big[ \xi^B  {\nabla}_B 	+  {\L}^{jk} [\xi] J_{jk} + 2 \s[\xi] \Big] q^{j)} = 0 ~.
\eea
Thus, given conformal Killing
vector superfields $\x^a_1, \x^a_2, \dots \x^a_n$, the operator
\bea
\label{ProductofSymms}
\mathfrak{D}^{(n)}:=  \mathfrak{D}^{(1)}_{\xi_1}  \mathfrak{D}^{(1)}_{\xi_2} 
\dots  \mathfrak{D}^{(1)}_{\xi_n}
\eea
satisfies \eqref{HigherSymmConstraint}. Therefore, the algebra of such symmetries contains the universal enveloping algebra of the conformal algebra of the background superspace. As was discussed in the previous section, \eqref{ProductofSymms} admits a decomposition as a sum of symmetry operators determined by their top component
\bea
\label{ProductofSymmsCanonicalForm}
\mathfrak{D}^{(n)} = \mathfrak{D}^{(n)}_{\z} + \dots + \z_0 ~,
\eea
only when the superspace is conformally-flat. Therefore, it is of interest to construct the symmetry operators $\mathfrak{D}^{(n)}_{\z}$ in backgrounds with vanishing super-Weyl tensor, $W^{\a \b} = 0$.

Here we will restrict our attention to the evaluation of $\mathfrak{D}^{(2)}_{\z}$. When acting on the hypermultiplet it takes the form
\bea
\label{Rank2HS}
\mathfrak{D}^{(2)}_\z q^i &=& \z^{a b} \nabla_{a} \nabla_{b} q^i - \hf \z^{a \a i} \nabla_{a} \nabla_{\a j} q^j + \z^{a i}{}_{j} \nabla_{a} q^j + \z^{a} \nabla_{a} q^i - \hf \z^{\a i} \nabla_{\a j} q^j \non \\
&+& \z^{i}{}_{j} q^j + \z q^i ~.
\eea
The unique solution compatible with \eqref{HigherSymmConstraint} and \eqref{HMSymmSW} is
\begin{subequations}
\label{Rank2CSSSoln}
\bea
\z^{a \a i } &=& - \frac{\ri}{16} (\tilde{\g}_b)^{\a \b} \nabla_{\b}^i \z^{ab} ~, \\
\z^{a ij} &=& - \frac{3}{80} \ri \nabla^{(i} \tilde{\g}_b \nabla^{j)} \z^{ab} ~, \\
\z^{a} &=& \frac{3}{4} \nabla_{b} \z^{ab} ~, \\
\z^{\a i} &=& -\frac{\ri}{20} (\tilde{\gamma}_a)^{\a \b} \nabla_{\b}^i \nabla_b \z^{ab} + \frac{1}{800} (\tilde{\g}_a)^{\a \b} (\tilde{\g}_b)^{\g \d} \nabla_{\g}^i \nabla_{(\b}^j \nabla_{\d) j} \z^{ab} ~, \\
\z^{ij} &=& - \frac{\ri}{80} \nabla^{(i} \tilde{\gamma}_{a} \nabla^{j)} \nabla_b \z^{ab} ~, \\
\z &=& - \frac{3}{20} \nabla_{a} \nabla_{b} \z^{ab} + \frac{1}{800} (\tilde{\g}_a)^{\a \b} (\tilde{\g}_b)^{\g \d} \nabla_{(\a}^i \nabla_{\b)i} \nabla^j_{(\g} \nabla_{\d)j} \z^{ab} ~.
\eea
\end{subequations}
In particular, we find that all parameters are expressed solely in terms of the conformal Killing tensor $\z^{ab}$.

For completeness, we also present the $\sSU(2)$ superspace form of $\mathfrak{D}^{(2)}_\z$. A routine degauging leads to
\bea
\label{Rank2SU2Op}
\mathfrak{D}_\z^{(2)}q^i &=& \z^{a b} \cD_{a} \cD_{b} q^i - \hf \hat{\z}^{a \a i} \cD_{a} \cD_{\a j} q^j + \hat{\z}^{a i}{}_{j} \cD_{a} q^j + \hat{\z^{a}} \cD_{a} q^i - \hf \hat{\z}^{\a i} \cD_{\a j} q^j \non \\
&+& \hat{\z}^{i}{}_{j} q^j + \hat{\z} q^i ~,
\eea
where have employed the definitions
\allowdisplaybreaks{
\begin{subequations}
\label{Rank2SU2Soln}
\bea
\hat{\z}^{a \a i } &=& -\frac{\ri}{16} (\tilde{\g}_b)^{\a \b} \cD_{\b}^i \z^{ab} ~, \\
\hat{\z}^{a ij} &=& - \frac{3}{80} \ri \cD^{(i} \tilde{\g}_b \cD^{j)} \z^{ab} - \frac{2}{5} C_{b}^{ij} \z^{ab} ~, \\
\hat{\z}^{a} &=& \frac{3}{4} \cD_{b} \z^{ab} ~, \\
\hat{\z}^{\a i} &=& - \frac{\ri}{20} (\tilde{\gamma}_a)^{\a \b} \cD_{\b}^i \cD_b \z^{ab} + \frac{1}{800} (\tilde{\g}_a)^{\a \b} (\tilde{\g}_b)^{\g \d} \cD_{\g}^i \cD_{(\b}^j \cD_{\d) j} \z^{ab} \non \\
&+& \frac{\ri}{25} (\tilde{\gamma}_a)^{\a \b} C_{b}^{ij} \cD_{\b j} \z^{ab} - \frac{11}{50} \ri (\tilde{\gamma}_a)^{\a \b} \mathcal{N}_{b \b}{}^i \z^{ab} + \frac{\ri}{10} (\tilde{\gamma}_a)^{\a \b}  \mathcal{C}_{b \b}{}^i \z^{ab} , \\
\hat{\z}^{ij} &=& -\frac{\ri}{80} \cD^{(i} \tilde{\gamma}_{a} \cD^{j)} \cD_b \z^{ab}  - \frac{\ri}{80} (\tilde{\gamma}_a)^{\a \b} \mathcal{C}_{b \a}{}^{(i} \cD_{\b}^{j)} \z^{ab} - \frac{3}{80} \ri (\tilde{\gamma}_a)^{\a \b} \mathcal{N}_{b \a}{}^{(i} \cD_{\b}^{j)} \z^{ab} \non \\
&+& \frac{\ri}{80} C_{a}^{k(i} \cD^{j)} \tilde{\g}_{b} \cD^{k} \z^{ab} - \frac{\ri}{10} (\tilde{\gamma}_a)^{\a \b} \cD_{\a}^{(i} \mathcal{C}_{b \b}{}^{j)} \z^{a b} + \frac{\ri}{10} (\tilde{\gamma}_a)^{\a \b} \cD_{\a}^{(i} \mathcal{N}_{b \b}{}^{j)} \z^{a b} \non \\
&+& \frac{5}{16} \ri (\tilde{\gamma}_a)^{\a \b}  \mathcal{C}_{b \a}{}^{(i} \cD_{\b}{}^{j)} \z^{a b} - \frac{5}{16} \ri (\tilde{\gamma}_a)^{\a \b}  \mathcal{N}_{b \a}{}^{(i} \cD_{\b}{}^{j)} \z^{a b} + \cD_{a} C_{b}^{ij} \z^{a b} ~, \non \\
\hat{\z} &=& - \frac{3}{20} \cD_{a} \cD_{b} \z^{ab} + \frac{1}{800} (\tilde{\g}_a)^{\a \b} (\tilde{\g}_b)^{\g \d} \cD_{(\a}^i \cD_{\b)i} \cD^j_{(\g} \cD_{\d)j} \z^{ab} + \frac{73}{800} \ri C_{a}^{ij} \cD_{(i} \tilde{\g}_{b} \cD_{j)} \z^{a b} \non \\
&-& \frac{\ri}{10} (\tilde{\g}_a)^{\a \b} (\tilde{\g}_b)^{\g \d} N_{\a \g} \cD_{(\b}^i \cD_{\d) i} \z^{a b}  + \frac{\ri}{50} (\tilde{\g}_a)^{\a \b} \cD_{\a}^i \mathcal{N}_{b \b}{}^{i} \z^{ab} + \frac{\ri}{10} (\tilde{\g}_a)^{\a \b} \cD_{\a}^i \mathcal{C}_{b \b}{}^{i} \z^{ab}  \non \\
&+& \frac{81}{10} C_{a}^{ij} C_{b ij} \z^{a b} - \frac{12}{5}  (\tilde{\g}_a)^{\a \b} (\tilde{\g}_b)^{\g \d} N_{\a \g} N_{\b \d} \z^{a b} + \frac{91}{400} \ri (\tilde{\g}_{a})^{\a \b} \mathcal{C}_{b \a}{}^{i} \cD_{\b i} \z^{a b} \non \\
&-& \frac{47}{80} \ri (\tilde{\g}_{a})^{\a \b} \mathcal{N}_{b \a}{}^{i} \cD_{\b i} \z^{a b} ~.
\eea
\end{subequations}}
It may be verified that \eqref{Rank2SU2Op} is a superconformal dimension-0 operator
\bea
\d_{\s} \mathfrak{D}^{(2)}_\z q^i = 2 \s \mathfrak{D}^{(2)}_\z q^i ~,
\eea
and yields a symmetry on conformally-flat superspace backgrounds
\bea
\cD_{\a}^{(i} \mathfrak{D}^{(2)}_{\z} q^{j)} = 0~.
\eea

As a result, we have shown the existence of the higher symmetry operator $\mathfrak{D}^{(2)}_{\z}$, which is completely determined in terms of $\z^{ab}$.
We expect that, in conformally-flat superspaces, this is true for symmetries of all orders; every $\mathfrak{D}^{(n)}_{\z}$ is uniquely determined in terms of its top component, the conformal Killing tensor superfield $\z^{a(n)}$, as was shown for the non-supersymmetric case in section \ref{section5.1}.

\section{Higher symmetries of the vector multiplet}
\label{section7}

The higher symmetry operators \eqref{HSGeneralForm} 
belong to a broader family of symmetry operators acting on tensor superfields of arbitrary index structure. Here we will generalise these operators by adding Lorentz dependent terms via an analysis of the higher symmetries of the vector multiplet. 

\subsection{Superconformal vector multiplet} 


Consider a vector multiplet coupled to conformal supergravity. Its dynamics is descried by the higher-derivative 
action constructed in \cite{BKNT}, which is a locally supersymmetric extension of $F \Box F$. 
The vector multiplet can be realised in terms of the field strength  $F^{\a i}$ subject 
to the Bianchi identities \cite{HSierraT,BKNT,Siegel79}
\bea
\label{BI}
\nabla_{\a}^{(i} F^{\b j)} - \frac{1}{4} \d_{\a}^{\b} \nabla_{\g}^{(i} F^{\g j)} = 0 ~, \quad \nabla_\a^i F_i^\a = 0 ~.
\label{7.1} 
\eea
The field strength  $F^{\a i}$ is a primary superfield of dimension $\frac{3}{2}$,
\bea
 S^\b_j F^{\a i} =0~,\qquad
 \bbD F^{\a i}= \frac{3}{2} F^{\a i}~. 
\eea
The equation of motion for the superconformal vector multiplet \cite{KNT} is
\bea 
\cG^{ij} := \nabla^a \nabla_a  X^{ij} - 2 Y_\a{}^\b{}^{ij} F_\b{}^\a 
+ \frac{5}{2} X^{\a (i} \overleftrightarrow{\nabla}_{\a\b} F^{\b j)} =0~,
\label{conformalVM}
\eea
where we have defined  $S {\overleftrightarrow{\nabla}}_a T := S \nabla_a T - (\nabla_a S) T$, for arbitrary 
superfields $S$ and $T$, and introduced the following descendants of $F^\a_i$:
\bea
X^{ij} := \frac{\ri}{4} \nabla^{(i}_\g F^{\g j)} ~, \qquad 
F_\a{}^\b := - \frac{\ri}{4} \Big( \nabla_\a^k F^\b_k - \frac{1}{4} \d_\a^\b  \nabla_\g^k F^\g_k \Big)
= - \frac{\ri}{4}  \nabla_\a^k F^\b_k ~.
\eea
The equation of motion \eqref{conformalVM} involves 
the torsion superfields $Y_\a{}^\b{}^{ij} $ and $X^{\a i}$ which are defined according to  \eqref{XYfields}.

Let $F^{\a i}$ be a solution of the equations
\eqref{7.1} and \eqref{conformalVM}. 
A superconformal dimension-0 operator $\mathfrak{D} $ is called a symmetry of these equations if 
$\mathfrak{D} F^{\a i} $ is also a solution. 
Given a conformal Killing vector superfield $\x^a $, the first-order operator $\mathfrak{D}^{(1)}_{\xi} $ defined 
by \eqref{3.12} is a symmetry. Higher-order symmetries of the equations
\eqref{7.1} and \eqref{conformalVM}
may be generated by considering products of
the first-order symmetries, 
\bea
 \mathfrak{D}^{(n)}:=  \mathfrak{D}^{(1)}_{\xi_1}  \mathfrak{D}^{(1)}_{\xi_2} 
 \dots  \mathfrak{D}^{(1)}_{\xi_n}~.
 \eea


\subsection{Supersymmetric Maxwell theory} 
 
In $\sSU(2)$ superspace, the off-shell vector multiplet is described by a superfield $F^{\a i}$ 
subject to the constraints
\bea
\label{OffShellVM}
\cD_{\a}^{(i} F^{\b j)} - \frac{1}{4} \d_{\a}^{\b} \cD_{\g}^{(i} F^{\g j)} = 0 ~, \quad \cD_\a^i F_i^\a = 0 ~.
\eea
These constraints are super-Weyl invariant provided 
$F^{\a i}$ is a primary superfield with the super-Weyl transformation 
\bea
\d_{\s} F^{\a i} = \frac{3}{2} \s F^{\a i} ~.
\eea
When the vector multiplet is placed on-shell, it obeys the additional equtation
\bea
\label{OnShellVM}
\cD_{\a}^{(i} F^{\a j)} = 0 \quad  \implies \quad \cD_{\a}^{(i} F^{\b j)} = 0
\eea

It is important to note that the equation \eqref{OnShellVM}  is not super-Weyl invariant, 
since the super-Weyl invariance has been fixed by imposing an appropriate gauge condition.
In the superconformal setting to Poincar\'e supergravity, the vector multiplet couples to the tensor compensator
$\F$ introduced in section \ref{Isometries}. The compensator appears in the superspace action for the vector multiplet \cite{LT-M12}, and the action is super-Weyl invariant. The corresponding equation of motion for the vector multiplet is 
\bea
\frac14 \F \cD_{\a}^{(i} F^{\a j)}  +\cD_{\a}^{(i} \F F^{\a j)} =0~.
\eea
Choosing the super-Weyl gauge 
\bea
\F =1
\label{7.10}
\eea
reducing the equation of motion to \eqref{OnShellVM}. In this subsection we make use of the super-Weyl gauge \eqref{7.10}.

We now turn to the analysis of the higher symmetries of this theory. The operator $\mathfrak{D}$ is said to be a symmetry of the (on-shell) vector multiplet if
\bea
\label{HSVMOnShell}
\cD_{\a}^{(i} \mathfrak{D} F^{\b j)} = 0 ~, \quad \cD_{\a}^{i} \mathfrak{D} F^{\a}_{i} = 0 ~.
\eea
The most general $n^{th}$-order symmetry operator for the vector multiplet is
\bea
\label{HSCanonicalFormVM}
\mathfrak{D}^{(n)} &=&
\sum_{k=0}^{n} \z^{a(k)} \mathcal{D}_{a_1} \dots \mathcal{D}_{a_k} + \sum_{k=0}^{n-1} \z^{a(k) \b}{}_{i} \mathcal{D}_{a_1} \dots \mathcal{D}_{a_k} \mathcal{D}_{\b}^i \non \\
&+& \sum_{k=0}^{n-1} \z^{a(k)ij} \mathcal{D}_{a_1} \dots \mathcal{D}_{a_k} J_{ij} + \frac{1}{2} \sum_{k=0}^{n-1} \z^{a(k),bc} \mathcal{D}_{a_1} \dots \mathcal{D}_{a_k} M_{bc} \non \\
&+& \frac{1}{2} \sum_{k=0}^{n-1} \z^{a(k-1) \a ,bc}{}_{i} \mathcal{D}_{a_1} \dots \mathcal{D}_{a_{k-1}} \cD_{\a}^i M_{bc}
+ \frac{1}{2} \sum_{k=0}^{n-2} \z^{a(k),bc ij} \mathcal{D}_{a_1} \dots \mathcal{D}_{a_k} M_{bc} J_{ij} ~,
\eea
where $\z^{a(k)}$ is a conformal Killing tensor \eqref{ConfKillingTensor}. When $n=1$, its action on $F^{\a i}$ reduces to
\bea
\mathfrak{D}^{(1)}_{\xi} F^{\a i}= \d_{\mathcal{K}[\xi]}  F^{\a i} = \big( \xi^{B} \mathcal{D}_{B} + K_{jk} [\xi] J^{jk} + \frac{1}{2} K^{bc} [\xi] M_{bc}\big) F^{\a i}~.
\eea

As the procedure to compute \eqref{HSCanonicalFormVM}, say for $n=2$, is analogous to that of \eqref{HSCanonicalForm} for the hypermultiplet, we will not pursure such analysis here. Instead, we will extract some non-trivial information regarding the structure of this operator for general $n$ via \eqref{HSVMOnShell}. Our analysis reveals the following restrictions on its parameters:
\bea
\z^{a(n-1) \b i} &=& \frac{\ri}{4(n+2)} \nabla_{\a}^i \z^{a(n-1)b} (\tilde{\g}_b)^{\a \b}~, \\
\z^{a(n-1) ij} &=& - \frac{\ri n (n+1)}{8(n+2)(n+3)} \nabla^{(i} \tilde{\gamma}_{b} \nabla^{j)} \z^{a(n-1) b} ~, \\
\z^{a(n-1)}{}_{\a}{}^{\b} &=& -\frac{\ri}{8(n+2)} \big(  (\tilde{\g}_b)^{\b \g} \cD_{\a}^{i} \cD_{\g i} - \frac{1}{4} \d_{\a}^{\b} \cD^{i} \tilde{\g}_b \cD_i \big) \z^{a(n-1)b} \non \\
&+& n \z^{a(n-1)b} N_{bcd} (\g^{cd})_{\a}{}^{\b} ~,
\eea
as well as the Killing condition for tensor superfields
\bea
\cD^{b} \z_{a(n-1)b} = 0 \quad \Longrightarrow \quad  \cD_{(a_1} \z_{a_2 \dots a_{n+1})} = 0~.
\label{7.15}
\eea

Equations \eqref{ConfKillingTensor} and \eqref{7.15} define  $\cN=(1,0)$ Killing tensor superfields in six dimensions.\footnote{The concept of a Killing tensor superfield was introduced for the first time in \cite{GKS} in the framework of $\cN=1$ AdS supersymmetry in four dimensions.} Given two Killing tensor superfields $\z_1^{a(m)}$ and $\z_2^{b(n)}$, it may be shown that their bracket, defined by \eqref{SNb}, is also Killing
\bea
\cD^{b} [ \z_1 , \z_2 ]_{a(m+n-2)b} = 0 ~.
\eea


\section{Maximally supersymmetric backgrounds}
\label{section8}

The existence of (conformal) Killing vector and tensor superfields places non-trivial restrictions 
on the superspace geometry. So far we have not examined the constraints imposed by such
conditions.
In this section 
the case of Killing vectors (where eq.~\eqref{constr-sigma} is imposed) is further elaborated on.
More results are given in appendix \ref{appendixB}, where we 
discuss how to obtain component results from superspace.

Here we restrict ourselves to the case of eight supercharges, i.e. maximally $\cN=(1,0)$ supersymmetric 
backgrounds, and derive constraints on the superspace geometry.
By a similar analysis\footnote{For any background admitting eight supercharges, 
if there is a  tensor superfield $T$ such that 
its bar-projection vanishes, $T|=0$, and this condition is supersymmetric,
then the entire superfield is zero, $T=0$. See \cite{KNT-M} for a more detailed discussion.} 
to \cite{KNT-M},
in such backgrounds it may be shown that 
\eqref{BB} implies
\bea
\cD_\a^i C_{a}{}^{kl}= 0~, \qquad 
\cD_\a^i W_{abc} = 0~, \qquad 
\cD_\a^i N_{abc} = 0~. 
\label{no-full-Cond}
\eea
Additionally, the Killing spinor equation \eqref{conf-Killing-0} reduces to
\be \cD_a \xi^\g_k = (\g_{ab})_\b{}^\g \xi^{\b j} C^b{}_{jk}
+ \hf (\g^{bc})_\b{}^\g \xi^\b_k ( W_{abc} + 2 N_{abc} ) \ .
\ee

Equation \eqref{no-full-Cond} leads to severe restrictions on the backgroud superspace geometry.
In particular, the integrability conditions $\{ \cD_{\a}^{ i}, \cD_{\b}^{ j} \}C_a^{kl}=0$,
$\{ \cD_{\a}^{ i}, \cD_{\b}^{ j} \}W_{abc}=0$, and $\{ \cD_{\a}^{ i}, \cD_{\b}^{ j} \}N_{abc}=0$ imply
the following differential equations 
\bsubeq\label{DT8}
\bea \cD_d W_{abc} &=& - 6 ( W_{d [a}{}^e + 2 N_{d[a}{}^e ) W_{bc]e} \ , \\
\cD_d N_{abc} &=& - 6 ( W_{d [a}{}^e + 2 N_{d[a}{}^e ) N_{bc]e} \ , \\
\cD_d C_a{}^{kl} &=& - 2 ( W_{dab} + 2 N_{dab} ) C^b{}^{kl}
- 6 C_d{}^{p(k} C_a{}_p{}^{l)}
\ ,
\eea
\esubeq
together with the algebraic conditions
\bsubeq\label{T8-2}
\bea
 (\g_{[a}{}^{de})_{\a\b} W_{bc] d} C_e{}^{ij} &=&0~,
 \\
  (\g_{[a}{}^{de})_{\a\b} N_{bc] d} C_e{}^{ij} &=& 0 ~, 
  \\
(\g_{abc})_{\a\b} C^{b il} C^{c jk}&=& - \frac{2}{3} N^{bcd} (\g_{bcd})_{\a\b} (\eps^{j(i} C_a{}^{l) k} + \eps^{k(i} C_a^{i) j}) ~.
\eea
\esubeq

Note that \eqref{DT8} can be compactly rewritten as
\be
\tilde\cD_d W_{abc} = 0~,~~~~~~
\tilde\cD_d N_{abc} = 0 ~,~~~~~~
\tilde \cD_d C_a{}^{kl} = 0 ~,
\ee
where we have defined
\be \tilde{\cD}_a := \cD_a - 3 C_a{}^{kl} J_{kl} + (W_{abc} + 2 N_{abc}) M^{bc} \ .
\label{tildeD}
\ee
A lengthy, though straightforward, analysis of the 
consistency of \eqref{DT8} and \eqref{T8-2} together with the superspace Bianchi identities leads to the following
algebraic constraints
\bsubeq \label{DT8-alg}
\bea
C_a{}^{kl}&=&C_aC^{kl}~,~~~~~~C^{ij}C_{ij}=2
~,
\label{DT8-alg-1}
\\
W_{abc}C_{d}&=& 0
~,
\label{DT8-alg-2}
\\
N_{abc}C_{d}&=& 0
~,
\label{DT8-alg-3}
\\
W^{\a\g}N_{\b\g}&=&
\frac{1}{4}
\d_{\b}^{\a}W^{\g\d}N_{\g\d}
~,
\label{DT8-alg-4}
\eea
\esubeq
as well as the conditions
\bea
\label{covConst}
\cD_A C_a=0
~,~~~
\cD_A C^{kl}=0
~,~~~
\cD_A N_{bcd}=0
~,~~~
\cD_A W_{bcd}=0
~.
\eea
It should be emphasised that, due to \eqref{DT8-alg-2} and \eqref{DT8-alg-3}, two branches of solutions exist, defined by: (i) $C_a = 0$; and (ii) $W_{abc} = N_{abc} = 0$.

Note that the algebraic constraint \eqref{DT8-alg-4}, due to the (anti-)self-duality conditions on $W_{abc}$ and $N_{abc}$,
 is equivalent to any of the following relations
\bea
N_{[a}{}^{de}W_{b]de}&=&0
\Longleftrightarrow
N_{a[b}{}^eW_{cd]e}=0
\Longleftrightarrow
W_{a[b}{}^eN_{cd]e}=0
\Longleftrightarrow
N_{[ab}{}^eW_{cd]e}=0
~,
\eea
while the following relations hold identically
\bsubeq
\bea
&W_{[a}{}^{de}W_{b]de}=0
\Longleftrightarrow
W_{a[b}{}^eW_{cd]e}=0
\Longleftrightarrow
W_{[ab}{}^eW_{cd]e}=0
\Longleftrightarrow
W_{[abc}W_{def]}=0
~,
\\
&N_{[a}{}^{de}N_{b]de}=0
\Longleftrightarrow
N_{a[b}{}^eN_{cd]e}=0
\Longleftrightarrow
N_{[ab}{}^eN_{cd]e}=0
\Longleftrightarrow
N_{[abc}N_{def]}=0
~.
\eea
\esubeq

By a routine calculation it may be shown that 
\begin{subequations}\label{one-forms}
\bea
N_{abc} &=& \a\Big( \o^{(0)}_{[a} \o^{(1)}_b \o^{(2)}_{c]} + \o^{(3)}_{[a} \o^{(4)}_b \o^{(5)}_{c]} \Big)~, \\
W_{abc} &=& \b\Big(  \o^{(0)}_{[a} \o^{(1)}_b \o^{(2)}_{c]} - \o^{(3)}_{[a} \o^{(4)}_b \o^{(5)}_{c]} \Big)~,
\eea
\end{subequations}
is a solution to \eqref{DT8-alg-4}, provided that $\o^{(i)}_{a} ,~ 0 \le i \le 5 $, 
are orthogonal one-forms, 
that is,
\bea
\o^{(i)}_a \o^{(j) a} = 
 0 ~, \quad \quad (i \neq j)
 ~.
\eea
This result was originally derived in \cite{FigueroaPapadopoulos} 
(see also \cite{Hitchin:2000jd}). These one-forms may be normalised to constitute an orthonormal basis, and then 
the expressions for $N_{abc} $ and $W_{abc} $ will, in general, involve overall factors
$\a$ and $\b$ as in \eqref{one-forms}.

In accordance with our analysis, for every maximally supersymmetric $\cN=(1,0)$ background 
the algebra of covariant derivatives 
is given by the following graded commutation relations:
\bsubeq \label{8.13}
\bea
\{ \cD_{\a}^{ i}, \cD_{\b}^{ j} \}&=&
-2 \ri \ve^{ij} (\g^{a})_{\a\b}\cD_{a}
+\Big(
-2\ri \ve^{ij} (\g^{a})_{\a\b}\big(W^{acd}+ 2 N^{acd}\big) 
+2\ri (\g^{acd})_{\a\b} C_{a}C^{ij} \Big) M_{cd}
\non\\
&&
+\Big(
6\ri \ve^{ij} (\g^{a})_{\a\b}C_{a}C^{kl}
+ \frac{8\ri}{3}(\g^{abc})_{\a\b} N_{abc}\ve^{i(k}\ve^{l)j}
\Big) J_{kl}
~,
\\
{[}\cD_{a} ,\cD_{\g }^{k}] &=& 
\Big(
({\g}_{ab})_{ \g}{}^{\d} C^{b}C^k{}_{l}
-\hf(W_{abc}+2N_{abc})({\g}^{bc})_{\g}{}^{\d}\d^k_l
\Big) 
\cD^l_{\d} 
~,
\\
{[}\cD_{a} ,\cD_b] &=&
\Big(8\d^{[c}_{[a}C_{b]}C^{d]}
-4 \d^{[c}_{[a}\d^{d]}_{b]}C^{e} C_{e}
\Big)
M_{cd}
\non\\
&&
+\Big(
8\d_{[a}^{[c}N^{d]ef}N_{b]ef}
+2\d_{[a}^{[c}W^{d]ef}W_{b]ef}
+16N_{[a}{}^{e[c}W_{b]e}{}^{d]}
\Big)M_{cd}~.
\label{8.13c}
\eea
\esubeq
The algebra is determined by the four tensors $C_a$, $C^{kl}$, $N_{bcd}$
and $W_{abc}$, which are covariantly constant, eq. \eqref{covConst}, 
and obey the algebraic constraints \eqref{self} and \eqref{DT8-alg}.
In conjunction with the Lorentz and $R$-symmetry commutation relations, 
$[M_{cd} , \cD_A] $ and $[J_{kl}, \cD_A]$, the graded commutation relations 
\eqref{8.13} define the most general superalgebras with eight supercharges, 
which are associated with the maximally supersymmetric backgrounds of 
$\cN= (1,0)$ Poincar\'e supergravity. These superalgebras were derived two years ago 
\cite{MedeirosFigueroaSanti} using sophisticated algebraic techniques. Here we have demonstrated that the superspace techniques allow one to derive these superalgebras via a simple calculation.

In accordance with the discussion in appendix \ref{appendixB}, 
the commutation relation   \eqref{8.13c} is equivalent to that of the spacetime covariant derivatives.
Therefore we can immediately read off the Ricci tensor, the scalar curvature and the Weyl tensor:
\bsubeq
\bea
R_{ab}&=&
-16C_{a}C_{b}
+16\eta_{ab}C^{c} C_{c}
-16N_a{}^{cd}N_{bcd}
-4W_a{}^{cd}W_{bcd}
-\frac{8}{3}\eta_{ab}N^{cde}W_{cde}
~,~~~~~~
\\
R&=&
80C^{c}C_{c}
-16N^{cde}W_{cde}
~,
\\
C_{abcd}
&=& 
 R_{abcd} 
- \hf \Big( 
\eta_{a[c} R_{d]b}
- \eta_{b[c} R_{d]a}
\Big)
+ \frac{1}{10} R \eta_{a[c} \eta_{d]b}
~,\non\\
&=&
\label{8.10c}
16N_{a[c}{}^{e}W_{d]be}
-16N_{b[c}{}^{e}W_{d]ae}
+\frac{16}{15}\eta_{a[c} \eta_{d]b}N^{efg}W_{efg}
~.
\eea
\esubeq
Note that the condition $W_{abc}=0$ implies that the superspace is conformally 
flat \cite{BKNT}. As a result of \eqref{8.10c} it is clear that  $W_{abc}=0$ implies $C_{abcd}=0$, though the reverse is not true in general.

It should be remarked that the algebra of covariant derivatives \eqref{8.13}
 takes a particularly simple form if the torsion-free covariant derivative $\cD_a$ is replaced with 
the torsionful one defined by
\eqref{tildeD}. One obtains
\bsubeq
\label{superalgebra}
\bea
\{ \cD_{\a}^{ i}, \cD_{\b}^{ j} \}&=&
-2 \ri \ve^{ij} (\g^{a})_{\a\b}\tilde{\cD}_{a}
+2\ri (\g^{acd})_{\a\b} C_{a}C^{ij} M_{cd}
- \frac{8\ri}{3}(\g^{abc})_{\a\b} N_{abc}J^{ij}
~,
\\
{[}\tilde{\cD}_{a} ,\cD_{\g }^{k}] &=&
\Big\{
\big[({\g}_{ab})_{ \g}{}^{\d} C^{b}
-3C_{a}\d_\g^\d
\big]C^k{}_{l}
-(W_{abc}+2 N_{abc})({\g}^{bc})_{\g}{}^{\d}\d^k_l
\Big\}
\cD^l_{\d} 
~,
\\
{[}\tilde{\cD}_{a} ,\tilde{\cD}_{b}{]}
&=& 
4(W_{ab}{}^{d}+2N_{ab}{}^{d})\tilde{\cD}_{d}
+\Big(8\d^{[c}_{[a}C_{b]}C^{d]}
-4 \d^{[c}_{[a}\d^{d]}_{b]}C^{e} C_{e}
\Big)
M_{cd}
~.
\eea
\esubeq
We see that the $R$-symmetry curvature vanishes if $N_{abc}=0$. The graded commutation relations take 
a remarkably simple form if $N_{abc}=0$ and $C_a=0$; the bosonic body of this superspace is a conformally flat   $\text{AdS}_3 \times S^3$ spacetime or a pp-wave, see later.

We now employ the above analysis  to identify all possible maximally supersymmetric spacetimes, 
which are the bosonic bodies of the superspaces with geometry \eqref{8.13}, or equivalently 
	\eqref{superalgebra}.
The most obvious solution is Minkowski space,
${\mathbb M}^6 \equiv \mathbb{R}^{5,1}$, which corresponds to the choice $C_{a} = 0$ and 
$N_{abc} = W_{abc} = 0$. When this is not the case, it follows from \eqref{DT8-alg-2} and \eqref{DT8-alg-3} that there are 
two disconnected branches of solutions, defined by $C_a\ne0$ or $C_a=0$.

Solutions belonging to the branch $C_a \neq 0$, which necessarily have $N_{abc}=W_{abc}=0$, 
are characterised by the existence of a parallel, nowhere vanishing vector field.
Thus, since $C^2 = C^a C_{a}$ is constant, the possible backgrounds are locally equivalent to the following three cases,
$\mathbb{R} \times S^5$ for $C^2 < 0$, $\text{AdS}_5 \times \mathbb{R}$ when $C^2 > 0$ and 
a pp-wave spacetime if $C^2 = 0$ \cite{MedeirosFigueroaSanti}.

When $C_a = 0$, the corresponding geometries are described by the covariantly constant three-forms 
$N_{abc}$ and $W_{abc}$, which decompose as the sum of two orthogonal simple forms \eqref{one-forms}.
If either of the corresponding simple forms are null, the background is a pp-wave spacetime \cite{Meessen}. 
When this is not the case, it follows that locally 
the spacetime decomposes into the product of two three-dimensional symmetric 
spaces. This can be inferred by the structure of the three-form fluxes given in \eqref{one-forms}. 
In particular, the possible solutions are  \cite{MedeirosFigueroaSanti}: 
(i) $\text{AdS}_3 \times S^3$; (ii) $\text{AdS}_3 \times \mathbb{R}^3$; and (iii) $\mathbb{R}^{2,1} \times S^3$.

In general, for backgrounds belonging to (i) the radii of the $\text{AdS}_3$ and $S^3$ do not
necessarily coincide
--- in particular, (ii) and (iii) are degenerate cases of (i).
Additionally, if $N_{abc}=0$ and $W_{abc}\ne0$, their radii must be equal
(proportional to $\beta$ in \eqref{one-forms}).
This background is one of the well-known solutions to minimal $\cN=(1,0)$ supergravity in six dimensions 
\cite{GutowskiMartelliReall}.
It is also an example of a superspace which is not superconformally flat, $W_{abc}\ne0$,
though its bosonic body is conformally flat, $C_{abcd}\equiv 0$.\footnote{The reader can consult 
\cite{Bandos:2002nn} for an interesting discussion of 
superconformal flatness of $\text{AdS}_p \times S^q$ superspaces based on coset constructions.}

So far we have not specified any conformal compensators $\X$. 
We have worked in the super-Weyl 
gauge \eqref{Xgauge1}, where $\bm \X$ is a descendant of the
 compensators $\X$ which is a singlet under the structure group and has the properties: 
(i) it is an algebraic function of $\X$; 
(ii)  it is nowhere vanishing; and 
(iii) it has a  non-zero super-Weyl weight $w_{\bm \X}$, 
$\d_\s {\bm \X} = w_{\bm \X}  \s {\bm \X}$.
Additional constraints on supergravity backgrounds often occur once a specific choice of 
compensators is made. 

Let us analyse the case of the compensators introduced in section \ref{Isometries}, 
specifically: the tensor multiplet $\F $ and the linear multiplet $G^{ij}= G^{ji}$. 
Then it is possible to identify $\bm \X$ with $\F$. In the super-Weyl gauge \eqref{7.10},
the tensor multiplet constraint \eqref{3.35} reduces to 
\bea
C_{a}^{ij}=0~.
\label{8.15}
\eea
Every Killing supervector field $\x^B$ must leave the linear compensator $G^{ij}$ invariant, 
\bea
\Big(\x^{B} \cD_{B}  + K^{kl}[\x]J_{kl}    \Big) G^{ij}=0~,
\eea
in accordance with  \eqref{3.26b}. In the case of a maximally supersymmetric background,
this equation implies that $G^{ij}$ is annihilated by the spinor covariant derivatives, 
\bea
 \cD_\a^iG^{jk}=0
 ~.
 \label{8.17}
 \eea
Now, the integrability condition $\{\cD_\a^i,\cD_\b^j\}G^{kl}=0$ leads to the constraint
\bea
N^{abc}(\g_{abc})_{\a\b}\big(\ve^{k(i}G^{j)l}+\ve^{l(i}G^{j)k}\big)=0
~,
\eea
which is solved by 
\bea
N_{abc}=0~.
\label{8.19}
\eea
We have shown that the conditions \eqref{8.15} and \eqref{8.19} hold for all maximally supersymmetric backgrounds of Poincar\'e supergravity with the tensor and linear compensators.

Instead of identifying $\bm \X=\F$ as has been done above, we can instead choose 
 $\bm \X =G$. Next, we impose the super-Weyl gauge
 \bea
G^2=1\quad \Longleftrightarrow \quad
G^i{}_k G^k{}_j=- \d^i_j
~.
\label{linear-1}
\eea
Then the analyticity constraint \eqref{3.36}
and  the super-Weyl gauge condition  \eqref{linear-1}  tell us that 
$G^{ij}$ is annihilated by the spinor covariant derivatives, eq. \eqref{8.17}, and thus the integrability condition $\{\cD_\a^i,\cD_\b^j\}G^{kl}=0$ must hold. The latter contains nontrivial information, in accordance with 
the anti-commutation relation \eqref{Algebra-1}. Specifically, the integrability condition tells us that 
the condition \eqref{8.19} holds. Every Killing supervector field $\x^B$ must leave the tensor compensator $\F$ invariant, 
\bea
\x^{B} \cD_{B}  \F=0~,
\eea
in accordance with  \eqref{3.26b}. In the case of a maximally supersymmetric background,
this equation implies that $\F$ is annihilated by the spinor covariant derivatives, 
and therefore 
\bea
\F = {\rm const}~.
\eea
As a result, the tensor multiplet constraint \eqref{3.35} reduces to \eqref{8.15}. 

We have discussed the two possible choices: (i) $\bm \X=\F$; and  
(ii) $\bm \X =G$. Both of them lead to the same maximally supersymmetric backgrounds, which 
are characterised by the conditions \eqref{8.15} and \eqref{8.19}. The superspace torsion is determined by 
the super-Weyl tensor $W_{abc}$ which is covariantly constant. Such a superspaces are the only 
maximally supersymmetric solutions of  Poincar\'e supergravity. Let us discuss this point in 
more detail.

The equations of motion for Poincar\'e supergravity have the simplest form in conformal superspace.
In this setting, the tensor compensator $\F$ and the linear compensator $G^{ij}$ obey the constraints
\begin{subequations}
\bea
 \de_\a^{(i}\de_{\b}^{j)}\Phi&=&0
 ~,
 \label{tensor-conf}\\
 \de_\a^{(i}G^{jk)}&=&0~.
 \label{8.24b}
 \eea
\end{subequations}
For more details, including the Poincar\'e supergravity action,
we refer the reader to \cite{KNT,Kuzenko:2017jdy,BNOPT-M}. 
The superfield equations of motion for $\cN=(1,0)$  Poincar\'e supergravity were derived in 
\cite{KNT}. They have the form 
\bea
{\mathbb W}^{\a i}=0
~,~~~ 
\de_\a^{(i}\de_{\b}^{j)}\Big(\frac{G}{\Phi}\Big)=0
~,
\label{sugra-EOM-1}
\eea
where ${\mathbb W}^{\a i}$ is the field strength of a composite vector multiplet
\bea {\mathbb W}^{\a i} &=& \frac{1}{G} \nabla^{\a\b} \U_\b^i
+ \frac{4}{G} (W^{\a\b} \U_\b^i + 10 \ri X^\a_j G^{ij})
- \frac{1}{2 G^3} G_{jk} (\nabla^{\a\b} G^{ij}) \U_\b^k \non\\
&&+ \frac{1}{2 G^3} G^{ij} F^{\a\b} \U_{\b j}
+ \frac{\ri}{16 G^5} \ve^{\a \b \g \d} \U_{\b j} \U_{\g k} \U_{\d l} G^{ij} G^{kl} \ ,
\eea
with $\U_\a^i := \frac{2}{3} \nabla_{\a j} G^{ij}$ 
and $F_{\a\b} := \frac{\ri}{4} \nabla_{[\a}^k \U^{\phantom{k}}_{\b] k}$. 
To make contact with our previous results, we now degauge to $\sSU(2)$ superspace.
Upon degauging,
the tensor multiplet constraint \eqref{tensor-conf} turns into  \eqref{3.35}, while the linear multiplet constraint
 \eqref{8.24b} takes the form \eqref{8.17}.
The second equation of motion for Poincar\'e supergravity in \eqref{sugra-EOM-1} becomes
\bea
\big(\cD_\a^{(i}\cD_{\b}^{j)}+4\ri C_{\a \b}^{ij}\big)\Big(\frac{G}{\Phi}\Big)=0
~. 
\label{max-EOM-C}
\eea
The analysis given above tells us that the following properties hold for all maximally supersymmetric backgrounds:   (i) both compensators $\F$ and $G^{ij}$ are covariantly constant; and (ii) the conditions \eqref{8.15} and \eqref{8.19} hold. Now eq. \eqref{max-EOM-C} is satisfied. 
The first equation of motion in \eqref{sugra-EOM-1},
${\mathbb W}^{\a i}=0$,
is also satisfied, since all maximally supersymmetric backgrounds have no 
covariant background spinor superfields.


\section{Conclusion}
\label{section9}

To conclude, we summarise the main results of this paper and outline some interesting areas for future work. Our main outcomes are as follows:
\begin{itemize}
	\item We have described the structure of (conformal) isometries of $\mathcal{N} = (1,0)$ supergravity backgrounds within the $\sSU(2)$ and conformal superspace formulations. In the infinitesimal case they were shown to form a closed algebra on any fixed supergravity background. Further, we detailed how these may be utilised to trivially read off the (conformal) Killing spinor equation at the component level. Its solutions may be uplifted to a unique (conformal) Killing vector superfield on $\mathcal{M}^{6|8}$.
	
	\item The conformal Killing spinor superfields $\e^\a$, which generate extended conformal supersymmetries, were introduced. In addition, their relation to the conformal Killing vector $\xi^a$ and tensor $\z^{a(n)}$ superfields was shown. The former parametrise the conformal isometries of superspace, while the latter are associated with the higher symmetries of the kinetic operators of on-shell multiplets. Additionally, it was proven that the conformal Killing tensors of a fixed superspace form a superalgebra with respect to the bracket \eqref{SNb}.
	
	\item We studied the higher symmetries of three on-shell models in curved backgrounds, namely: (i) the conformal scalar field; (ii) the hypermultiplet; and (iii) the non-conformal vector multiplet. In our analysis of (i) we have, for the first time, derived the explicit form of every higher symmetry operator on curved backgrounds. For (ii), it was proven that the conformal Killing tensor superfields $\z^{a(n)}$ generate all (non-trivial) symmetries of their kinetic operators. Finally, in the case of (iii), we deduced that its higher symmetries are parametrised by Killing tensor superfields, which were also introduced in this work \eqref{7.15}.
	
	\item The maximally supersymmetric backgrounds of $\mathcal{N} = (1,0)$ supergravity in six dimensions were classified. Our analysis leads to the superalgebra \eqref{8.13}, or equivalently 
	\eqref{superalgebra}, which contains three distinct branches. Further, their corresponding spacetime backgrounds are derived, reproducing the results of \cite{MedeirosFigueroaSanti,Meessen,GutowskiMartelliReall}.
\end{itemize}

Interesting open problems include the following:

\begin{itemize}
	\item Our approach to the higher symmetries of the conformal d'Alembertian in section \ref{section5.1} may be immediately generalised to the study of more complex conformal field theories. In particular, it would be interesting to extend this analysis to Maxwell electrodynamics in four dimensions.
	
	\item We believe that, as was shown for the conformal d'Alembertian, every higher symmetry operator for the hypermultiplet and vector multiplet is uniquely determined in terms of its top component. It would be interesting to prove this explicitly.
	
	\item As an extension of our analysis of the higher symmetries of the (massless) hypermultiplet in section \ref{section6}, it would be interesting to study the higher symmetries of the massive hypermultiplet on 
	$d=4$,  $\mathcal{N}=2$ and $d=5$, $\mathcal{N}=1$ supergravity backgrounds. 
\end{itemize}


\noindent
{\bf Acknowledgements:}\\
We are grateful to Joseph Novak for collaboration at early stages of this project. We thank Paul Howe for useful comments on the manuscript. 
The work of SMK is supported in part by the Australian 
Research Council, project No. DP200101944.
The research of U.L. has been partially supported by the 2236 Co-Funded Brain Circulation Scheme2 (CoCirculation2) of 
T\"UBITAK (Project No:120C067)\footnote{However the entire responsibility of the publication belongs to the owners of the 
publication. The financial support received from T\"UBITAK does not mean that the content of the publication is approved in 
a scientific sense by T\"UBITAK.}
and by Längmanska Fonden.
The work of ESNR is supported by the Hackett Postgraduate Scholarship UWA,
under the Australian Government Research Training Program.
The work of GT-M is supported by the Australian Research Council (ARC)
Future Fellowship FT180100353, and by the Capacity Building Package of the University
of Queensland.


\appendix

\section{Conventions} \label{AppendixA}

\subsection{Spinors in six dimensions}
\label{appendixA.1}
Our 6D notation and conventions are similar to to those of \cite{LT-M12},
with a few minor modifications. All relevant details are summarized here.

The Minkowski metric is
$\eta_{ab} = \textrm{diag}(-1,1,1,1,1,1)$, the Levi-Civita tensor
$\veps_{abcdef}$ is normalised by $\veps_{012345} = -\veps^{012345} = 1$,
and the Levi-Civita tensor with world indices 
is given by $\eps^{mnpqrs} := \eps^{abcdef} e_a{}^m e_b{}^n e_c{}^p e_d{}^q e_e{}^r e_f{}^s$.

We exclusively use four component spinors in the body of the paper,
but it is useful to relate these to eight component spinor conventions.
The  Dirac $8 \times 8$ matrices $\Gamma^a$ and the charge conjugation matrix
$C$ obey the relations
\begin{gather}
\{\G_a, \G_b\} = -2 \eta_{a b} \mathbbm{1}~, \quad
(\G^a)^\dag = -\G_a~, \quad
C \G_a C^{-1} = -\G_a^T~, \eol
C^\dag C = \mathbbm{1}~, \qquad C = C^{\rm T} = C^*~.
\end{gather}
In particular, $\Gamma_a C^{-1}$ is antisymmetric. The chirality matrix
$\Gamma_*$ is defined by
\begin{align}
\Gamma_{[a} \Gamma_b \Gamma_c \Gamma_d \Gamma_e \Gamma_{f]} = \veps_{abcdef} \Gamma_*~.
\end{align}
As a consequence of the above conditions, one can show that
\begin{align}\label{eq:GammaReality}
\Gamma^a = B (\Gamma^a)^* B^{-1}~, \qquad B = \Gamma_* \Gamma_0 C^{-1}~.
\end{align}
The charge conjugate $\Psi^c$ of a Dirac spinor is conventionally defined by
\begin{align}
\bar\Psi \equiv \Psi^\dag \Gamma_0 =: (\Psi^c)^T C \qquad \implies \quad
\Psi^c = - \Gamma_0 C^{-1} \Psi^* = -\Gamma_* B \Psi^*~.
\end{align}
Because $B^* B = -\mathbbm{1}$,
charge conjugation is an involution only for objects with an even number of spinor indices,
so it is not possible to have Majorana spinors in six dimensions.
One can instead have a symplectic Majorana condition when the spinors possess an $\sSU(2)$ index.
Conventionally this is denoted
\begin{align}\label{eq:SympMaj}
(\Psi_i)^c = \Psi^i \quad \implies \quad
\Psi^i = -\Gamma_0 C^{-1} (\Psi_i)^* = -\Gamma_* B (\Psi_i)^*
\end{align}
for a spinor of either chirality. We raise and lower $\sSU(2)$ indices $i=\1,\2$
using the conventions
\begin{align}
\Psi^i = \eps^{i j} \Psi_j~, \qquad \Psi_i = \eps_{i j} \Psi^j~, \qquad \eps^{\1\2} = \eps_{\2\1} = 1~.
\end{align}

We employ a Weyl basis for the gamma matrices so that
an eight-component Dirac spinor $\Psi$ decomposes into a four-component
left-handed Weyl spinor $\psi^\alpha$ and a four-component right-handed spinor $\chi_\alpha$
so that
\begin{align}\label{eq:ChiralDecomp1}
\Psi =
\begin{pmatrix}
\psi^\alpha \\
\chi_\alpha
\end{pmatrix}~, \qquad
\Gamma_* =
\begin{pmatrix}
\delta^\a{}_\b & 0 \\
0 & -\delta_\a{}^\b
\end{pmatrix}~, \qquad \alpha=1,\cdots, 4~.
\end{align}
The spinors $\psi^\alpha$ and $\chi_\alpha$ are valued in the
two inequivalent fundamental representations 
of $\mathfrak{su}^*(4) \cong \mathfrak{so}(5,1)$.
We further take
\begin{align}
\Gamma^a =
\begin{pmatrix}
0 & (\tilde\gamma^a)^{\alpha\beta} \\
(\gamma^a)_{\alpha\beta} & 0 
\end{pmatrix}~,\qquad
C =
\begin{pmatrix}
0 & \delta_\alpha{}^\beta \\
\delta^\alpha{}_\beta & 0
\end{pmatrix}~.
\end{align}
The Pauli-type $4 \times 4$ matrices $(\gamma^a)_{\alpha\beta}$ 
and $(\tilde\gamma^a)^{\alpha\beta}$ are antisymmetric and related by
\begin{align}
(\tilde\gamma^a)^{\alpha\beta} = \frac{1}{2} \veps^{\a\b\g\d} (\gamma^a)_{\g\d}~, \qquad
(\gamma^a)^* = \tilde\gamma_a~,
\end{align}
where $\veps^{\a\b\g\d}$ is the canonical antisymmetric symbol of $\mathfrak{su}^*(4)$.
They obey
\bsubeq
\bea (\g^a)_{\a\b} (\tilde{\g}^b)^{\b\g}
+ (\g^b)_{\a\b} (\tilde{\g}^a)^{\b\g} &=& - 2 \eta^{ab} \d^\g_\a \ , \\
(\tilde{\g}^a)^{\a\b} (\g^b)_{\b\g}
+ (\tilde{\g}^b)^{\a\b} (\g^a)_{\b\g} 
&=& - 2 \eta^{ab} \d^\a_\g \ ,
\eea
\esubeq
and as a consequence of \eqref{eq:GammaReality},
\begin{align}
(\gamma^a)_{\alpha\beta} = B_{\alpha}{}^\dgamma B_\beta{}^{\ddelta}
\big((\gamma^a)_{\gamma \delta}\big)^*~, \quad
(\tilde\gamma^a)^{\alpha\beta} = B^{\alpha}{}_\dgamma B^\beta{}_{\ddelta}
\big((\tilde\gamma^a)^{\gamma \delta}\big)^*~, \quad
B =
\begin{pmatrix}
B^{\alpha}{}_\dbeta & 0 \\
0 & B_{\alpha}{}^\dbeta
\end{pmatrix}~.
\label{eq:PauliReality}
\end{align}
A dotted index denotes the complex conjugate representation in $\mathfrak{su}^*(4)$.
It is natural to use the $B$ matrix to define bar conjugation on a
four component spinor via
\begin{align}
\bar\psi^\alpha = B^\alpha{}_\dbeta (\psi^\beta)^*~, \qquad
\bar \chi_\alpha = B_\alpha{}^\dbeta (\chi_\beta)^*~,
\label{A.12}
\end{align}
with the obvious extension to any object with multiple spinor indices.
For example, $\overline{(\gamma^a)_{\alpha\beta}} = (\gamma^a)_{\alpha\beta}$
using \eqref{eq:PauliReality} and similarly for $\tilde\gamma^a$.
We also note that, as a consequence of $B^* B = -\mathbbm{1}$,
\be
\overline{\overline{\psi^{\a}}} = -\psi^\a,
\ee
with the natural extension to any tensor carrying an odd number of spinor indices.
A symplectic Majorana spinor $\Psi_i$, decomposed as in \eqref{eq:ChiralDecomp1}
and obeying \eqref{eq:SympMaj}, has Weyl components that obey
\begin{align}\label{eq:SympMaj4c}
\overline{\psi^{\alpha i}} = \psi^\alpha_{i}~, \qquad
\overline{\chi_{\alpha i}} = \chi_\alpha^{i}~.
\end{align}
The Grassmann coordinates $\q^\alpha_i$ and the parameters $\eta_\alpha^i$ of $S$-supersymmetries
are both symplectic Majorana-Weyl using this definition. 

We define the antisymmetric products of two or three Pauli-type matrices as
\bsubeq
\begin{alignat}{2}
\g_{ab} &:= \g_{[a} \tilde{\g}_{b]} := \hf (\g_a \tilde{\g}_b - \g_b \tilde{\g}_a) \ , &\quad
\tilde{\g}_{ab} &:= \tilde{\g}_{[a} \g_{b]}  = -(\g_{ab})^T\ , \\
\g_{abc} &:= \g_{[a} \tilde{\g}_b \g_{c]} \ , &\quad \tilde{\g}_{abc} &:= \tilde{\g}_{[a} \g_b \tilde{\g}_{c]} \ .
\end{alignat}
\esubeq
Note that $\g_{ab}$ and $\tilde\g_{ab}$ are traceless, whereas $\g_{abc}$ and
$\tilde\g_{abc}$ are symmetric. Further antisymmetric products obey
\begin{subequations}
	\begin{alignat}{2}
	\g_{abc} &= -\frac{1}{3!} \eps_{abcdef} \g^{def}~, &\qquad
	\tilde \g_{abc} &= \frac{1}{3!} \eps_{abcdef} \tilde \g^{def}~, \\
	\g_{abcd} &= \frac{1}{2} \eps_{abcdef} \g^{ef}~, &\qquad
	\tilde \g_{abcd} &= -\frac{1}{2} \eps_{abcdef} \tilde \g^{ef}~, \\
	\g_{abcde} &= \eps_{abcdef} \g^f~, &\qquad
	\tilde \g_{abcde} &= -\eps_{abcdef} \tilde \g^f~, \\
	\g_{abcdef} &= -\eps_{abcdef}~, &\qquad
	\tilde \g_{abcdef} &= \eps_{abcdef}~.
	\end{alignat}
\end{subequations}

Making use of the completeness relations
\begin{subequations}
	\begin{align}
	(\gamma^a)_{\a\b} (\tilde\gamma_{a})^{\g\d} &= 4\, \delta_{[\a}{}^\g \delta_{\b]}{}^{\d}~, \\
	(\gamma^{ab})_\a{}^\b (\gamma_{ab})_\g{}^\d &= - 8\,\delta_{\a}{}^\d \delta_{\g}{}^{\b}
	+ 2\, \delta_{\a}{}^\b \delta_{\g}{}^{\d}~, \\
	(\gamma^{abc})_{\a\b} (\tilde\gamma_{abc})^{\g\d} &= 48\, \delta_{(\a}{}^\g \delta_{\b)}{}^{\d}~, \\
	(\gamma^{abc})_{\a\b} (\tilde\gamma_{abc})_{\g\d} &= (\gamma^{abc})^{\a\b} (\tilde\gamma_{abc})^{\g\d} = 0~,
	\end{align}
\end{subequations}
it is straightforward to establish natural isomorphisms between tensors of $\mathfrak{so}(5,1)$
and matrix representations of $\mathfrak{su}^*(4)$.
Vectors $V^a$ and antisymmetric matrices $V_{\a\b} = - V_{\b\a}$ 
are related by
\be 
V_{\a\b} := (\g^a)_{\a\b} V_a \quad \Longleftrightarrow  \quad V_a = \frac{1}{4} (\tilde{\g}_a)^{\a\b} V_{\a\b} \ .
\ee
Antisymmetric rank-two tensors $F_{ab}$ are related to traceless matrices $F_\a{}^\b$ 
via
\bea 
F_\a{}^\b := - \frac{1}{4} (\g^{ab})_\a{}^\b F_{ab} \ , \quad F_\a{}^\a = 0 \quad
\Longleftrightarrow 
\quad F_{ab} = \hf (\g_{ab})_\b{}^\a F_\a{}^\b = - F_{ba} \ .
\label{A.18}
\eea
Self-dual and anti-self-dual rank-three antisymmetric tensors $T^{(\pm)}_{abc}$,
\be \frac{1}{3!} \eps^{abcdef} T_{def}^{(\pm)} = \pm T^{(\pm)abc} \ ,
\ee
are related to symmetric matrices $T_{\a\b}$ and $T^{\a\b}$ 
via
\bsubeq
\bea
T_{\a\b} &:=& \frac{1}{3!} (\g^{abc})_{\a\b} T_{abc} = T_{\b\a} \quad \Longleftrightarrow \quad 
T_{abc}^{(+)} = \frac{1}{8} (\tilde{\g}_{abc})^{\a\b} T_{\a\b} \ , \\
T^{\a\b} &:=& \frac{1}{3!} (\tilde{\g}^{abc})^{\a\b} T_{abc} = T^{\b\a} \quad 
\Longleftrightarrow \quad
T^{(-)}_{abc} = \frac{1}{8} (\g_{abc})_{\a\b} T^{\a\b} \ .
\eea
\esubeq

\subsection{The $\mathcal{N}=(1,0)$ superconformal algebra} \label{AppendixA.2}

The bosonic sector of the $\cN = (1, 0)$ superconformal algebra contains the translation ($P_{a}$), Lorentz ($M_{ab}$), 
special conformal ($K_{a}$), dilatation ($\mathbb{D}$) and $\sSU(2)$ generators ($J_{ij}$), where 
$a, b = 0 , 1 , 2 , 3 , 4, 5$ and $i , j = \underline{1} , \underline{2}$. 
Their algebra is
\bsubeq
\begin{align} [M_{a b} , M_{c d}] &= 2 \eta_{c [a} M_{b] d} - 2 \eta_{d [ a} M_{b ] c} \ , \\
[M_{a b}, P_c] &= 2 \eta_{c [a} P_{b]} \ , \quad [\mathbb{D} , P_a] = P_a \ , \\
[M_{a b} , K_c] &= 2 \eta_{c [a} K_{b]} \ , \quad [\mathbb{D} , K_a] = - K_{a} \ , \\
[K_a , P_b] &= 2 \eta_{a b} {\mathbb D} + 2 M_{ab} \ , \\
[J^{ij} , J^{kl}] &= \eps^{k(i} J^{j) l} + \eps^{l (i} J^{j) k} \ ,
\end{align}
\esubeq
with all other commutators vanishing. 

Its superconformal generalisation is obtained by extending the 
translation generator to $P_A = (P_a , Q_\a^i)$ and the special conformal generator to
$K^A = (K^a , S^\a_i)$.
The fermionic generator $Q_\a^i$ obeys the algebra
\bsubeq \label{SCA1}
\begin{align} \{ Q_\a^i , Q_\b^j \} &= - 2 \ri \eps^{ij} (\g^{c})_{\a\b} P_{c} \ , \quad [Q_\a^i , P_a ] = 0 \ , \quad [{\mathbb D} , Q_\a^i ] = \hf Q_\a^i \ , \\
[M_{ab} , Q_\g^k ] &= - \hf (\g_{ab})_\g{}^\d Q_\d^k \ , \quad [J^{ij} , Q_\a^k ] = \eps^{k (i} Q_\a^{j)} \ ,
\end{align}
\esubeq
while the generator $S^\a_i$ obeys the algebra
\bsubeq\label{SCA2}
\begin{align} \{ S^\a_i , S^\b_j \} &= - 2 \ri \eps_{ij} (\tilde{\g}^{c})^{\a\b} K_{c} \ , \quad [S^\a_i , K_a ] = 0 \ , \quad [{\mathbb D} , S^\a_i ] = - \hf S^\a_i~, \\
[M_{ab} , S^\g_k ] &= \hf (\g_{ab})_\d{}^\g S^\d_k  \ , \quad [J^{ij} , S^\a_k ] = \d_k^{(i} S_\a^{j)} \ ,
\end{align}
\esubeq
Finally, the (anti-)commutators of $K^{A}$ and $P_B$ are
\bsubeq\label{SCA3}
\begin{align} [ K_a , Q_\a^i ] &= - \ri (\g_a)_{\a\b} S^{\b i} \ , \quad [S^\a_i , P_a ] = - \ri (\tilde{\g}_a)^{\a\b} Q_{\b i} \ , \\
\{ S^\a_i , Q_\b^j \} &= 2 \d^\a_\b \d^j_i \mathbb D - 4 \d^j_i M_\b{}^\a + 8 \d^\a_\b J_i{}^j \ .
\end{align}
\esubeq


\section{The conformal Killing supervector fields of $\mathbb{M}^{6|16}$}
\label{reductionAppendix}

The aim of this appendix is to study the structure of conformal Killing supervector fields 
of $\mathcal{N} = (2,0)$ Minkowski superspace in six dimensions. Such analyses were previously conducted in \cite{Park98,GrojeanMourad}. By employing this construction, we will explicitly prove our earlier claim that the proposed conformal Killing spinor superfields \eqref{ConfKillingSpinor} naturally arise from an $\mathcal{N} = (2,0) \longrightarrow (1,0)$ superspace reduction.

We recall that $\mathcal{N}=(2,0)$ Minkowski superspace, $\mathbb{M}^{6|16}$, is parametrised by the coordinates $z^{A} = (x^a , \theta^\a_I)$, where $a=0,1,\cdots,5$, $\alpha=1,\cdots,4$ and  $I=\underline1,\cdots,\underline4$. Its covariant derivatives take the form 
\bea
\partial_a = \frac{\partial}{\partial x^a} ~, \quad D_{\a}^I = \frac{\partial}{\partial \theta^\a_I} - \ri (\g^a)_{\a \b} \theta^{\b I} \partial_a ~,
\eea
and satisfy the algebra
\bea
\{ D_{{} \alpha}^{ I}, D_{{} \beta}^{ J} \} =
-2 \ri \O^{IJ} \partial_{\a \b} ~, \quad \big[ \partial_a , D_{\a}^I \big] = 0 ~, \quad \big[ \partial_a , \partial_b \big] = 0 ~.
\eea
Here $\O^{IJ} = - \O^{JI}$ is an invariant tensor of the $\mathcal{N} = (2,0)$ R-symmetry group $\mathsf{USp}(4)$. It is convenient to choose a basis for $\O^{IJ}$ such that it takes the form
\bea
\O^{IJ} =
\left(\begin{array}{cc}
	\ve^{ij}  & 0 \\
	0 & \ve^{\hat i \hat j} \\
\end{array}\right) ~, \qquad i, \hat i = \underline{1} , \underline{2}.
\eea

We will say that the real supervector field 
\bea
\xi = \bar{\xi} = \xi^a \partial_a + \xi^\a_I D_\a^I ~,
\eea
is conformal Killing if it satisfies
\bea
\label{flatCKV}
\big[ \xi , D_{\a}^I \big] = - (D_{\a}^I \xi^\b_J) D_{\b}^J~.
\eea
This constraint implies the fundamental equation
\bea
\label{CKV(2,0)}
D_{\a}^I \xi^a = - 2 \ri (\g^a)_{\a \b} \xi^{\b I} ~,
\eea
which yields
\bea
\xi^{\a}_I = - \frac{\ri}{12} (\tilde{\g}_{a})^{\a \b} D_{\b I}\xi^a ~.
\eea
By a routine computation, we may bring \eqref{flatCKV} to the form
\bea
\big[ \xi , D_{\a}^I \big] = - \o_{\a}{}^{\b} [\xi] D_{\b}^I + \L^I{}_J [\xi] D_{\a}^J - \frac{1}{2} \s [\xi] D_{\a}^I,
\eea
where we have made use of the definitions
\bea
\o_{\a}{}^{\b} [\xi] &=& - \frac{1}{4} (\g^{ab})_{\a}{}^{\b} \partial_a \xi_b ~, \non \\
\L^{I}{}_{J} [\xi] &=& - \frac{1}{4} \big[ D_{\a}^I \xi^{\a}_J - \frac{1}{4} \d^I_J D_{\a}^K \xi^\a_K \big] ~, \non \\
\s [\xi] &=& \frac{1}{6} \partial^a \xi_a ~.
\eea
It is clear that the above parameters generate Lorentz, R-symmetry and scaling transformations, respectively.

We now briefly consider the problem of performing a reduction to $\mathcal{N}=(1,0)$ Minkowski superspace. Without loss of generality, we will assume that this coincides with the section of 
$\mathbb{M}^{6|16}$ defined by
$\theta^{\a}_I = (\theta^{\a}_i , \theta^{\a}_{\hat i}) = (\theta^{\a}_i , 0)$. It then follows that, upon such a reduction, every solution to \eqref{flatCKV} decomposes into an $\mathcal{N}=(1,0)$ conformal Killing supervector field, a spinor superfield and an additional triplet of scalar superfields defined by
\bea
\e^\a_{\hat{i}} = \xi^{\a}_{\hat i} |_{\theta_{\hat i}^\a = 0} ~,
\qquad
\lambda^{\hat{i}\hat{j}} := \frac{1}{4} D_\alpha^{(\hat{i}}\xi^{\alpha \hat{j})}|_{\theta_{\hat i}^\a = 0}
~.
\eea
The spinor $\e^\a_{\hat{i}}$ and $\lambda^{\hat{i}\hat{j}}$ generate non-manifest extended superconformal symmetries in $\mathbb{M}^{6|8}$, where $\lambda^{\hat{i}\hat{j}}$ generates
the hidden $\mathsf{SU}(2)$ R-symmetry within $\mathsf{USp}(4)$.
By making use of the fundamental equation \eqref{CKV(2,0)}, one may show that $\e^\a_{\hat{i}}$ satisfies
\bea
D_\a^i \e^{\b}_{\hat j} = \frac{1}{4} \d_\a^\b D_{\g}^i \e^{\g}_{\hat j}~,
\eea
which implies that $\e^\a_{\underline{1}}$ and $\e^\a_{\underline{2}}$ are conformal Killing spinor superfields \eqref{ConfKillingSpinor}. As a result, we have proven our original claim.


\section{Bosonic backgrounds} \label{appendixB}

In this appendix, we introduce a formalism to study 
6D $\cN=(1,0)$ supersymmetric backgrounds 
starting from a superspace perspective. Our analysis will be restricted to bosonic backgrounds, meaning that the following conditions hold
\bea
\label{BB}
\cD_\a^i C_{a}{}^{kl}|= 0~, \qquad 
\cD_\a^i W_{abc}| = 0~, \qquad 
\cD_\a^i N_{abc}| = 0~. 
\eea
Following \cite{GGRS,WZ,GKLR}, 
 the bar-projection of a superfield is defined as usual:
\bea
U|:= U(x,\q)\big|_{\q =0}~,
\eea 
for any superfield $U(z)=U(x,\q)$. The coordinates $x^{m}$ parametrise a curved spacetime $\cM^6$,
the bosonic body of the superspace $\cM^{6|8}$. 
The bar-projection of the superspace covariant derivatives 
is defined similarly by:
\bea
\cD_{{A}}| = E_{{A}}{}^{{M}}| \,\pa_{{M}}
- \hf \,\O_{{A}}{}^{bc}|\,M_{bc}
- \F_{A}{}^{kl}|\,J_{kl}~.
\eea

Due to \eqref{BB},
one can completely gauge away the gravitini 
such that the projection of the vector covariant derivatives 
takes the simple form
\bea
\cD_a|=\bD_a \quad \Longleftrightarrow \quad \psi_{m}{}^\a_i =0 \ ,
\label{gauge-gravitini}
\eea
where 
\bea
\bD_a=e_a
-\hf \o_a{}^{bc} M_{bc}
-\phi_a{}^{kl} J_{kl},~~~~~~
e_a:=e_a{}^{m} \pa_{m}
\eea
is a spacetime covariant derivative with Lorentz $(\o_{a}{}^{bc})$ and $\sSU(2)_R$ $(\phi_{a}{}^{kl})$ connections.
In what follows, 
the gauge \eqref{gauge-gravitini} will be assumed. 
The covariant derivatives $\bD_a$ obey 
\be
 [\bD_a , \bD_b] =- \hf \cR_{ab}{}^{cd} M_{cd} - \cR_{ab}{}^{kl} J_{kl} 
\ ,~~~~~~
\cR_{ab}{}^{cd} = R_{ab}{}^{cd}| \ , \quad \cR_{ab}{}^{kl} = R_{ab}{}^{kl}| \ .
\ee
For convenience, we also make the definitions
\be c_a{}^{kl} = C_a{}^{kl}| \ , \quad w_{abc} = W_{abc}| \ , \quad n_{abc} = N_{abc}| \ .
\ee

An important feature of such backgrounds \eqref{BB} is that every conformal Killing vector superfield \eqref{ConfKillingVec} can be uniquely decomposed as a sum of even and odd ones. We will say that the conformal Killing supervector $\xi^{A}$ is even if
\bea
\label{even}
v^{a} := \xi^{a}| \neq 0 ~, \quad \xi_i^\a| = 0~.
\eea
or odd if
\bea
\label{odd}
\xi^{a}|  = 0 ~, \quad \e_i^{\a} := \xi_i^{\a} \neq 0~.
\eea
The fields $v^{a}$ and $\e_i^{\a}$ encode complete information about the parent conformal Killing vector superfield.
This appendix is devoted to the study of symmetries they induce at the spacetime level.


\subsection{Conformal Killing vectors}

Let $\xi^{A}$ be an even conformal Killing supervector field \eqref{even}. By bar-projecting \eqref{3.5} we obtain
\bea
\bD_{a} v_{b} = k_{ab}[v] + \eta_{ab} \o[v] ~,
\label{B.10}
\eea
where we have defined
\bea
k_{ab}[v] := K_{ab}[\xi] | = \bD_{[a} v_{b]} ~, \quad
\o[v] := \s[\xi] | = \frac{1}{6} \bD^{a} v_a ~.
\eea
In particular, it follows from \eqref{B.10} that $v^a$ is a conformal Killing vector field
\bea
\bD_{(a} v_{b)} = \frac{1}{6} \eta_{ab} \bD_{c} v^c ~.
\eea
By employing the results of section \ref{section3}, it may be shown that every conformal Killing vector field on $\mathcal{M}^6$ may be lifted to a unique even conformal Killing vector superfield on $\mathcal{M}^{6|8}$.

We also note that, at the component level, $\sSU(2)_R$ transformations are generated by 
\bea
k^{ij} [v] := K^{ij}[\xi]| ~,
\eea
which satisfies the differential equation
\bea
\label{B.14}
\bD_{a} k^{ij} [v] = \mathcal{R}_{ab}{}^{ij} v^b + v^b \bD_{b} c_a{}^{ij} + k_{a}{}^{b} [v] c_b^{ij} + 2 k^{(i}{}_{k} [v] c_{a}^{j)k} + \o [v] c_{a}^{ij} ~.
\eea

It should be remarked that in the case of Poincar\'e supergravity, we must supplement the constraints above with
\bea
\s[\xi] = 0 \implies \o[v] | = 0~,
\eea
which implies that $v^a$ is a Killing vector field
\bea
\bD_{(a} v_{b)} = 0~.
\eea
Further, by making use of \eqref{3.17} equation \eqref{B.14} reduces to
\bea
\bD_{a} k^{ij} [v] = \mathcal{R}_{ab}{}^{ij} v^b ~.
\eea


\subsection{Conformal Killing spinors}

Our analysis in this subsection will be restricted 
to those backgrounds which admit 
at least one conformal supersymmetry. 
Such spacetimes are associated with a superspace possessing
an odd conformal Killing supervector field $\x^{A}$ \eqref{odd},
from which one can identify a conformal Killing spinor $\e^{\a}_i$
as the bar projection of $\x^\a_i$.
The spinor $\e^\a_i$ generates a $Q$-supersymmetry transformation, while
$S$-supersymmetry transformations are parametrised by
\be \eta_\a^i := \cD_\a^i \s[\xi]| \ .
\ee

With the previous assumptions at hand, bar-projecting equation \eqref{conf-Killing-0} gives
\bea
\bD_a \e^\g_k 
= (\g_{ab})_\b{}^\g \e^{\b j} c^b{}_{jk}
+ \hf (\g^{bc})_\b{}^\g \e^\b_k (w_{abc} + 2 n_{abc})
+ \frac{\ri}{2} (\tilde\g_a)^{\b\g} \eta_{\b k} \ . \label{confKillspinEq}
\eea
Moreover, eq. \eqref{3.38} implies
\bea
\eta_{\b k} = - \frac{\ri}{3} \bD_{\b\d} \e^\d_k + \frac{4 \ri}{3} \e^{\d j} c_{\b\d}{}_{jk}
- \frac{\ri}{6} (\g^{abc})_{\d\b} \e^\d_k (w_{abc} + 2 n_{abc}) \ ,
\eea
and \eqref{confKillspinEq} becomes
\be \bD_a \e_k^\g - c_a{}_k{}^j \e^\g_j
- \hf (w_{abc} + 2 n_{abc}) (\g^{bc})_\b{}^\g \e^\b_k
= \frac{\ri}{2} (\tilde{\g}_a)^{\b\g} \big( \eta_{\b k} - 2 \ri (\g_b)_{\b\d} \e^{\d j} c^b{}_{jk} \big)
~,
\ee
or, equivalently,
\be \hat{\bD}_a \e_k^\g 
= \frac{\ri}{2} (\tilde{\g}_a)^{\b\g} \big( \eta_{\b k} - 2 \ri (\g_b)_{\b\d} \e^{\d j} c^b{}_{jk} \big) 
\equiv 
\frac{\ri}{2} (\tilde{\g}_a)^{\b\g} \hat\eta_{\b k}\ ,
\ee
where we have defined
\be \hat{\bD}_a := \bD_a + c_a{}^{kl} J_{kl} - (w_{abc} + 2 n_{abc}) M^{bc} \ ,
\ee
and
\be \hat \eta_{\a k} := - \frac{\ri}{3} (\g^a)_{\a\b} \hat\bD_a \e^\b_k
~.
\ee

The conformal Killing spinor equation then takes the particularly simple form
\bea
\hat\bD_a \e^\g_k = - \frac{1}{6} (\tilde{\g}_a)^{\g\b} (\g^b)_{\b\d} \hat{\bD}_b \e^\d_k
~~~~~~
\Longleftrightarrow
~~~~~~
\hat\bD_{\a \b} \e^\g_k = - \frac{2}{3} \d_{[\a}{}^{\g} \hat \bD_{\b] \d} \e^{\d}_k
\eea
In particular, we see that the gamma-traceless part of $\hat\bD_a \e^\g_k$ is identically zero.

Associated with a non-zero \emph{commuting} spinor $\e^\a_i$ is the 6-vector
\be V_a = (\g_a)_{\a\b} \eps^{ij} \e^\a_i \e^\b_j \ ,
\ee
which proves to be a conformal Killing vector field when $\e^\a_i$ is a solution to \eqref{confKillspinEq}:
\be \bD_{(a} V_{b)} = \frac{1}{6} \eta_{ab} \bD^c V_c \ .
\ee
Furthermore, it is a null vector
\be V^2 := V^a V_a = 0 \ .
\ee

By construction, the following identities hold
\be 
\d (\cD_\a^i C_a{}^{jk}) = 0 \ , \quad \d (\cD_\a^i W_{abc}) = 0 \ , \quad \d (\cD_\a^i N_{abc}) = 0 \ , \label{FermcompTosionCond}
\ee
which implies the conditions \eqref{BB} are superconformal. The 
bar-projection of the above conditions imply the following constraints
\bsubeq\label{3dev-sigma}
\bea
(\tilde{\g}_a)^{\b\g} \cD_\a^i \cD_{\b}^{(j} \cD_{\g}^{k)} \s|
&=& 4 \ri \e^\b_l \Big(- [\cD_\a^i , \cD_\b^l] C_a{}^{jk}| 
- 2 \ri \eps^{il} \bD_{\a\b} c_a{}^{jk}
- 4 \ri (\g_{abc})_{\a\b} c^b{}^{il} c^c{}^{jk} \non\\
&&~~~~~~~
+ 4 \ri \eps^{il} (\g^b)_{\a\b} (w_{a bc} + 2 n_{abc}) c^c{}^{jk} 
- 12 \ri \eps^{il} c_{\a\b}{}^{p (j} c_a{}_p{}^{k)} \non\\
&&~~~~~~~
- \frac{8 \ri}{3} N^{bcd} (\g_{bcd})_{\a\b} \big(
\eps^{j(i} c_a{}^{l)k}
+ \eps^{k(i} c_a{}^{i) j} \big)
\Big) 
\non\\
&&
+ 8 \ri (\g_{ab})_\a{}^\g \eta_\g^i c^b{}^{jk} 
- 32 \ri \eta_{\a l} \big( \eps^{j(i} c_a{}^{l)k} 
+ \eps^{k(i} c_a{}^{l) j} 
\big)
+ 8 \ri \eta_\a^i c_a{}^{jk}
~,~~~~~~~~~
\eea
\bea
0 &= &
\e^\b_j \Big( - \hf [\cD_\a^i , \cD_\b^j ] W_{abc}|  
- \ri \eps^{ij} \bD_{\a\b} w_{abc}
+ 6 \ri (\g_{[a}{}^{de})_{\a\b} w_{bc]d} c_e{}^{ij}
 \non\\
&&~~~~
+ 6 \ri \eps^{ij} (\g_d)_{\a\b} 
(w^{de}{}_{[a} + 2 n^{de}{}_{[a}) w_{bc]e}
 \Big) 
\non\\
&&- 3 \eta_\b^i (\g^d{}_{[a})_\a{}^\b w_{bc] d}
+ \eta_\a^i w_{abc}
~,
\eea
\bea
\frac{\ri}{32} (\tilde{\g}_{abc})^{\g\d} \cD_\a^i \cD_\g^k \cD_{\d k} \s|
&=&
\e^\b_j \Big( - \hf [\cD_\a^i , \cD_\b^j ] N_{abc}|  
- \ri \eps^{ij} \bD_{\a\b} n_{abc} 
+ 6 \ri (\g_{[a}{}^{de})_{\a\b} n_{bc]d} c_e{}^{ij}
\non\\
&&~~~~~~
+ 6 \ri \eps^{ij} (\g_d)_{\a\b} 
(w^{de}{}_{[a} + 2 n^{de}{}_{[a}) n_{bc]e}
 \Big) 
\non\\
&&- 3 \eta_\b^i (\g^d{}_{[a})_\a{}^\b n_{bc] d}
+ \eta_\a^i n_{abc}
~.
\eea
\esubeq
Restrictions on higher mass-dimension component parameters may be obtained from the 
invariance of higher-order spinor derivatives of  $C_a{}^{jk}$, $W_{abc}$ and $N_{abc}$. These results exemplify 
how results in components can be efficiently obtained from a superspace setting.


In the case of Poincar\'e  supergravities, the equations given above must be supplemented by the additional condition
\be
\s[\xi] = 0 \implies \eta_\a^i = 0 \ ,
\ee
which is a consequence of \eqref{constr-sigma}.
The conformal Killing spinor equation \eqref{confKillspinEq} becomes
\be
\hat\bD_a \e^\g_k = (\tilde{\g}_a)^{\b\g} (\g_b)_{\b\r} \e^{\r j} c^b{}_{jk} \ ,
\ee
which implies
\be \hat\bD_{\d\r} \e^\r_k
= 6 (\g_b)_{\d\r} \e^{\r j} c^b{}_{jk} \ ,
\ee
and
\be \e^{\d k} \hat\bD_{\d\r} \e^\r_k = 0
~.
\ee
This implies that
\be \hat\bD^a V_a = 0 = \bD^a V_a \ ,
\ee
thus $V_a$ is a Killing vector field
\bea
\bD_{(a} V_{b)} = 0 \ .
\eea



\section{From conformal to $\sSU(2)$ superspace} 
\label{degauging}

As is well known, $\sSU(2)$ superspace exists as a gauge-fixed version of conformal superspace. The process of moving from the latter to the former is known as `degauging' and we outline it here extending previous analysis in $3\leq D\leq 5$,
see \cite{Butter4DN=1,Butter4DN=2,BKNT-M1,BKNT-M5D}.

The first step in this
procedure is to eliminate the dilatation connection. Under an infinitesimal special conformal gauge transformation
the one-form $B = E^a B_a + E^\a_i B_\a^i$ transforms as
\begin{align}
\d_{K}(\L) B = -2 E{}^a \L_a - 2   E{}^\a_i \L_\a^i \ . \label{GFsuperWeyl}
\end{align}
Thus, in exchange for a loss of unconstrained special conformal gauge freedom,\footnote{There exists a class of combined local dilatations and special conformal transformations preserving the gauge $B=0$. These exactly reproduce the super-Weyl transformations \eqref{WeylTransInf}, see e.g. \cite{KLT-M,Butter4DN=1,KPR}. } one can gauge away $B$.
\be B_A = 0 \ . \label{GCond}
\ee 
As a result, the special conformal connection becomes auxiliary and must be manually extracted from $\nabla_{A}$.

The \emph{degauged} covariant derivatives are given by
\be
\mathscr{D}_A := \nabla_A + \mathfrak{F}_{AB} K^B
	= E_A - \frac{1}{2} \Omega_A{}^{bc} M_{bc} - \Phi_A{}^{ij} J_{ij} \ .
\ee
Since their structure group is $\sSO(5, 1) \times \sSU(2)_{R}$
it is clear that they are $\sSU(2)$ 
superspace covariant derivatives. 
They satisfy the algebra
\bea
[\mathscr{D}_A,\mathscr{D}_B\}
=
-\bT_{AB}{}^C\mathscr{D}_C
- \frac{1}{2} \bR_{AB}{}^{cd} M_{cd} - \bR_{AB}{}^{kl} J_{kl}
~.
\eea
The degauged special 
conformal connections $\mathfrak{F}_A{}^B$ provide new
contributions to the torsion, and by extension to the other curvatures.

We use different symbols for the degauged derivatives and the ${\sSU(2)}$ ones of section \ref{SU2-superspace}
since, as we will see, they 
satisfy slightly different torsion constraints.
Since the vielbein, Lorentz, and $\sSU(2)$ connections are exactly those
of conformal superspace, it is easy to give expressions for the new torsion and curvature
tensors in terms of their conformal counterparts. This can be done by using the expression of the conformal superspace
torsion and curvature two-forms in terms of the vielbein and connection superfields \cite{BKNT}
\begin{subequations} \label{torCurExp}
\bea
\scT^a &=& \rd E^a + E^b \wedge \Omega_b{}^a + E^a \wedge B \ , \\
\scT{}^\a_i &=& \rd E^\a_i 
+ E^\b_i \wedge \Omega_\b{}^\a 
+ \hf E^\a_i \wedge B - E^{\a j} \wedge \Phi_{ji} 
- \ri \, E^c \wedge \mathfrak{F}_{\b i} (\tilde{\g}_c)^{\a\b} \ ,~~~~~~~~~~~ \\
\sRD &=& \rd B + 2 E^a \wedge \mathfrak{F}_a + 2 E^\a_i \wedge \mathfrak{F}_\a^i \ ,
\label{RDconf} \\
\sRM^{ab} &=& \rd \Omega^{ab} + \Omega^{ac} \wedge \Omega_c{}^b 
- 4 E^{[a} \wedge \mathfrak{F}^{b]} 
+ 2 E^\a_j \wedge \mathfrak{F}_\b^j (\g^{ab})_\a{}^\b \ , \\
\sRJ^{ij} &=& \rd \Phi^{ij} - \Phi^{k (i} \wedge \Phi^{j)}{}_k - 8 E^{\a (i} \wedge \mathfrak{F}_{\a}^{j)} \ , \\
\sRK^a &=& \rd \mathfrak{F}^a 
+ \mathfrak{F}^b \wedge \Omega_b{}^a 
- \mathfrak{F}^a \wedge B 
- \ri \mathfrak{F}_\a^k \wedge \mathfrak{F}_{\b k} (\tilde{\g}^a)^{\a\b} \ , \\
\sRS_\a^i &=& \rd \mathfrak{F}_\a^i 
- \mathfrak{F}_\b^i \wedge \Omega_\a{}^\b
- \hf \mathfrak{F}_\a^i \wedge B 
- \mathfrak{F}_\a^j \wedge \Phi_j{}^i 
- \ri E^{\b i} \wedge \mathfrak{F}^c (\g_c)_{\a\b}    \ .
\eea
\end{subequations}
For example, in the gauge $B=E^AB_A\equiv 0$, one finds the torsion tensors are related by
\be\label{eq:degaugedTorsion}
\bT^a = \scT^a \ , \quad
\bT^\a_i = \scT{}^\a_i 
+ \ri E^c \wedge \mathfrak{F}_{\b i} (\tilde{\g}_c)^{\a\b}
~.
\ee

By investigating \eqref{eq:degaugedTorsion}, one can extract the structure of the torsion constraints in the degauged geometry. We find that these are all the same as
for the covariant derivatives $\cD_A$, except that
\be  
\bT_a{}_{\b (j}{}^\b{}_{k)} \neq 0 
\ , \quad 
\bT_{a b}{}^c \neq 0 
\ . 
\label{unNatTorsionComps}
\ee
In $\sSU(2)$ superspace geometry of section \ref{SU2-superspace}, both of these torsions are required to vanish.
As we will see, these conditions can be satisfied by redefining the degauged vector covariant derivative. Then, the resulting geometry exactly reproduces the $\sSU(2)$ superspace geometry of section \ref{SU2-superspace}.

To elaborate further, we must analyse the additional superfields introduced
by the special conformal connections $\mathfrak{F}_A{}^B$.
In the gauge \eqref{GCond} the dilatation curvature,
eq.~\eqref{RDconf}, is given by
\be 
\sRD_{AB} = 2 \mathfrak{F}_{AB}
 - 2 \mathfrak{F}_{BA} (-1)^{\eps_A \eps_B} \ .
\ee
The vanishing of the dilatation curvature at dimension-1, see \eqref{algb-000},
 constrains the special conformal connection as
\be 
\mathfrak{F}_\a^i{}_\b^j = - \mathfrak{F}_\b^j{}_\a^i 
=
 - \frac{\ri}{4} A_{\a\b}{}^{ij} + \ri \eps^{ij} Y_{\a \b}  \ ,
\label{dim-1-frakF}
\ee
where the superfields $A_{\a\b}{}^{ij}$, and $Y_{\a \b}$ satisfy 
\be 
A_{\a \b}{}^{ij} = (\g^a)_{\a\b} A_a{}^{ij}
=A_{\a \b}{}^{ji}
=-A_{\b \a}{}^{ji}
 \ , 
\quad 
Y_{\a \b} = Y_{\b \a} 
=\frac{1}{6}(\g^{abc})_{\a\b} Y_{abc}
\ .
\ee

At this point it is possible to derive the degauged 
algebra of covariant derivatives. An efficient way to do this is to consider 
a weight-zero primary superfield $U_0$ transforming as a tensor in some representation 
of the remainder of the superconformal algebra. For example, 
to determine the anti-commutator of spinor derivatives we consider
\be
\{ \mathscr{D}_\a^i , \mathscr{D}_\b^j \} U_0 = \{ \nabla_\a^i , \nabla_\b^j \} U_0
	+ \mathfrak{F}_\a^i{}_C [K^C , \nabla_\b^j \} U_0
	+ \mathfrak{F}_\b^j{}_C [K^C , \nabla_\a^i \} U_0  \ .
\ee
The resulting algebra is
\bea
\{ \mathscr{D}_\a^i , \mathscr{D}_\b^j \}  &=&
- 2 \ri \eps^{ij} (\g^a)_{\a\b} \mathscr{D}_a  
+4\ri\eps^{ij} Y^{bcd}(\g_b)_{\a\b} M_{cd}
+2\ri \ve^{ij}(\g^a)_{\a\b}A_a{}^{kl}   J_{kl}
\non\\
&&
+\frac{\ri}{2} A_b{}^{ij}(\g^{bcd})_{\a\b} M_{cd}
+\frac{8\ri}{3} (\g^{abc})_{\a \b}Y_{abc}    J^{ij}
~.
\eea
To match 
\eqref{Algebra-1} it is necessary to make the following identifications
\bea
A_a{}^{ij}=4C_a{}^{ij}
~,~~~~~~
Y_{abc}=-N_{abc}
~,
\eea
and
\bea
\mathscr{D}_\a^i=\cD_\a^i~,~~~~~~
\mathscr{D}_a={\cD_a }
+{W}_{abc} M^{bc}
+C_{a}^{kl}J_{kl} 
~.
\label{bD-cD}
\eea

Next, we compute $[\mathscr{D}_a,\mathscr{D}_\b^j]$ at dimension-1. One finds
\be
[ \mathscr{D}_a , \mathscr{D}_\b^j ] U_0 = [ \nabla_a , \nabla_\b^j ] U_0
+ \mathfrak{F}_a{}_C [K^C , \nabla_\b^j \} U_0
- \mathfrak{F}_\b^j{}_C [K^C , \nabla_a \} U_0 
~,
\non
\ee
which implies
\bea
[ \mathscr{D}_a , \mathscr{D}_\b^j ] 
&=&
C_a{}^{j}{}_{k}\mathscr{D}_{\b}^{ k}
+C^b{}^{j}{}_{k}(\g_{ab})_{\b}{}^{\d}\mathscr{D}_{\d }^{k}
-W_{acd}(\g^{cd})_\g{}^{\d}\mathscr{D}_\d^j
-N_{acd}(\g^{cd})_\b{}^\d \mathscr{D}_{\d}^{j}
+\cdots
\non
\eea

Finally, we turn to $\mathfrak{F}_a{}_\b^j$.
Since at mass dimension-3/2, \eqref{RDconf}  implies
\bea 
\sRD_{a}{}_\b^j &=& 
2 \mathfrak{F}_{a}{}_\b^j
 - 2 \mathfrak{F}_\b^j{}_{a} 
 ~,
\eea
by 
employing the dilatation curvatures of eq. \eqref{def-W}, 
one obtains
\bsubeq
\bea
\mathfrak{F}_\b^j{}_{a} 
&=&
 \mathfrak{F}_{a}{}_\b^j
+ \frac{1}{16}  \mathscr{D}_{\b}^i W^{\a\b}
=
 \mathfrak{F}_{a}{}_\b^j
+ \frac{5\ri}{8} X^{\b i}
~.
\label{frakF32-1}
\eea
\esubeq

By examining the expressions for the conformal superspace torsion and curvatures of eq.~\eqref{torCurExp}, one can obtain
\begin{subequations}
\bea
\bT_{AB}{}^c&=&
\scT_{AB}{}^c
  \ , 
\\
\bT_{AB}{}^\g_k &=& 
\scT_{AB}{}^\g_k
-\ri \d^c_{A}(\tilde{\g}_c)^{\g\d}\mathfrak{F}_{B,}{}_{\d k} 
+\ri \d^c_{B}(\tilde{\g}_c)^{\g\d}\mathfrak{F}_{A,}{}_{\d k} 
 \ ,~~~~~~~~~~~ 
\\
\bR_{AB}{}^{cd} 
&=&
\sRM_{AB}{}^{cd}
- 4 \d_{A}^{[c}\mathfrak{F}_{B,}{}^{d]} 
+ 4 \d_{B}^{[c}\mathfrak{F}_{A,}{}^{d]} 
+2\d_{A}{}^\g_k (\g^{cd})_\g{}^\d \mathfrak{F}_{B,}{}_\d^k(-1)^{\ve_B} 
\non\\
&&
-2\d_{B}{}^\g_k (\g^{cd})_\g{}^\d \mathfrak{F}_{A,}{}_\d^k
~,~~~~~~~~~~~ 
\\
\bR_{AB}{}^{kl}
&=&
\sRJ_{AB}{}^{kl}
-8\d_{A}{}^{\r}_p\,\mathfrak{F}_{B,}{}_{\r}^{(k}\ve^{l)p} (-1)^{\ve_B} 
+8 \d_{B}{}^{\r}_p\,\mathfrak{F}_{A,}{}_{\r}^{(k}\ve^{l)p} 
 \ ,
\eea
\end{subequations}
as well as the following conditions on the special conformal connections
\begin{subequations}
\bea
\sRS_{AB}{}_\g^k &=& 
2\mathscr{D}_{[A}\mathfrak{F}_{B)}{}_\g^k 
+\bT_{AB}{}^D \mathfrak{F}_{D,}{}_\g^k
+\ri\ve^{kl}\d_{A}{}^{\d}_{l} (\g_c)_{\g\d}\mathfrak{F}_{B,}{}^c (-1)^{\ve_B}
\non\\
&&
-\ri\ve^{kl}\d_{B}{}^{\d}_{l} (\g_c)_{\g\d}\mathfrak{F}_{A,}{}^c 
\ ,
\label{3/2Scurv}
\\
\sRK_{AB}{}^c &=&
2\mathscr{D}_{[A}\mathfrak{F}_{B)}{}^c
+\bT_{AB}{}^D \mathfrak{F}_{D,}{}^c
+ \ri \mathfrak{F}_{A,}{}_{\g}^{ k}\mathfrak{F}_{B,}{}_{\d k} (\tilde{\g}^c)^{\g\d}(-1)^{\ve_B}
\non\\
&&
- \ri \mathfrak{F}_{B,}{}_{\g}^{ k}\mathfrak{F}_{A,}{}_{\d k} (\tilde{\g}^c)^{\g\d}(-1)^{\ve_{A}\ve_B+\ve_A}
 \ .~~~~~~~~~~~~
\eea
\end{subequations}

At dimension-3/2 only the $S$-curvature equation, eq.~\eqref{3/2Scurv}, with $A ={}_\a^i$ and $B ={}_\b^j$
gives nontrivial constraints.
In particular, by using  $\sRS_{\a}^i{}_\b^j{}_\g^k=0$ and \eqref{frakF32-1}, one obtains 
\bea
0 &=&
\mathscr{D}_{\a}^i\mathfrak{F}_{\b}^j{}_\g^k 
+\mathscr{D}_{\b}^j\mathfrak{F}_{\a}^i{}_\g^k 
+2\ri\ve^{ij}(\g^a)_{\a\b} \mathfrak{F}_{a}{}_\g^k
-\ri\ve^{ki}(\g^a)_{\g\a} \mathfrak{F}_{a}{}_{\b}^j
-\ri\ve^{kj}(\g^a)_{\g\b} \mathfrak{F}_{a}{}_{\a}^i
\non\\
&&
+\frac{\ri}{8}\ve^{ij}\ve_{\a\b\g\d}\mathscr{D}^k_\r W^{\d\r}
~.
\label{BI-000001}
\eea
Its solution implies the differential constraints
\bea
\mathscr{D}_{(\a}^{(i}C_{\b)\g}{}^{jk)}=0
~,~~~~~~
\mathscr{D}_{(\a}^{i} N_{\b\g)}=0
~,
\label{dim-3/2-diff-constr}
\eea
which indicates that the decomposition into irreducible and nontrivial tensors of the spinor derivatives of dimension-1 torsions
is
\bsubeq
\bea
\mathscr{D}_{\gamma}^{ k } C_{a}{}^{ij} &=&
(\gamma_a)_{\gamma \delta}\mathcal C^{\delta}{}^{kij}
-\varepsilon^{k(i}\mathcal C_{a\, \gamma}{}^{ j)} 
- \varepsilon^{k(i}(\gamma_a)_{\gamma \delta}\mathcal C^{\delta j)}
~,
\\
\mathscr{D}_{\gamma}^{ k} N_{abc}&=&
-\frac{3}{4} (\g_{[ab})_{\g}{}^{\b}\cN_{c]}{}_{\b}{}^{k}
~,
\\
\mathscr{D}_{\gamma}^{ k} W_{abc}
&=&
\ri (\g_{abc})_{\a\b}X_{\gamma}{}^{\alpha \beta}{}^k 
+ \ri (\g_{abc})_{\g\d}X^{\d}{}^k
~.
\eea
\esubeq
Equation \eqref{BI-000001} then implies
\bea
\mathfrak{F}_{a}{}_{\b}^j
=
-\frac{5\ri}{24}(\g_a)_{\b\d}X^{\d j}
-\frac{1}{2}(\g_a)_{\b\d}\cC^{\d j}
+\frac{1}{2}\cC_{a}{}_{\b}{}^{j} 
-\frac{1}{2}\cN_{a}{}_{\b}{}^{j}
~.
\eea

To conclude the analysis at dimension-3/2 we derive the corresponding torsion and curvatures.
For the dimension-3/2 torsion it holds that
\bea
\bT_{ab}{}^\g_k &=& 
\scT_{ab}{}^\g_k
-2\ri (\tilde{\g}_{[a})^{\g\d}\mathfrak{F}_{b]}{}_{\d k} 
~,
\eea
which leads to
\bea
\bT_{ab}{}^\g_k &=&
(\g_{ab})_\b{}^\a 
\Big(
X_{\a}{}^{\b\g}{}_k 
- \frac{3}{4} \d^\g_\a X^\b_k 
\Big)
+\ri(\tilde{\g}_{[a})^{\g\d}
\Big(
\cN_{b]\d}{}_{k}
-\cC_{b] \d}{}_{k} 
\Big)
\non\\
&&
+ ({\g}_{ab})_{\r}{}^{\g}
\Big(
\frac{5}{12}X^{\r}_{ k}
-\ri\cC^{\r}_ {k}
\Big)
~.
\eea
The dimension-3/2 Lorentz curvature can be computed by using
\bea
\bR_{a}{}_\b^j{}^{cd} 
&=&
\sRM_{a}{}_\b^j{}^{cd}
- 4 \d_{a}^{[c}\mathfrak{F}_{\b}^j{}^{d]} 
-2\d_\b^\g\d^j_k (\g^{cd})_\g{}^\d \mathfrak{F}_{a}{}_\d^k
~,~~~~~~~~~~~ 
\eea
which becomes
\bea
\bR_{a}{}_\b^j{}^{cd} 
&=&
2\ri(\g_a)_{\b\g} (\g^{cd})_\d{}^\r X_\r{}^{\g\d}{}^j
+(\g_a{}^{cd})_{\b\g}
\Big(
\frac{2\ri}{3}X^{\g j}
+\cC^{\g j}
\Big)
+\d_{a}^{[c}(\g^{d]})_{\b\g}\Big(
- \frac{4\ri}{3} X^{\g j}
+4\cC^{\g j}
\Big)
\non\\
&&
+2\d_{a}^{[c}\Big(
\cN^{d]}{}_{\b}{}^{j}
-\cC^{d]}{}_{ \b}{}^{j} 
\Big)
+(\g^{cd})_\b{}^\g\Big(
\cN_{a\g}{}^{j}
-\cC_{a \g}{}^{j} 
\Big)
~.
\eea
Finally, the $\sSU(2)_R$ curvature derives from
\bea
\bR_{a}{}_\b^j{}^{kl}
&=&
\sRJ_{a}{}_\b^j{}^{kl}
+8\d_\b^\r\d^j_p\,\mathfrak{F}_{a}{}_{\r}^{(k}\ve^{l)p} 
\eea
which implies
\bea
\bR_{a}{}_\b^j{}^{kl}
=
\Big\{
(\g_a)_{\b\g}\Big(
\frac{10\ri}{3}  X^{\g (k}
-4\cC^{\g (k}\Big)
-4\Big(
\cN_{a\b}{}^{(k}
-\cC_{a \b}{}^{(k} 
\Big)
\Big\}
\ve^{l)j}
~.
\eea
One can then prove that these results coincide with the dimension-3/2 results of section \eqref{SU2-superspace}
upon using \eqref{bD-cD} and identifying
\bea
X_{\gamma k}{}^{\alpha \beta}=-\frac{\ri}{4}\mathcal W_{\gamma k}{}^{\alpha \beta}
~,~~~~~~
X^{\beta}_k=-\frac{\ri}{4}\mathcal W^{\beta}_k
~.
\eea

It is straightforward to continue the degauging procedure and
obtain results
at dimensions higher than 3/2. 
We will not pursue such an analysis here.


\begin{footnotesize}

\end{footnotesize}


\begin{thebibliography}{66}

\bibitem{BSVanP} 
  E.~Bergshoeff, E.~Sezgin and A.~Van Proeyen,
  ``Superconformal tensor calculus and matter couplings in six dimensions,''
  Nucl.\ Phys.\ B {\bf 264}, 653 (1986)
  Erratum: [Nucl.\ Phys.\ B {\bf 598}, 667 (2001)].

\bibitem{deWvHVP1}
 B.~de Wit, J.~W.~van Holten and A.~Van Proeyen,
 ``Transformation rules of $N=2$ supergravity multiplets,''
Nucl.\ Phys.\  B {\bf 167}, 186 (1980).
  
\bibitem{deRvHdeWVP} 
  M.~de Roo, B.~de Wit,   J.~W.~van Holten and A.~Van Proeyen,
  ``Chiral superfields in $N=2$ supergravity,''
  Nucl.\ Phys.\ B {\bf 173}, 175 (1980).  

\bibitem{deWvHVP2}
  B.~de Wit, J.~W.~van Holten and A.~Van Proeyen,
  ``Structure of $N=2$ supergravity,''
  Nucl.\ Phys.\ B {\bf 184}, 77 (1981)
  Erratum: [Nucl.\ Phys.\ B {\bf 222}, 516 (1983)].

\bibitem{deWPVP} 
  B.~de Wit, R.~Philippe and A.~Van Proeyen,
  ``The improved tensor multiplet in $N=2$ supergravity,''
  Nucl.\ Phys.\ B {\bf 219}, 143 (1983).

\bibitem{deWLPSVP} 
  B.~de Wit, P.~G.~Lauwers, R.~Philippe, S.~Q.~Su and A.~Van Proeyen,
  ``Gauge and matter fields coupled to $N=2$ supergravity,''
  Phys.\ Lett.\ B {\bf 134}, 37 (1984).

\bibitem{deWLVP}
B.~de Wit, P.~G.~Lauwers and A.~Van Proeyen,
``Lagrangians of $N=2$ supergravity-matter systems,''
Nucl.\ Phys.\  B {\bf 255}, 569 (1985).


\bibitem{CVanP}
F.~Coomans and A.~Van Proeyen,
``Off-shell N=(1,0), D=6 supergravity from superconformal methods,''
JHEP \textbf{02}, 049 (2011)
[erratum: JHEP \textbf{01}, 119 (2012)]
[arXiv:1101.2403 [hep-th]].


\bibitem{BCSVanP} 
  E.~Bergshoeff, F.~Coomans, E.~Sezgin and A.~Van Proeyen,
  ``Higher derivative extension of $6D$ chiral gauged supergravity,''
  JHEP {\bf 1207}, 011 (2012)
  [arXiv:1203.2975 [hep-th]].

\bibitem{LVP}
E.~Lauria and A.~Van Proeyen,
{\it $N=2$ Supergravity in $D=4,5,6$ Dimensions},
Lect. Notes Phys. \textbf{966}, Springer, 2020
[arXiv:2004.11433 [hep-th]].

\bibitem{BSS1} 
  E.~Bergshoeff, A.~Salam and E.~Sezgin,
  ``A supersymmetric $R^2$-action in six dimensions and torsion,''
  Phys.\ Lett.\ B {\bf 173}, 73 (1986).  
  
\bibitem{BSS2} 
  E.~Bergshoeff, A.~Salam and E.~Sezgin,
  ``Supersymmetric $R^2$ actions, conformal invariance and Lorentz Chern-Simons term in 6 and 10 dimensions,''
  Nucl.\ Phys.\ B {\bf 279}, 659 (1987).  
  
\bibitem{BR} 
  E.~Bergshoeff and M.~Rakowski,
  ``An off-shell superspace $R^2$-action in six dimensions,''
  Phys.\ Lett.\ B {\bf 191}, 399 (1987).  

\bibitem{VanP} 
A.~Van Proeyen,
``Superconformal symmetry and higher-derivative Lagrangians,''
Springer Proc.\ Phys.\  {\bf 153}, 1 (2014) [arXiv:1306.2169 [hep-th]].



\bibitem{LT-M12} 
W.~D.~Linch III and G.~Tartaglino-Mazzucchelli,
  ``Six-dimensional supergravity and projective superfields,''
  JHEP {\bf 1208}, 075 (2012)
  [arXiv:1204.4195 [hep-th]].

\bibitem{BKNT}
D.~Butter, S.~M.~Kuzenko, J.~Novak and S.~Theisen,
``Invariants for minimal conformal supergravity in six dimensions,''
JHEP \textbf{12}, 072 (2016)
[arXiv:1606.02921 [hep-th]].

\bibitem{Nahm}  W.~Nahm,
 ``Supersymmetries and their representations,''
  Nucl.\ Phys.\ B {\bf 135}, 149 (1978).

\bibitem{Howe1}
P.~S.~Howe,
``A superspace approach to extended conformal supergravity,''
Phys.\ Lett.\  B {\bf 100}, 389 (1981).

\bibitem{Howe2}
P.~S.~Howe,
``Supergravity in superspace,''  Nucl.\ Phys.\  B {\bf 199}, 309 (1982).

\bibitem{GGRS}
 S.~J.~Gates Jr., M.~T.~Grisaru, M.~Ro\v{c}ek and W.~Siegel,
{\it Superspace, or One Thousand and One Lessons in Supersymmetry},
Benjamin/Cummings (Reading, MA), 1983,
arXiv:hep-th/0108200.
     

\bibitem{KLRT-M2}
S.~M.~Kuzenko, U.~Lindstr\"om, M.~Ro\v cek and G.~Tartaglino-Mazzucchelli,
``On conformal supergravity and projective superspace,''
JHEP {\bf 0908}, 023 (2009)
[arXiv:0905.0063 [hep-th]].

    
  \bibitem{KLR}
A. Karlhede, U. Lindstr\"om and M. Ro\v cek,
``Self-interacting tensor multiplets in N=2 superspace,''
Phys.\ Lett.\ B {\bf 147}, 297 (1984).

\bibitem{LR1}
U.~Lindstr\"om and M.~Ro\v{c}ek,
``New hyperk\"ahler  metrics  and new supermultiplets,''
  Commun.\ Math.\ Phys.\  {\bf 115}, 21 (1988).
  
\bibitem{LR2}
U.~Lindstr\"om and M.~Ro\v{c}ek,  
 ``N=2 super Yang-Mills theory in projective superspace,''
Commun.\ Math.\ Phys.\  {\bf 128}, 191 (1990).   

\bibitem{K06} 
  S.~M.~Kuzenko,
  ``On compactified harmonic/projective superspace, 5D superconformal theories, 
  and all that,'' Nucl.\ Phys.\ B {\bf 745}, 176 (2006)
  [hep-th/0601177].
  
\bibitem{K07}
S.~M.~Kuzenko, ``On superconformal projective hypermultiplets,''
JHEP {\bf 0712}, 010 (2007) [arXiv:0710.1479].
    


\bibitem{KT-M08}
  S.~M.~Kuzenko and G.~Tartaglino-Mazzucchelli,
  ``Super-Weyl invariance in 5D supergravity,''
  JHEP {\bf 0804}, 032 (2008)
   [arXiv:0802.3953 [hep-th]].
   
\bibitem{HIPT}
P.~S.~Howe, J.~M.~Izquierdo, G.~Papadopoulos and P.~K.~Townsend,
``New supergravities with central charges and Killing spinors in 2+1 dimensions,''
Nucl.\ Phys.\  B {\bf 467}, 183 (1996)  [arXiv:hep-th/9505032].
    
\bibitem{KLT-M}
 S.~M.~Kuzenko, U.~Lindstr\"om and G.~Tartaglino-Mazzucchelli,
``Off-shell supergravity-matter couplings in three dimensions,''
JHEP {\bf 1103}, 120 (2011)  [arXiv:1101.4013 [hep-th]].

\bibitem{Butter14} 
  D.~Butter,
  ``New approach to curved projective superspace,''
  Phys.\ Rev.\ D {\bf 92}, no. 8, 085004 (2015)
  [arXiv:1406.6235 [hep-th]].
  
  \bibitem{Butter15} 
  D.~Butter,
  ``Projective multiplets and hyperk\"ahler cones in conformal supergravity,''
  JHEP {\bf 1506}, 161 (2015)
  [arXiv:1410.3604 [hep-th]].

\bibitem{KT-M_5D2}
S.~M.~Kuzenko and G.~Tartaglino-Mazzucchelli,
  ``Five-dimensional superfield supergravity,''
 Phys.\ Lett.\  B {\bf 661}, 42 (2008)
  [arXiv:0710.3440 [hep-th]].
 
\bibitem{KT-M_5D3} 
S.~M.~Kuzenko and G.~Tartaglino-Mazzucchelli,
  ``5D supergravity and projective superspace,''
JHEP {\bf 0802}, 004 (2008)
  [arXiv:0712.3102 [hep-th]].

\bibitem{KLRT-M1}
S.~M.~Kuzenko, U.~Lindstr\"om, M.~Ro\v cek and G.~Tartaglino-Mazzucchelli,
``4D N=2 supergravity and projective superspace,'' 
JHEP {\bf 0809}, 051 (2008) [arXiv:0805.4683].

\bibitem{GT-M_2D44}
G.~Tartaglino-Mazzucchelli,
``2D N = (4,4) superspace supergravity and bi-projective superfields,''
JHEP \textbf{04}, 034 (2010)
[arXiv:0911.2546 [hep-th]]; 
``On 2D N=(4,4) superspace supergravity,''
Phys. Part. Nucl. Lett. \textbf{8}, 251-261 (2011)
[arXiv:0912.5300 [hep-th]].

\bibitem{Butter4DN=1} 
  D.~Butter,
  ``N=1 conformal superspace in four dimensions,''
  Annals Phys.\  {\bf 325}, 1026 (2010)
  [arXiv:0906.4399 [hep-th]].
  
\bibitem{Butter4DN=2} 
  D.~Butter,
  ``N=2 conformal superspace in four dimensions,''
  JHEP {\bf 1110}, 030 (2011)
  [arXiv:1103.5914 [hep-th]].
  
\bibitem{BKNT-M1} 
  D.~Butter, S.~M.~Kuzenko, J.~Novak and G.~Tartaglino-Mazzucchelli,
  ``Conformal supergravity in three dimensions: New off-shell formulation,''
  JHEP {\bf 1309}, 072 (2013)
  [arXiv:1305.3132 [hep-th]].
  
\bibitem{BKNT-M5D} 
  D.~Butter, S.~M.~Kuzenko, J.~Novak and G.~Tartaglino-Mazzucchelli,
  ``Conformal supergravity in five dimensions: New approach and applications,''
  JHEP {\bf 1502}, 111 (2015)
  [arXiv:1410.8682 [hep-th]].


\bibitem{KTvN1} 
M.~Kaku, P.~K.~Townsend and P.~van Nieuwenhuizen,
``Gauge theory of the conformal and superconformal group,''
Phys.\ Lett.\  B {\bf 69}, 304 (1977).

\bibitem{KTvN2}
M.~Kaku, P.~K.~Townsend and P.~van Nieuwenhuizen,
``Properties of conformal supergravity,''
Phys. Rev. D \textbf{17}, 3179 (1978).

\bibitem{KU}  T. Kugo and S. Uehara, N = 1 superconformal tensor calculus: multiplets with external Lorentz indices and spinor derivative operators, Prog. Theor. Phys. {\bf 73},  235 (1985).

  \bibitem{ISZ05} 
  E.~A.~Ivanov, A.~V.~Smilga and B.~M.~Zupnik,
  ``Renormalizable supersymmetric gauge theory in six dimensions,''
  Nucl.\ Phys.\ B {\bf 726}, 131 (2005)
  [hep-th/0505082].


\bibitem{BNT-M17}
D.~Butter, J.~Novak and G.~Tartaglino-Mazzucchelli,
``The component structure of conformal supergravity invariants in six dimensions,''
JHEP \textbf{05}, 133 (2017)
[arXiv:1701.08163 [hep-th]].

\bibitem{NOPT-M}
J.~Novak, M.~Ozkan, Y.~Pang and G.~Tartaglino-Mazzucchelli,
``Gauss-Bonnet supergravity in six dimensions,''
Phys. Rev. Lett. \textbf{119},  no.11, 111602 (2017) 
[arXiv:1706.09330 [hep-th]].

\bibitem{BNOPT-M}
  D.~Butter, J.~Novak, M.~Ozkan, Y.~Pang and G.~Tartaglino-Mazzucchelli,
  ``Curvature squared invariants in six-dimensional ${\cal N} = (1,0)$ supergravity,''
  JHEP {\bf 1904}, 013 (2019) 
  [arXiv:1808.00459 [hep-th]].


\bibitem{HL20}
P.~S.~Howe and U.~Lindstr\"om,
``Local supertwistors and conformal supergravity in six dimensions,''
[arXiv:2008.10302 [hep-th]].

\bibitem{Howe:2020xrg}
P.~S.~Howe and U.~Lindstr\"om,
``Superconformal geometries and local twistors,''
[arXiv:2012.03282 [hep-th]].


\bibitem{BSVanP2}
E.~Bergshoeff, E.~Sezgin and A.~Van Proeyen,
``(2,0) tensor multiplets and conformal supergravity in D = 6,''
Class.\ Quant.\ Grav.\  {\bf 16}, 3193 (1999)
[hep-th/9904085].

\bibitem{Lott:2001st}
J.~Lott,
``The Geometry of supergravity torsion constraints,''
math/0108125 [math-dg].

\bibitem{SokatchevAA}
  E.~Sokatchev,
  ``Off-shell six-dimensional supergravity in harmonic superspace,''
Class.\ Quant.\ Grav.\  {\bf 5}, 1459-1471 (1988).

\bibitem{BK} I.~L.~Buchbinder and S.~M.~Kuzenko,
{\it Ideas and Methods of Supersymmetry and
Supergravity or a Walk Through Superspace}, IOP, Bristol, 1995
(Revised Edition: 1998).

\bibitem{K15Corfu} 
S.~M.~Kuzenko,
``Supersymmetric spacetimes from curved superspace,''
PoS CORFU {\bf 2014}, 140 (2015) [arXiv:1504.08114 [hep-th]].

\bibitem{KLRST-M} 
  S.~M.~Kuzenko, U.~Lindstr\"om, M.~Ro\v{c}ek, I.~Sachs and G.~Tartaglino-Mazzucchelli,
  ``Three-dimensional $\mathcal{N} =$ 2 supergravity theories: From superspace to components,''
  Phys.\ Rev.\ D {\bf 89}, no. 8, 085028 (2014)
  [arXiv:1312.4267 [hep-th]].

\bibitem{BIL} 
  D.~Butter, G.~Inverso and I.~Lodato,
  ``Rigid 4D $ \mathcal{N}=2 $ supersymmetric backgrounds and actions,''
  JHEP {\bf 1509}, 088 (2015)
  [arXiv:1505.03500 [hep-th]].

\bibitem{KNT-M} 
  S.~M.~Kuzenko, J.~Novak and G.~Tartaglino-Mazzucchelli,
  ``Symmetries of curved superspace in five dimensions,''
  JHEP {\bf 1410}, 175 (2014)
  [arXiv:1406.0727 [hep-th]].
 
\bibitem{FV1} 
  E.~S.~Fradkin and M.~A.~Vasiliev,
  ``Candidate to the role of higher spin symmetry,''
  Annals Phys.\  {\bf 177}, 63 (1987).

\bibitem{FV2} 
  E.~S.~Fradkin and M.~A.~Vasiliev,
  ``Superalgebra of higher spins and auxiliary fields,''
  Int.\ J.\ Mod.\ Phys.\ A {\bf 3}, 2983 (1988).

\bibitem{Vasiliev88} 
  M.~A.~Vasiliev,
  ``Extended higher spin superalgebras and their realizations in terms of quantum operators,''
  Fortsch.\ Phys.\  {\bf 36}, 33 (1988).
 
\bibitem{KV1} 
S.~E.~Konstein and M.~A.~Vasiliev,
  ``Massless representations and admissibility condition for higher spin superalgebras,''
  Nucl.\ Phys.\ B {\bf 312} (1989) 402. 
 
\bibitem{KV2}
S.~E.~Konstein and M.~A.~Vasiliev,
  ``Extended higher spin superalgebras and their massless representations,''
  Nucl.\ Phys.\ B {\bf 331},  475 (1990).  

\bibitem{BKM} C. P. Boyer, E. G. Kalnins and W. Miller Jr., ``Symmetry and separation of variables for the Helmholtz and Laplace equations,'' Nagoya Math. J. {\bf 60}, 35 (1976).

\bibitem{Nikitin} A.~G.~Nikitin, ``Generalized Killing tensors of arbitrary rank and order,''
Ukrainian Math. J. {\bf 43}, 734 (1991).

\bibitem{NikitinP}
A.~G.~Nikitin, O.~I.~Prylypko, ``Generalized Killing tensors and
symmetry of Klein-Gordon-Fock equations,''
Preprint, Akad. Nauk UkrSSR, Inst. Math., 90.26, 2--60, Kiev (1990);
arXiv:math-ph/0506002.

\bibitem{BSSS} 
V.~G.~Bagrov, B.~F.~Samsonov,  A.~V.~Shapovalov and I.~V.~Shirokov, ``Identities on solutions of the wave equation in the enveloping algebra of the conformal group,''
Theor.\ Math.\ Phys.\  {\bf 83}, 347 (1990)
[Teor.\ Mat.\ Fiz.\  {\bf 83}, 14 (1990)].
    
\bibitem{ShSh}
A.~V.~Shapovalov and I.~V.~Shirokov,
``Symmetry algebras of linear differential equations,''
Theor.\ Math.\ Phys.\  {\bf 92}, 697 (1992)
[Teor.\ Mat.\ Fiz.\  {\bf 92}, 3 (1992)].

\bibitem{ShV} 
  O.~V.~Shaynkman and M.~A.~Vasiliev,
  ``Higher spin conformal symmetry for matter fields in (2+1)-dimensions,''
  Theor.\ Math.\ Phys.\  {\bf 128}, 1155 (2001)
  [Teor.\ Mat.\ Fiz.\  {\bf 128}, 378 (2001)]
  [hep-th/0103208].


\bibitem{Eastwood}
M.~G.~Eastwood,
``Higher symmetries of the Laplacian,''
Annals Math. \textbf{161}, 1645-1665 (2005)
[arXiv:hep-th/0206233 [hep-th]].


\bibitem{Vasiliev2004}
 M.~A.~Vasiliev,
  ``Higher spin superalgebras in any dimension and their representations,''
  JHEP {\bf 0412}, 046 (2004)  [hep-th/0404124].
  
  

\bibitem{HL2}
P.~S.~Howe and U.~Lindstr\"om,
``Super-Laplacians and their symmetries,''
JHEP \textbf{05}, 119 (2017)
[arXiv:1612.06787 [hep-th]].


\bibitem{HL1}
P.~S.~Howe and U.~Lindstr\"om,
``Notes on super Killing tensors,''
JHEP \textbf{03}, 078 (2016)
[arXiv:1511.04575 [hep-th]].


\bibitem{HL3}
P.~S.~Howe and U.~Lindstr\"om,
``Some remarks on (super)-conformal Killing-Yano tensors,''
JHEP \textbf{11}, 049 (2018)
[arXiv:1808.00583 [hep-th]].


\bibitem{KR}
S.~M.~Kuzenko and E.~S.~N.~Raptakis,
``Symmetries of supergravity backgrounds and supersymmetric field theory,''
JHEP \textbf{04}, 133 (2020)
[arXiv:1912.08552 [hep-th]].
   
\bibitem{Gates6D}
S.~J.~Gates Jr.,
``Superconformal transformations and six-dimensional space-time,''
Nucl.\ Phys.\ B {\bf 162}, 79 (1980).


\bibitem{Park98} 
  J.~H.~Park,
``Superconformal symmetry in six dimensions and its reduction to four dimensions,''
Nucl.\ Phys.\ B {\bf 539}, 599 (1999)
[hep-th/9807186].

\bibitem{HSierraT}
P.~S.~Howe, G.~Sierra and P.~K.~Townsend,
``Supersymmetry in six dimensions,''
Nucl.\ Phys.\ B {\bf 221} (1983) 331.
  
\bibitem{BSSokatchev}
E.~Bergshoeff, E.~Sezgin and E.~Sokatchev,
``Couplings of selfdual tensor multiplet in six-dimensions,''
Class. Quant. Grav. \textbf{13}, 2875-2886 (1996)
[arXiv:hep-th/9605087 [hep-th]].

\bibitem{GKS} 
S.~J.~Gates Jr., S.~M.~Kuzenko and A.~G.~Sibiryakov,
``Towards a unified theory of massless superfields of all superspins,''
Phys.\ Lett.\ B {\bf 394}, 343 (1997)
[hep-th/9611193].

\bibitem{Weir} G.J. Weir, ``Conformal Killing tensors in reducible spaces,'' J. Math. Phys. 
{\bf 18}, 1782 (1977).

\bibitem{Thompson} 
G. Thompson, ``Killing tensors in spaces of constant curvature,'' 
J. Math. Phys. {\bf 27}, 2693 (1986).

\bibitem{Siegel79}
W.~Siegel,
``Superfields in higher-dimensional spacetime,''
Phys. Lett. B \textbf{80}, 220-223 (1979).

\bibitem{KNT}
S.~M.~Kuzenko, J.~Novak and S.~Theisen,
``Non-conformal supercurrents in six dimensions,''
JHEP \textbf{02}, 030 (2018)
[arXiv:1709.09892 [hep-th]].

\bibitem{FigueroaPapadopoulos}
J.~M.~Figueroa-O'Farrill and G.~Papadopoulos,
``Pl\"ucker type relations for orthogonal planes,''
J. Geom. Phys. \textbf{49}, 294 (2004)
[arXiv:math/0211170 [math.AG]].

\bibitem{Hitchin:2000jd}
N.~J.~Hitchin,
``The geometry of three-forms in six dimensions,''
J. Diff. Geom. \textbf{55},   no.3, 547-576 (2000)
[arXiv:math/0010054 [math.DG]].

\bibitem{MedeirosFigueroaSanti}
P.~de Medeiros, J.~Figueroa-O'Farrill and A.~Santi,
``Killing superalgebras for Lorentzian six-manifolds,''
J. Geom. Phys. \textbf{132}, 13-44 (2018)
[arXiv:1804.00319 [hep-th]].


\bibitem{Meessen}
P.~Meessen,
``A Small note on P P wave vacua in six-dimensions and five-dimensions,''
Phys. Rev. D \textbf{65}, 087501 (2002)
[arXiv:hep-th/0111031 [hep-th]].

\bibitem{GutowskiMartelliReall}
J.~B.~Gutowski, D.~Martelli and H.~S.~Reall,
``All supersymmetric solutions of minimal supergravity in six-dimensions,''
Class. Quant. Grav. \textbf{20}, 5049-5078 (2003)
[arXiv:hep-th/0306235 [hep-th]].

\bibitem{Bandos:2002nn}
I.~A.~Bandos, E.~Ivanov, J.~Lukierski and D.~Sorokin,
``On the superconformal flatness of AdS superspaces,''
JHEP \textbf{06}, 040  (2002)
[arXiv:hep-th/0205104 [hep-th]].

\bibitem{Kuzenko:2017jdy}
S.~M.~Kuzenko, J.~Novak and S.~Theisen,
``New superconformal multiplets and higher derivative invariants in six dimensions,''
Nucl. Phys. B \textbf{925} (2017), 348-361
[arXiv:1707.04445 [hep-th]].

\bibitem{GrojeanMourad}
C.~Grojean and J.~Mourad, ``Superconformal six-dimensional (2,0) theories in superspace,''
Class. Quant. Grav. \textbf{15}, 3397-3409 (1998)
[arXiv:hep-th/9807055 [hep-th]].

\bibitem{WZ}
J.~Wess and B.~Zumino,
``The component formalism follows from the superspace formulation of supergravity,''
Phys. Lett. B \textbf{79}, 394 (1978).

\bibitem{GKLR} S.~J.~Gates Jr., A.~Karlhede, U.~Lindstr\"om and M.~Ro\v{c}ek, 
``$N=1$ superspace geometry of extended supergravity,'' Nucl. Phys. B \textbf{243}, 221 (1984). 

\bibitem{KPR}
S.~M.~Kuzenko, M.~Ponds and E.~S.~N. Raptakis,
``New locally (super)conformal gauge models in Bach-flat backgrounds,''
JHEP \textbf{2008}, 068 (2020) [arXiv:2005.08657 [hep-th]].

\end{thebibliography}
\end{document}